\documentclass[twocolumn]{aastex63}
\hypersetup{breaklinks}
\usepackage{comment}
\usepackage{amsmath,amstext}

\newcommand{\caii}{\ion{Ca}{2}}
\newcommand{\caiihk}{\ion{Ca}{2} H \& K}

\newcommand{\rphk}{\ensuremath{R'_{\rm HK}}}
\newcommand{\lrphk}{\ensuremath{\log{\rphk}}} 
\newcommand{\halpha}{\ensuremath{\mbox{H}\alpha}}

\newcommand{\teff}{\ensuremath{T_{\mbox{\scriptsize eff}}}}
\newcommand{\logg}{\ensuremath{\log g}}
\newcommand{\msun}{\ensuremath{\mbox{M}_{\odot}}}

\newcommand{\prot}{\ensuremath{P_{\mbox{\scriptsize rot}}}}

\newcommand{\kms}{\ensuremath{\mbox{km s}^{-1}}}
\newcommand{\mps}{\ensuremath{\mbox{m s}^{-1}}}

\newcommand{\soren}{S\o ren Meibom}

\newcommand{\mas}{\ensuremath{\mbox{mas yr}^{-1}}}

\newcommand{\ith}{\ensuremath{^{\rm th}}}
\newcommand{\nd}{\ensuremath{^{\rm nd}}}
\newcommand{\gbr}{\ensuremath{(G_{\rm BP} - G_{\rm RP})}}
\def\amin{\ifmmode^{\prime}\else$^{\prime}$\fi}
\newcommand{\lapprox }{{\lower0.8ex\hbox{$\buildrel <\over\sim$}}}
\newcommand{\gapprox }{{\lower0.8ex\hbox{$\buildrel >\over\sim$}}}

\received{April 10, 2020}
\accepted{September 28, 2020}
\shorttitle{\sc{Advancing Gyrochronology with Ruprecht~147}}
\shortauthors{Curtis et al.}

\begin{document}

\title{\sc{When Do Stalled Stars Resume Spinning Down?\\Advancing Gyrochronology with Ruprecht~147}}

\newcommand{\ames}{NASA Ames Research Center, Moffett Field, CA 94035, USA}
\newcommand{\aarhus}{Stellar Astrophysics Centre, Department of Physics and Astronomy, Aarhus University, \\ Ny Munkegade 120, DK-8000 Aarhus C, Denmark}
\newcommand{\amnh}{Department of Astrophysics, American Museum of Natural History, Central Park West, New York, NY, USA}
\newcommand{\berkeley}{Astronomy Department, University of California, Berkeley, CA, USA}
\newcommand{\cahill}{Cahill Center for Astrophysics, California Institute of Technology, Pasadena, CA, USA}
\newcommand{\caltech}{Department of Astronomy, California Institute of Technology,
Pasadena, CA, USA}
\newcommand{\cehw}{Center for Exoplanets and Habitable Worlds, Department of Astronomy \& Astrophysics, 
    The Pennsylvania State University, \\ 
    525 Davey Laboratory, University Park, PA 16802, USA}
\newcommand{\cfa}{Center for Astrophysics $\vert$ Harvard \& Smithsonian, 60 Garden Street, Cambridge, MA 02138, USA}
\newcommand{\chicago}{Department of Astronomy and Astrophysics, University of Chicago, 
  5640 S. Ellis Ave., Chicago, IL 60637, USA}
\newcommand{\columbia}{Department of Astronomy, Columbia University, 550 West 120\ith\ Street, New York, NY 10027, USA}
\newcommand{\dunlap}{Dunlap Institute for Astronomy and Astrophysics, University of Toronto, 50 St. George Street, Toronto, Ontario M5S 3H4, Canada}
\newcommand{\exeter}{University of Exeter, Department of Physics \& Astronomy, Stocker Road, Exeter, EX4 4QL, UK}
\newcommand{\hawaii}{Institute for Astronomy, University of Hawai`i, 2680 Woodlawn Drive, Honolulu, HI 96822, USA}
\newcommand{\flatiron}{Center for Computational Astrophysics, Flatiron Institute, 162 5\ith\ Avenue, Manhattan, NY, USA}
\newcommand{\sac}{Stellar Astrophysics Centre, Department of Physics and Astronomy, Aarhus University, \\ Ny Munkegade 120, DK-8000 Aarhus C, Denmark}
\newcommand{\sagan}{NASA Sagan Fellow}
\newcommand{\seti}{SETI Institute, 189 Bernardo Avenue, Mountain View, CA 94043, USA}
\newcommand{\sydney}{Sydney Institute for Astronomy (SIfA), School of Physics, University of Sydney, NSW 2006, Australia}
\newcommand{\texas}{Department of Astronomy, The University of Texas at Austin, Austin, TX 78712, USA}
\newcommand{\unc}{Department of Physics and Astronomy, University of North Carolina, Chapel Hill, NC 27599, USA}
\newcommand{\westwash}{Department of Physics \& Astronomy, Western Washington University, Bellingham, WA 98225-9164, USA}
\newcommand{\yale}{Department of Astronomy, Yale University, 52 Hillhouse Avenue, New Haven, CT 06511, USA}

\correspondingauthor{Jason Lee Curtis}
\email{jasoncurtis.astro@gmail.com}
\author[0000-0002-2792-134X]{Jason Lee Curtis}
\affiliation{\columbia}
\affiliation{\amnh}

\author[0000-0001-7077-3664]{Marcel A.~Ag\"{u}eros}
\affiliation{\columbia}

\author[0000-0001-9590-2274]{Sean P.~Matt}
\affiliation{\exeter}

\author[0000-0001-6914-7797]{Kevin R.~Covey}
\affiliation{\westwash}

\author[0000-0001-7371-2832]{Stephanie T.~Douglas}
\affiliation{\cfa}

\author[0000-0003-4540-5661]{Ruth Angus}
\affiliation{\amnh}
\affiliation{\flatiron}
\affiliation{\columbia}

\author{Steven H.~Saar}
\affiliation{\cfa}

\author[0000-0002-3656-6706]{Ann Marie Cody}
\affiliation{Bay Area Environmental Research Institute, 625 2\nd\ Street, Ste. 209, Petaluma, CA 94952, USA}

\author[0000-0001-7246-5438]{Andrew Vanderburg}
\affiliation{\texas}
\affiliation{\cfa}

\author[0000-0001-9380-6457]{Nicholas M. Law}
\affiliation{\unc}

\author[0000-0001-9811-568X]{Adam L.~Kraus}
\affiliation{\texas}

\author[0000-0001-9911-7388]{David W.~Latham}
\affiliation{\cfa}


\author[0000-0002-1917-9157]{Christoph Baranec}
\affiliation{Institute for Astronomy, University of Hawai`i at M\={a}noa, 640 N. A`oh\={o}k\={u} Pl., Hilo, HI 96720-2700, USA}

\author[0000-0002-0387-370X]{Reed Riddle}
\affiliation{Division of Physics, Mathematics, and Astronomy, California Institute of Technology, Pasadena, CA 91125, USA}

\author[0000-0002-0619-7639]{Carl Ziegler}
\affiliation{\dunlap}

\author[0000-0001-9214-5642]{Mikkel N.~Lund}
\affiliation{\aarhus}

\author[0000-0002-5286-0251]{Guillermo Torres}
\affiliation{\cfa}


\author{\soren}
\affiliation{\cfa}

\author[0000-0002-6137-903X]{Victor Silva Aguirre}
\affiliation{\aarhus}

\author[0000-0001-6160-5888]{Jason T.~Wright}
\affiliation{Penn State Extraterrestrial Intelligence Center / Center for Exoplanets and Habitable Worlds /  Department of Astronomy \& Astrophysics, 525 Davey Laboratory, The Pennsylvania State University, University Park, PA, 16802, USA}



\begin{abstract}
Recent measurements of rotation periods (\prot) in the benchmark open clusters 
Praesepe (670~Myr), NGC~6811 (1~Gyr), and NGC~752 (1.4~Gyr)  demonstrate that, after converging onto a tight sequence 
of slowly rotating stars in mass--period space, stars 
temporarily stop spinning down. These data also show that the duration of this epoch of stalled spin-down increases toward lower masses. 
To determine when stalled stars resume spinning down, 
we use data from the \textit{K2} mission and the Palomar Transient Factory to measure \prot\ 
for 58 dwarf members of the 2.7-Gyr-old cluster Ruprecht~147, 
39 of which satisfy our criteria designed to remove short-period or near-equal-mass binaries. 
Combined with the \textit{Kepler} \prot\ data for the approximately coeval cluster NGC~6819 (30 stars with $M_\star > 0.85$~\msun), 
our new measurements more than double the number of $\approx$2.5~Gyr benchmark rotators
and extend this sample down to $\approx$0.55~\msun. 
The slowly rotating sequence for this joint sample appears relatively flat ($22$$\pm$$2$ days)
compared to sequences for younger clusters.
This sequence also intersects the \textit{Kepler} intermediate period gap, demonstrating that this gap was not created by a lull in star formation. 
We calculate the time at which stars resume spinning down, 
and find that 0.55~\msun\ stars remain stalled for at least 1.3~Gyr. To accurately age-date low-mass stars in the field, gyrochronology formulae must be modified to account for 
this stalling timescale. 
Empirically tuning a core--envelope coupling model with open cluster data 
can account for most of the apparent stalling effect. However, alternative explanations, e.g., a temporary reduction in the magnetic braking torque, cannot yet be ruled out.
\end{abstract}



\section{Introduction} \label{s:intro}

Observations of Sun-like stars revealed that they spin down over time via magnetic braking.
This led to the development of gyrochronology, a promising age-dating technique that uses rotation periods (\prot) as a clock \citep{Barnes2003}.\footnote{We define Sun-like broadly, to include all stars with radiative cores and convective envelopes; i.e., $0.4 \lesssim M_\star \lesssim 1.3$~\msun, or spectral types early M to late F. These stars have solar-like dynamos and undergo magnetic braking.}
Recent measurements of \prot\ in the benchmark open clusters  
Praesepe \citep[670~Myr;][]{Douglas2017}, 
the Hyades \citep[730~Myr;][]{Douglas2019}, 
NGC~6811 \citep[1.0~Gyr;][]{Curtis2019}, and NGC~752 \citep[1.4~Gyr;][]{Agueros2018}, however,
have shown that the formula describing the process of stellar spin-down cannot be as simple as it once appeared. 
Instead of \prot\ evolving continuously as a power law with a braking index that is constant in time and common to all stars 
(i.e., $\prot \propto t^n$, with \citeauthor{skumanich1972}~\citeyear{skumanich1972} originally proposing $n = 0.5$), 
it is now clear that, after converging onto a slowly rotating sequence that illustrates the tight relationship between mass and \prot\ in clusters older than $\approx$100~Myr, stars temporarily stop spinning down. Furthermore, based on where the period sequences for these benchmark clusters overlap, 
\citet{Agueros2018} concluded that 
the duration of this epoch of stalled spin-down increases toward lower stellar masses.

Most empirical gyrochronology relations \citep[e.g.,][]{barnes2007, Barnes2010} 
and angular momentum evolution models \citep[e.g.,][]{vanSaders2013, Matt2015, Gallet2015, Lanzafame2015}, 
which predate these new observational findings, 
do not account for the phenomenon of stalling,
so ages inferred for low-mass stars with these models will likely be incorrect.
It is therefore imperative that we extend the benchmark sample of cluster rotators to lower masses and older ages. This will allow us to constrain the timescale for stalled spin-down and to eventually repair empirical gyrochronology models, so that accurate ages can be inferred for low-mass stars in the field.
These new data are also needed to tune free parameters in the more physically motivated class of gyrochronology models 
\citep[for a recent example using our new \prot\ data for NGC~6811 
from \citeauthor{Curtis2019}~\citeyear{Curtis2019}, see][]{Spada2019}.

\begin{figure}\begin{center}
\includegraphics[trim=1.0cm 0.5cm 0.0cm 0.7cm, clip=True,  width=3.4in]{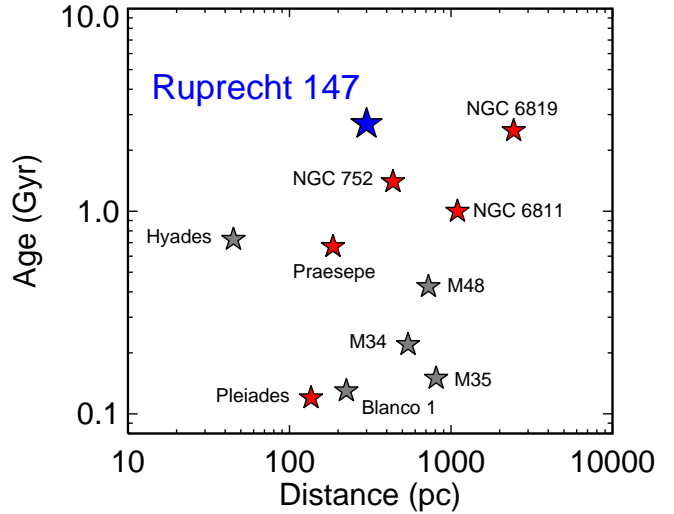}
  \caption{Age vs.~distance for a selection of benchmark open clusters with \prot\ data. Red stars indicate  clusters used in this study; gray stars represent 
  other notable clusters with \prot\ data
  mentioned in this paper. Ruprecht~147 and NGC~6819 have similar ages, but Ruprecht~147 is much closer to Earth and its stars are $\approx$77$\times$ brighter. This allows us to measure \prot\ for Ruprecht~147 stars 
  with much lower masses than was possible with the \textit{Kepler} survey of NGC~6819.
   \label{f:family}}
\end{center}\end{figure}

We present measurements of \prot\ for the 2.7-Gyr-old open cluster Ruprecht~147, the oldest nearby open cluster (see Figure~\ref{f:family}).
In Section~\ref{s:147}, we discuss the properties of and membership criteria for this cluster, 
and then identify binaries among, and estimate stellar parameters for, the targets of this study.
In Section~\ref{s:rot}, we analyze photometric time series data from NASA's \textit{K2} mission \citep{Howell2014}
and the Palomar Transient Factory \citep[PTF;][]{nick2009,rau2009}, and measure \prot\ for 58 dwarf members with masses as low as $M_\star \approx 0.55$~\msun. 
In Section~\ref{s:sample}, we identify 35 of these as benchmark rotators, 
including 23 likely single stars,
discuss the impact of binarity on the color--period distribution,
and present a rotation catalog for Ruprecht~147.

Our sample of rotators in Ruprecht 147 contains few stars with masses $\gapprox$1~\msun. 
To remedy this, we combine our sample with \prot\ measurements for the approximately coeval (2.5~Gyr) cluster 
NGC~6819, which was surveyed during the primary \textit{Kepler} mission \citep{Meibom2015}.
Because stars in NGC~6819 were monitored 10-16$\times$ longer than those in Ruprecht 147, 
and because the \textit{Kepler} light curves contain negligible systematics compared to our \textit{K2} Campaign~7 data, 
\citet{Meibom2015}
were able to recover weaker amplitude signals in the NGC~6819 data than we could detect in the Ruprecht~147 light curves.
This enabled \citet{Meibom2015} to measure \prot\ for early G dwarfs in NGC~6819, 
although the cluster's larger distance modulus 
restricted their overall \prot\ sample to $M_\star > 0.85$~\msun. 
In this study, we contribute rotation periods for 20 benchmark rotators with masses smaller than this lower limit for NGC 6819.

Pairing the rotation data for NGC~6819 and Ruprecht~147 allows us to describe stellar rotation at 2.5-2.7~Gyr as a function of mass from late F down to M1 spectral types.
Section~\ref{s:25} describes this procedure, which involves calculating the average interstellar reddening 
toward NGC~6819 relative to Ruprecht~147, and estimating their relative ages with gyrochronology 
using the portions of each sample that overlap significantly in mass (i.e., mid-to-late G dwarfs).

The slowly rotating sequence for this joint sample appears relatively flat 
compared to sequences for younger clusters.
In Section~\ref{s:dis}, we use this sample to determine when stars resume spinning down after enduring the temporary stalling epoch. We find that the lowest mass stars in our sample resumed spinning down only 700~Myr ago. 
This is consistent with the new mass-dependent core--envelope coupling timescale derived by \citet{Spada2019}. 
We also note that the Ruprecht 147 \prot\ sequence intersects the \textit{Kepler} intermediate period gap
\citep{AmyKepler}; 
i.e., the color dependence of the gap is different than the color dependence of rotation periods at any one age.
This demonstrates that the gap was not created by a temporary lull in the star formation rate 600~Myr ago.

We conclude in Section~\ref{s:concl}.

\section{Properties, Membership, Multiplicity, and Stellar Parameters for Ruprecht~147} \label{s:147}

\begin{figure*}[!t]
\begin{center}
\includegraphics[trim=1.1cm 0.3cm 0.0cm 0.4cm, clip=True,  width=3.45in]{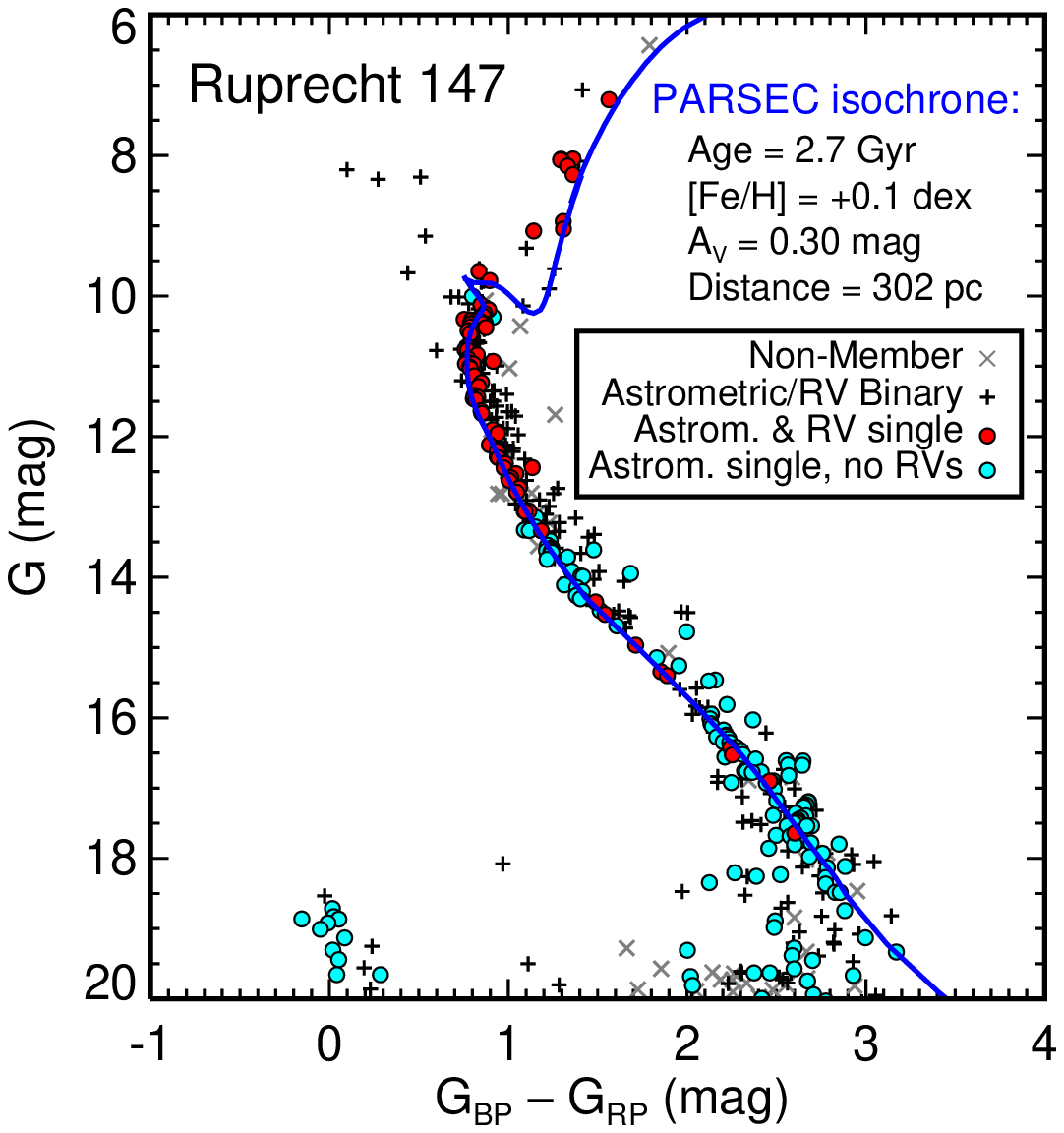}
\includegraphics[trim=1.1cm 0.3cm 0.0cm 0.4cm, clip=True,  width=3.45in]{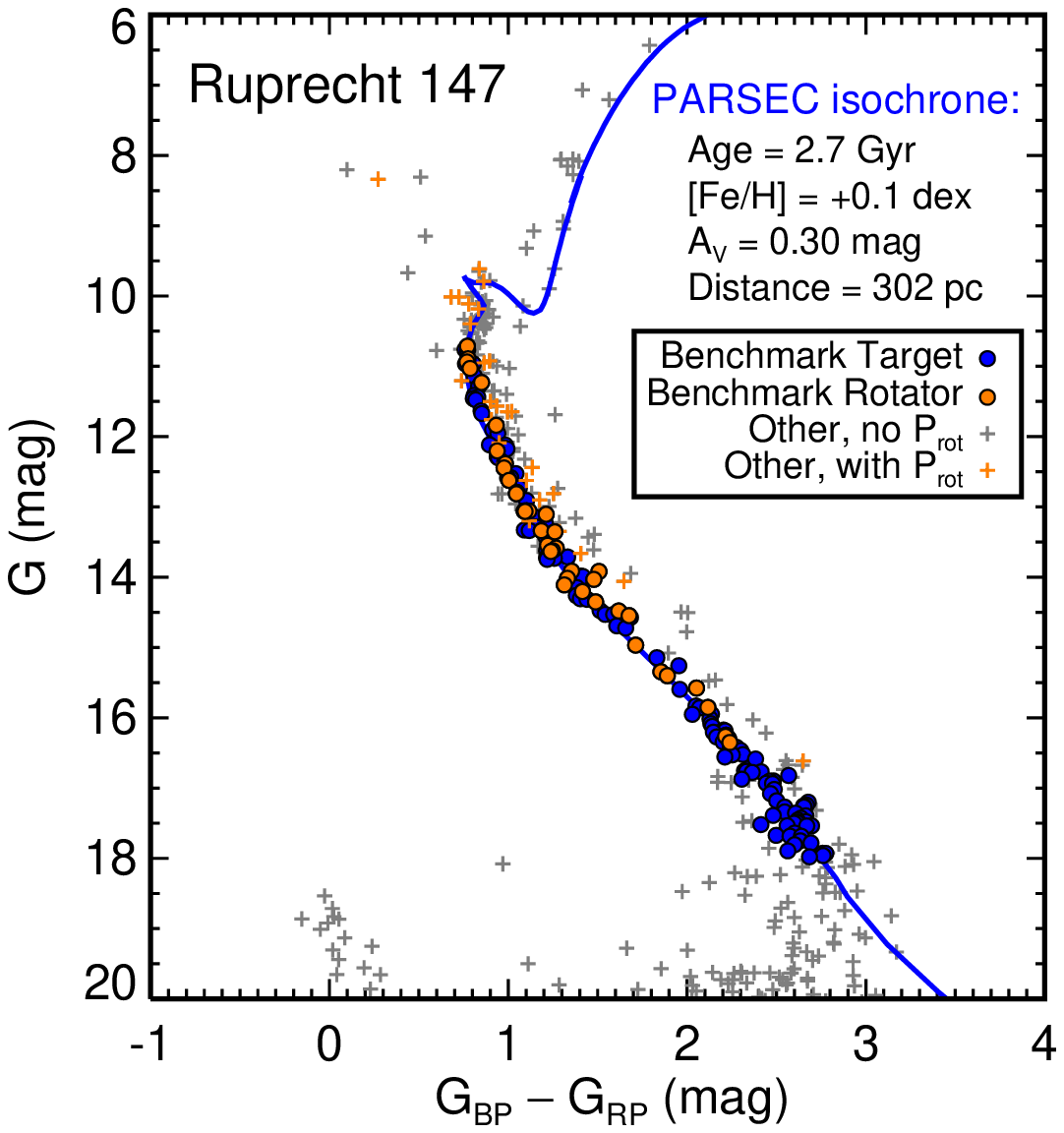}
  \caption{\textit{Gaia}~DR2 CMDs for Ruprecht~147. 
    (\textit{Left}) Stars with astrometry and RVs consistent with single-star membership are highlighted in red (77 stars).
    Cyan points indicate astrometrically-single stars lacking RVs (145 stars). 
    Stars with parallaxes and/or RVs in violation of our membership criteria are marked with ``$\times$'' symbols (43 stars). 
    All other stars marked with ``$+$'' symbols are assumed to be either astrometric and/or RV binaries (175 stars).
    The PARSEC isochrone solution we adopt is overlaid: 
    2.7~Gyr, [Fe/H] = +0.10~dex, $A_V = 0.30$, $(m - M)_0 = 7.40$.
  (\textit{Right}) This CMD highlights the gyrochronology benchmark targets, which are photometrically-single dwarfs (10.5$<G<$18) that are not EBs, SB2s, or short-period SB1s (161 blue points).
  This includes 39 benchmark dwarfs with measured rotation periods (orange points). Also shown are non-benchmark stars (gray pluses, 211 stars) and non-benchmark rotators (orange pluses, 29 stars).
  \label{f:cmd}}
\end{center}\end{figure*}

\subsection{Fundamental cluster properties}

We adopt an age of 2.7$\pm$0.2~Gyr for Ruprecht 147, 
based on the analysis of the masses and radii for three eclipsing binary systems \citep[EBs;][]{Torres2018, Torres2019, Torres2020}.
This value is consistent with ages obtained from fitting isochrones to color--magnitude diagrams \citep[CMDs;][]{Curtis2013, Curtis2016PhD}.

Based on spectra for six single solar twins observed with the MIKE spectrograph \citep{MIKE} on the 6.5-m Magellan Clay Telescope at Las Campanas Observatory and with the High Resolution Echelle Spectrometer \citep[HIRES;][]{HIRES} on the 10-m Keck telescope, which we analyzed with Spectroscopy Made Easy \citep[SME;][]{valenti2005, Brewer2015}, we adopt a metallicity for the cluster of [Fe/H] = $+0.10\pm0.03$~dex \citep[see Table 5.4 in][]{Curtis2016PhD}. This is consistent with all other analyses in the literature \citep{redgiants, Curtis2013, Bragaglia2018, Casamiquela2019}.

We calculated the average interstellar reddening by comparing the \textit{Gaia} DR2 colors,
\gbr,\footnote{We applied extinction coefficients 
calculated by the PARSEC isochrone service \citep{parsec} using the 
\citet{DR2phot2} passbands: \url{http://stev.oapd.inaf.it/cgi-bin/cmd}. $A_G \approx 0.86 \, A_V, E(G_{\rm BP} -G_{\rm RP}) \approx 0.415 \, A_V$.} 
with spectroscopic temperatures (\teff) derived from high-resolution spectroscopy 
using SME 
for nearby, unreddened field stars \citep{valenti2005} and the same six solar twin members of Ruprecht~147 used to measure the cluster metallicity.
We found a reddening value of $E(B-V) = 0.099 \pm 0.010$, or equivalently an extinction value of 
$A_V = 0.31 \pm 0.03$
assuming a standard $R_V = A_V / E(B-V) = 3.1$ relationship.
This new value is consistent with CMD isochrone fitting 
\citep[$A_V = 0.30$ in Figure~\ref{f:cmd} in this work; 
$A_V = 0.25$$\pm$$0.05$ from][]{Curtis2013} and
EB analyses \citep[$A_V = 0.35$$\pm$$0.09$;][]{Torres2018}. For this work, we adopt $A_V = 0.30$.

We adopt a distance modulus of $(m - M)_0 = 7.40$ ($d \approx 302$~pc).\footnote{$(m - M)_0$ refers to the un-reddened 
distance modulus representing only the physical distance. When a photometric band is listed in the subscript, 
this indicates the total difference between the apparent and absolute magnitude; i.e., 
$(m - M)_V = A_V + 5\,\log_{10}(d/10)$, where $d$ is in pc.} 
This is consistent with the CMD isochrone fitting \citep[$(m - M)_0 = 7.35$$\pm$$0.1$;][]{Curtis2013} and 
EB results \citep[$(m - M)_0 = 7.26$$\pm$$0.13$;][]{Torres2018}.
The cluster parallax from the second \textit{Gaia} data release \citep[DR2;][]{GaiaDR2} is $\varpi = 3.2516$$\pm$$0.0038$
\citep{DR2HRD}; 
if the DR2 parallaxes are indeed systematically biased toward smaller values \citep[e.g.,][]{DR2astrom,KeivanWillie2018,Zinn2018,Sahlholdt2018},
the parallax would increase to between 3.28 and 3.33 mas, 
corresponding to $(m - M)_0 = 7.42$ or 7.39 mag, respectively.
This bias does not alter the age or interstellar reddening values, which 
were determined independently of distance.

\subsection{Cluster membership} \label{s:mem}

Our pre-\textit{Gaia} target list included $>$1000 candidates 
based on proper motions from PPMXL \citep{ppmxl} and catalogs produced by the United States Naval Observatory \citep[e.g., NOMAD, UCAC2;][]{nomad, UCAC2}, 
and on photometry from 
CFHT/MegaCam \citep[$g', r', i'$;][]{Curtis2013, Curtis2016PhD}, 
UKIRT/WFCAM \citep[$J, K$;][]{Curtis2016PhD}, 
and the Two Micron All Sky Survey \citep[2MASS $J, H, K_S$;][]{2MASS}.
From this preliminary candidate list, \citet{Curtis2016PhD} identified 150 stars with radial velocities (RVs) consistent with membership.\footnote{We have measured RVs for cluster candidates using the Hamilton echelle spectrometer on the 120-in Shane telescope at Lick Observatory \citep{Hamilton},
the East-Arm Echelle \citep{EastArm} on the Hale 200-in at Palomar Observatory,
the Hectochelle multiobject spectrograph \citep{hectochelle, gabor} on the 6.5-m telescope at MMT Observatory \citep{mmt},
the MIKE and Magellan Echellette  \citep[MagE;][]{MAGE} spectrographs on the  6.5-m Clay Telescope at Las Campanas Observatory, and the Tillinghast Reflector Echelle Spectrograph \citep[TRES;][]{andytres} on the 1.5-m Tillinghast telescope at Fred Lawrence Whipple Observatory \citep[see Chapters~2.2 and 3.3.2;][]{Curtis2016PhD}. 
We supplemented  these data with archival RVs from the High Accuracy Radial velocity Planet Searcher \citep[HARPS;][]{HARPS} on the 3.6-m telescope at La Silla Observatory 
\citep[PI Minniti; Run IDs 091.C-0471(A) and 095.C-0947(A), accessible from the public archive by][]{HARPSarchive2020}, 
and from the \textit{Gaia} Radial Velocity Spectrometer \citep{GaiaRVS}. \label{foot:rv}}

With \textit{Gaia} DR2, we now also have high-precision astrometry and photometry for  stars reaching down to $G \approx 20$ mag.
We expanded our candidate list by merging \textit{Gaia}-based membership lists from the literature, including 
234 stars from the \citet{DR2HRD}, 
191 stars from \citet{CG2018}, and
259 stars from \citet{Olivares2019}. 

However, even with \textit{Gaia} data,  constructing a complete cluster catalog is not straightforward, 
as binaries can severely bias the {\it Gaia} astrometry. 
Indeed, in rare cases,
unresolved binaries can cause the astrometric and photometric solutions to fail entirely, 
preventing a star's inclusion in \textit{Gaia}~DR2.
For example, the \citet{Curtis2013} cluster member CWW~87\footnote{Also known as EPIC~219659980,
2MASS J19160785$-$1610360, and NOMAD 0738-0795586. PPMXL proper motions and our RVs support its membership of the cluster.}
does not appear in \textit{Gaia} DR2. A high-resolution Robo-AO \citep{roboAO} image for this star shows a companion at 0\farcs42 with a SDSS $i$-band contrast of $\Delta i = 0.57$$\pm$$0.05$ 
\citep[see figure~3.10 in][]{Curtis2016PhD},\footnote{We acquired high-spatial-resolution imaging for 130 cluster members and candidates with Robo-AO in 2013 while it was on the Palomar 60-in telescope
\citep[Chapters 2.6 and 3.3.1 of][]{Curtis2016PhD}.} 
which we suspect is responsible for this star's exclusion. 
Such cases are why assembling a complete membership catalog for Ruprecht 147 is beyond the scope of this paper.

\begin{deluxetable*}{lllcl}
\tabletypesize{\scriptsize}
\tablecaption{Description of data in the Ruprecht 147 catalog. \label{t:prot}}
\tablewidth{0pt}
\tablehead{
\colhead{Column} & \colhead{Format} & \colhead{Units} & \colhead{Example} & \colhead{Description}
}
\startdata
\sidehead{\textit{Identifiers:}} 
DR2Name & string & \nodata & 4183944182414043136 & \textit{Gaia} DR2 Source ID \\
Twomass & string & \nodata & J19172940$-$1611577 & 2MASS Source ID \\
EPIC    & string & \nodata & 219651610 & \textit{K2} EPIC ID \\
CWW     & string & \nodata & 108 & ID from \citet{Curtis2013} \\
NOMAD   & string & \nodata & 0738-0797617 & NOMAD ID \citep{nomad} \\
\sidehead{\textit{Astrometry:}} 
RA      & double & degrees & 289.37254 & Right ascension in decimal degrees \\
Dec     & double & degrees & $-$16.199489 & Declination in decimal degrees \\
pmra    & float  & \mas    & $-0.946$ & Right ascension proper motion \\
pmde    & float  & \mas    & $-$26.775 & Declination proper motion \\
epm     & float  & \mas    & 0.071 & Proper motion error \\
astrom\_source & string  & \nodata & Gaia DR2 & Source of astrometry (Gaia DR2 or PPMXL) \\
plx     & float  & mas     & 3.3023 & Parallax \\
eplx    & float  & mas     & 0.0304 & Parallax error \\
epsi    & float  & \nodata & 0.000 & Astrometric excess noise ($\epsilon_i$) \\
sepsi    & float  & \nodata & 0.000 & Significance of astrometric excess noise ($D$)\\
ruwe    & float  & \nodata & 1.093 & Re-normalised Unit-Weight Error \\
\sidehead{\textit{Photometry:}} 
Gmag    & float  & mag     & 14.3257 & \textit{Gaia} DR2 $G$ magnitude \\
bp\_rp  & float  & mag     & 1.47600 & \textit{Gaia} DR2 color: \gbr \\
e\_br & float  & mag     & 0.004 & \textit{Gaia} DR2 Photometric error: $\sqrt(\sigma_{\rm BP}^2+\sigma_{\rm RP}^2)$ \\
Jmag    & float  & mag     & 12.474 & 2MASS $J$ magnitude \\
Kmag    & float  & mag     & 11.767 & 2MASS $K_S$ magnitude \\
e\_jk & float  & mag     & 0.031 & 2MASS Photometric error: $\sqrt(\sigma_{\rm J}^2+\sigma_{\rm K}^2)$ \\
\sidehead{\textit{Radial Velocities:}} 
RV\_Gaia        & float   & \kms     & \nodata & \textit{Gaia} DR2 radial velocity \\
e\_RV\_Gaia     & float   & \kms     & \nodata & \textit{Gaia} DR2 radial velocity error \\
RV\_R147Project & float   & \kms     & 42.15   & Median radial velocity from non-Gaia data \\
nRV             & integer & \nodata  & 5       & Number of non-Gaia RV measurements \\
rv\_ei        & float   & \nodata  & 1.7     & RV variability to error ratio \citep[$e/i$;][]{Geller2008}\\
dRV             & float   & \kms     & 0.6     & RV deviation from cluster  \\
RVdelT          & float   & days     & 1886    & RV epoch baseline \\
\sidehead{\textit{Membership and Binarity:}} 
Member         & string  & \nodata  & Yes   & 1$^{\rm st}$ character: ``Y'' (Yes), ``P'' (Probable), ``N'' (Non-Member) \\
Photo\_Binary  & string  & \nodata  & No    & if $|{\rm dcmd}|>0.4$ mag then ``Yes'', else ``No'' \\
Wide\_Binary   & string  & \nodata  & No    & ``Yes'' if (pmd$>$2 \mas) or (AO=Yes) or (RUWE$>$1.4), else ``No'' \\
Spec\_Binary   & string  & \nodata  & No    & SB2, SB1-Short, SB1-Long, No, N/A (see comment for criteria) \\
rad            & float   & degrees  & 0.7   & Radial RA/Dec coordinate distance  \\
drad           & float   & pc       & 5.0   & 3D distance from cluster center \\
pmd            & float   & \mas     & 0.1   & Proper motion deviation from cluster \\
dplx           & float   & mas      & 0.05  & Parallax deviation from cluster \\
dcmd           & float   & mag      & 0.04  & Photometric excess in $G$ \\      
AO             & string  & \nodata  & No    & Robo-AO detection? (Yes, No, N/A)  \\
\sidehead{\textit{Stellar Properties:}} 
Teff   & integer & K       & 4525 & Effective temperature \\
Mass   & float   & \msun   & 0.74 & Mass \\
SpT    & string  & \nodata & K4 & Spectral type \\
\sidehead{\textit{Rotation data:}} 
K2\_PTF\_Data & string & \nodata & S\_Y & Observed by \textit{K2}? (S=superstamp, A=aperture, N=No) PTF? (Y/N) \\
Prot       & float  & days    & 20.4     & Rotation period. Negative values indicate \prot\ is not trusted.  \\
Prot\_Source    & string & \nodata & Both & Light curve used to measure \prot: PTF, K2, or both \\
sigma\_LC  & float  & ppt     &  4.8     & Photometric noise for light curve \\
Rvar\_LC   & float  & ppt     & 20.3     & Photometric amplitude for light curve\\
Near\_mag      & float   & mag      & 18.3 & $G$ magnitude of brightest neighbor within 12$''$ \\
Near\_rad      & float   & arcsec   & 7.0 & Distance to brightest neighbor within 12$''$ \\
Benchmark & string & \nodata & Yes    & No/Yes/Yes-Rapid\_Outlier/Yes-Prot\_Secondary? \\
Notes    & string & \nodata & \nodata & Notes on \textit{K2} light curve or the target\\
\enddata
\tablecomments{The table is available for download in the online journal. To reproduce our benchmark rotator sample, query ``Prot'' $>$ 0 and ``Benchmark'' = ``Yes''.}
\end{deluxetable*}

Our final list of cluster members contains 440 stars.\footnote{Two of these stars lack five-parameter astrometric solutions 
from \textit{Gaia}, so we adopt the PPMXL proper motions for them.
Nine stars lack \textit{Gaia} color, \gbr, 
but they were not observed with \textit{K2} and rotation periods were not detected with PTF, so 
this does not affect our study.} 
This list is provided as a machine-readable table in the online journal, and its contents are described in Table~\ref{t:prot}.
The CMD for this list is displayed in Figure~\ref{f:cmd}.
Because only Sun-like stars experience magnetic braking, we also limit our list of stars of interest to those with $G > 10.5$, corresponding to stars with  
$M_\star \lesssim 1.4$~\msun.
Of these, 258 are main-sequence dwarfs with $G$~$<$~18~mag; 
the remainder are white dwarfs, red giants, blue stragglers, or dwarfs too faint for us to measure their \prot. 
Below we assess the binarity/multiplicity of each of these stars.

\subsection{Stellar multiplicity} \label{s:bin}

\subsubsection{Spectroscopic binaries}\label{s:sb1}
We identified seven EBs and 23 double-lined spectroscopic binaries 
\citep[SB2s; see Chapters 3.1.4.4 and 4 in][]{Curtis2016PhD}. Because these systems have short orbital periods, 
the rotational evolution of these stars can be affected by tidal interactions, 
limiting their use in constraining gyrochronology \citep{Meibom2005, Douglas2019}.
Furthermore, if absorption lines are detectable for multiple stars in an optical spectrum (thus identifying it as an SB2), then the light curve will also certainly be sensitive to the binary components as well. For both reasons, we reject SB2s and EBs from our ``benchmark sample''---those stars which appear to be single, or effectively single like long-period, high-contrast binaries.

We identify single-lined spectroscopic binaries (SB1) by observing 
variability in RV measurements.
Following the methodology developed by the WIYN Open Cluster Study
\citep{Geller2008},
we calculate the $e/i$ statistic (called ``rv\_ei'' in Table~\ref{t:prot}), 
where $e$ represents the 
variance in the RV data set 
and $i$ is the expected measurement 
precision, 
and flag 13 stars with 
$e/i > 4$ as SB1s (not counting SB2s or EBs). 
There are an additional six stars with \textit{Gaia} RV errors greater than 4~\kms, 
which we assume are due to RV variations, 
and we therefore classified these as SB1s as well.

Candidate long-period binaries can be identified by calculating 
the absolute RV deviation from the cluster median, 
and flagging those greater than some threshold value 
based on the expected intrinsic cluster dispersion 
($\approx$0.5~\kms) and the 
measurement precision.
We initially adopted a larger 5~\kms\ value and found six RV outliers: 
each only has RVs from \textit{Gaia}, 
and given the relatively 
low RV errors (0.2-3.4~\kms) from multiple measurements (2-10 each), 
and relatively large RV deviations (7-107~\kms), 
these are likely non-members, and are discussed further in  Section~\ref{s:nm}.

Table~\ref{t:prot} lists spectroscopic binary classifications in the ``spec\_binary'' column, and are assigned in this order: 
``N/A'' if no RVs are available; 
``RV-NM'' if the median RV is systematically offset from R147 by ``dRV''$>$10~\kms;
``SB1-Long'' if either the RVs exhibit long-term trends, 
  have measured orbital periods $>$100~days, 
  or  5$<$``dRV''$<$10; 
  while satisfying astrometric and photometric membership criteria;
``SB1-Short'' if ``rv\_ei''$>$4 or ``e\_RV\_Gaia''$>$4~\kms;   
``SB2'' according to the spectrum and cross-correlation function shapes; 
else ``No'' if RV data indicate the star is likely single. 
See Appendix~\ref{a:othernotes} for two exceptions.

\subsubsection{Astrometric binaries} \label{s:ab}
The \textit{Gaia} DR2 proper motions are precise enough 
that certain binaries can be identified by a star's 
moderate deviation from the cluster's average value.
If the cluster's internal velocity dispersion is 
$\sigma_V = 0.5$~\kms, at 300~pc this would 
be equal to 0.35~\mas.
The \textit{Gaia} proper motion error does increase
toward fainter magnitudes,
but this is insignificant over the brightness range of the 
rotator sample we construct in the next section
(e.g., the error for the faintest rotator 
in our sample is only $\sigma_\mu = 0.16$~\mas), 
and so we ignore this.
Since we do not know the internal velocity dispersion yet, 
we conservatively 
flag stars that deviate from the cluster by 
$\Delta \mu > 2$~\mas\ as astrometric binary candidates, 
based on the distribution of $\Delta \mu$ for our target list. 

Excess astrometric noise can indicate if a source deviates significantly from the single-star model used to derive the astrometric parameters for the \textit{Gaia} DR2 catalog. 
Candidate wide binaries can therefore be identified by selecting sources with poor astrometric solutions. 
The renormalized unit weight error (RUWE)\footnote{RUWE values were downloaded from \url{http://gaia.ari.uni-heidelberg.de/singlesource.html}}
accounts for the strong dependencies of the astrometric noise on color and magnitude.\footnote{For a description of RUWE, see \url{http://www.rssd.esa.int/doc_fetch.php?id=3757412}}
Single stars with good astrometric solutions should have RUWE $\approx$1.
We classify 32 dwarfs with RUWE $>1.4$ as 
candidate binaries.
Many of these also appear as photometric binaries in the CMD, but this is not required as even high-mass-ratio companions can impart measurable astrometric perturbations.

\subsubsection{Visual binaries resolved with adaptive optics} \label{s:ao}

\begin{deluxetable}{llccl}
\tablecaption{Cluster gyrochronology targets with companions detected in our Robo-AO observations \label{t:ao}}
\tabletypesize{\scriptsize}
\tablewidth{0pt}
\tablehead{
\colhead{EPIC ID} & \colhead{NOMAD ID} & \colhead{RUWE} & \colhead{Ang.~Sep.} &  \colhead{Contrast} 
}
\startdata
219633753 & 0737$-$0795762 & 4.4  & 0\farcs7 & $\Delta m =2.7$ mag \\
219678096 & 0738$-$0796304 & 0.9  & 1\farcs6 & $\Delta i =0.0$ mag \\
219664556 & 0738$-$0795948 & 24.6 & 0\farcs2 & $\Delta i =0.6$ mag \\
219777155 & 0741$-$0796328 & 2.4  & 0\farcs5 & $\Delta i =2.4$ mag \\
219366731 & 0730$-$0979511 & 79.5 & 0\farcs6 & $\Delta m =1.6$ mag
\enddata
\tablecomments{The NOMAD IDs \citep{nomad} are useful for looking up the Robo-AO images presented in figures 3.9 and 3.10 in \citet{Curtis2016PhD}.
The third column provides the \textit{Gaia} DR2 Renormalised Unit Weight Error (RUWE), which should be valued near 1 for single stars; 
we flag stars with RUWE $>1.4$ as candidate wide binaries.
The last column gives the difference between the magnitude of the detected neighbor and of the target, where $i$ is the SDSS filter and $m$ denotes the Robo-AO 600~nm long pass filter.
For reference, 0\farcs5 at 300~pc projects to 150~AU.}
\end{deluxetable}

We have observed 130 cluster stars with Robo-AO \citep{roboAO}, of which 50 are dwarfs less massive than the 1.4~\msun\ cutoff for this study.
These data were analyzed following \citet{Ziegler2018}.
Companions/neighbors were detected for the five targets in Table~\ref{t:ao} 
\citep[see also Chapters 2.6 and 3.3.1 of][]{Curtis2016PhD}.

The four stars with neighbors within $1''$ all have RUWE $>$2,
confirming that this parameter is useful for 
identifying unresolved wide binaries in \textit{Gaia} DR2.
Both components for the nearly equal mass wide binary EPIC~219678096 
were resolved in \textit{Gaia} DR2, and have separate entries in that and our catalog.

\subsubsection{Photometric binaries} \label{s:pb}
Binaries can also appear brighter than the single-star main sequence.
We use a 2.7~Gyr PARSEC isochrone with [Fe/H] = +0.1~dex, 
$A_V = 0.30$, and ($m - M)_0 = 7.4$ to measure the $G$-band excess in the 
\gbr\ vs. $G$ CMD (``dcmd'' in Table~\ref{t:prot}). We also inspect \gbr\ vs. $M_G$ and $(G \, - \,$2MASS\;$K_S)$  vs. $G$ CMDs.
If a star is within 0.4 mag of the single-star sequence in at least one of these diagrams, it is classified as photometrically single. 
We flag 70 stars as photometric binaries
from our list of 258 dwarf targets. 
Identifying photometric binaries/multiples is important because, like SB2s, the spot modulation signals from both/all stars are 
blended in the light curves, and those signals will interfere with each other and can confuse the periodicity analysis.
In such situations, even if multiple periods can be distinguished, we cannot confidently assign each to the appropriate 
binary component.

\subsubsection{Non-Members} \label{s:nm}

Ruprecht 147 likely has tidal tails and a diffuse halo containing many additional members that are currently dispersing into the Galaxy
\citep{R147dissolve}. \citet{Kounkel2019} identified
candidates eight degrees away from the cluster center. Focusing on their 178 brighter stars  
with precise and reliable astrometry, 
70\% are within 20~pc of the cluster center, 90\% are contained to 85~pc, and the most distant is at 490~pc. Out of the 85 stars with \textit{Gaia} RVs with errors under 4~\kms\ (to filter out short-period SB1s), 70 share Galactic $UVW$ velocities to within $<5$~\kms\ of our Ruprecht 147 membership, reaching out to 54~pc. 
Most do not yet have RVs, 
which will be even more critical to corroborate the membership of such evaporated low-mass stars than in the vicinity of the core where proper motions alone are often sufficient.
It is therefore premature to incorporate them into our analysis, and we focus our present study on those that we can place the highest confidence in membership. 
In the future, 
especially after subsequent \textit{Gaia} data releases provide even higher precision astrometry, and RVs for fainter stars, we will revisit the topic of the evaporating membership and examine their magnetic activity and rotational behavior relative to the distribution found for our bona fide single-star members.

For this reason, stars with reliable parallaxes (RUWE$<$1.4) indicating that they are $>$100~pc away from the core are classified as non-members.
Some stars have high-quality 
five-parameter astrometric solutions (coordinates, proper motions, parallaxes), 
and have RVs from either Gaia and/or our database.
For those which are also not already classified as SB2s or short-period SB1s, 
we calculate Galactic $UVW$ velocities, and re-classify stars with large discrepancies as non-members ($\Delta UVW > 10$~\kms). Those with differential values ranging between 5-10~\kms\ are classified as possible members and candidate binaries, irrespective of their distance.

For example, Gaia DR2 4089241111304212992 (\gbr\ = 1.25 and $G = 11.66$),
appears to be a photometric binary or even triple in the CMD. However, with $\varpi = 2.5$~mas, 
it is 100~pc away from the core. Although the proper motion and RVs suggest membership, 
transforming these into Galactic $UVW$ reveals its 3D space motion is 13~\kms\ discrepant from R147. We therefore re-classify it as a non-member. 

In Table~\ref{t:prot}, the first character of the ``member'' column will read either ``Y'' (yes, these are considered members; 395 stars), ``P'' (these are probably members; 2 stars), or ``N'' (no, these are not likely members; 43 stars). 

\subsection{Stellar properties} \label{s:stellarproperties}
We estimated \teff\ and spectral type for members of Ruprecht 147 following the procedure described by \citet{Curtis2019}.
Specifically, we calculated \teff\ from the dereddened \textit{Gaia} DR2 color $\gbr_0$ using a 
color--temperature relation we constructed using nearby benchmark stars from 
\citet{Brewer2016}, \citet{Boyajian2012}, and \citet{Mann2015}; 
see Appendix~\ref{a:temp} for details.
We then interpolated the stellar properties table in \citet{kraus2007} to 
estimate spectral types from our photometric \teff\ values.

In \citet{Curtis2019}, we also estimated stellar masses by interpolating 
this same table. 
However, this procedure yields biased results for the high-mass end of the 
Ruprecht~147 sample,
where $M_\star > 1.2$~\msun\ stars have 
evolved substantially away from the zero-age main sequence toward cooler temperatures. 
For such stars, color inaccurately biases 
estimated masses toward lower values.
For this study, we instead use a PARSEC isochrone appropriate for Ruprecht~147 (2.7 Gyr, [Fe/H] = +0.10 dex, $A_V = 0.30$, $(m - M)_0 = 7.4$) 
to estimate masses from \textit{Gaia} photometry.


Figure~\ref{f:cmd} presents the \textit{Gaia}~DR2 CMD for the Ruprecht~147 members (440 stars) 
and highlights stars with astrometry and RVs consistent with single-star membership (222 stars).
Approximately half of our membership list are candidate binaries, which is consistent with the stellar multiplicity seen in the solar neighborhood \citep{MultiReview, Raghavan2010}.
The figure also  includes a version of the CMD that highlights our gyrochronology benchmark targets (161 stars), 
which are photometrically single dwarfs (10.5$<G<$18), excluding EBs, SB2s, and short-period SB1s.
In Section~\ref{s:sample}, we report rotation periods for 40 of these benchmark stars, which are also highlighted in this CMD.
\section{Measuring  Rotation in Ruprecht~147} \label{s:rot}

\subsection{Measuring \prot\ with \textit{K2} light curves}
Our team
petitioned to adjust the pointing for \textit{K2}'s Campaign 7 so that it covered Ruprecht~147, 
which we then proposed to monitor 
(GO proposal 7035).\footnote{\url{https://keplerscience.arc.nasa.gov/data/k2-programs/GO7035.txt}} 
Our GO program was allocated 1086 individual apertures for candidate members. A series of contiguous apertures, a ``superstamp,'' was created to tile the inner cluster core in response to a different proposal, and covered 96 additional candidates from our preliminary membership list \citep{K2SUPERSTAMP}.

Our target list was designed in 2015 to maximize completeness in anticipation of the high-precision {\it Gaia} astrometry and photometry that would become available three years later. For this reason, a large number of objects on the GO~7035 target list are now known to be non-members.

Of our 258 potential gyrochronology targets,
105 have \textit{K2} data: 70 stars were allocated individual apertures, and 35 are in the superstamp. 
Fifty-one other members of Ruprecht~147 also have \textit{K2} data,
including evolved stars, blue stragglers, and members with $G$~$>$~18~mag.

All targets that were allocated individual apertures have light curves 
produced by the \textit{K2} team with the 
Pre-search Data Conditioning Simple Aperture Photometry pipeline 
\citep[PDCSAP;][]{pdcsap1, pdcsap2}.\footnote{\url{http://keplerscience.arc.nasa.gov/pipeline.html}}\textsuperscript{,}\footnote{Our preliminary 
results were based on \textit{K2} Data Release 9, 
whereas this manuscript uses Data Release 36:
\url{https://keplerscience.arc.nasa.gov/k2-data-release-notes.html\#k2-campaign-7}}
These targets also have light curves that were produced by community-created pipelines, 
including 
\texttt{EVEREST} \citep[EPIC Variability
Extraction and Removal for Exoplanet Science Targets;][]{EVEREST1,EVEREST2} 
and \texttt{K2SFF}  \citep{k2andrew}.

We produced light curves for the superstamp targets using a moving aperture procedure with a 2-pixel radius circular aperture \citep[see also][]{Rebull2018}.
These superstamp light curves show a common systematic: stars tend to brighten over the course of the Campaign. 
We median-combined the normalized light curves for our 
superstamp targets, and then divided out this common systematic from each light curve. 
A typical example is provided in Appendix~\ref{a:amc} to illustrate this procedure.
The set of light curves is available for download.

With these high-precision light curves,
we discovered the sub-Neptune K2-231\,b transiting a solar twin \citep[EPIC 219800881;][]{PlanetR147},
the warm brown dwarf CWW~89\,Ab transiting another solar twin \citep[EPIC 219388192;][]{BDposter2016, Beatty2018},\footnote{This system was independently discovered  
by \citet{Nowak2017}.}
and six EBs \citep[][]{Curtis2016PhD}, 
three of which have been precisely characterized 
\citep[EPICs 219394517, 219568666, and 219552514;][]{Torres2018, Torres2019, Torres2020}.

We visually inspected all of the  pipeline-generated light curves available for every target.\footnote{Only cadences with \texttt{SAP\_QUALITY = 0} are used.}
We did not detect spot-modulation variability in light curves for targets fainter than 
$G = 16.5$ mag. These light curves suffered from low $S/N$, so it is possible that the spot modulation amplitudes are too weak in comparison to the photometric noise (see Appendix~\ref{a:noise}). It is also possible that the \prot\ for these stars are too long compared to the duration of Campaign 7.

For the 151 non-WD stars with \textit{K2} data, 
we computed auto-correlation functions \citep[ACF; e.g.,][]{AmyKepler} and 
Lomb--Scargle  periodograms \citep[LS;][]{Scargle1982, press1989}. 
For stars that clearly show rotational modulation, 
we also measured \prot\ by fitting the timing of successive local maxima sets and minima sets (e.g., see the discussion of EPIC 219333882 in Appendix~\ref{a:timing})
This final, visual method is important for correcting cases where the automated analyses detected half-period harmonics. It also allows us to identify data quality problems or, e.g., light curves affected by significant spot evolution midway through the Campaign. In most cases, the three methods yielded \prot\ 
consistent to within 10\% (after doubling the \prot\ for 
those cases where the LS periodogram favored the half-period harmonic). 

This approach is similar to the one we employed for NGC~6811 using \textit{Kepler} data \citep{Curtis2019}. The key difference is that for Ruprecht~147, we prefer the ACF \prot\ to the LS \prot. The ACF more accurately recovers \prot\ in the spot modulation patterns, which more often double dip at the longer \prot\ found in Ruprecht 147 stars compared to stars in younger clusters which more often exhibit sinusoidal modulation patterns 
\citep{Basri2018}. 

We measured preliminary periods for 68 stars, including 59 of the 105 main-sequence dwarfs ($G>10.5$, ignoring red giants, blue stragglers, and stars at the main sequence turnoff). 
Light curves for 11 other stars showed variability consistent with spot modulation, but we were unable to unambiguously assign a \prot\ because the signal was confused by data quality problems, 
by interference with a neighbor blended in the \textit{K2} pixels, and/or by spot evolution.
We do not report periods for the six EBs, as these are being 
analyzed separately \citep{Curtis2016PhD, Torres2018, Torres2019, Torres2020}.

Figure~\ref{f:k2} shows the results of our light curve analysis for one mid-K dwarf, EPIC~219651610; similar figures for all non-WD candidate members with \textit{K2} data are provided in the online journal (151 stars). Table~\ref{t:prot} presents data for the entire 440-star membership catalog, including results for all \textit{K2} targets.
We estimate a typical \prot\ uncertainty of 10\% for all targets with reported values; see Appendix~\ref{a:eprot} for details.

See \citet{GB20} for an independent analysis of the \textit{K2} data for Ruprecht~147 stars allocated individual apertures. 

\subsection{Measuring \prot\ with light curves from the Palomar Transient Factory}

\begin{figure*}\begin{center}
\includegraphics[trim=0.0cm 0cm 0.0cm 0cm, clip=True,  width=6.5in]{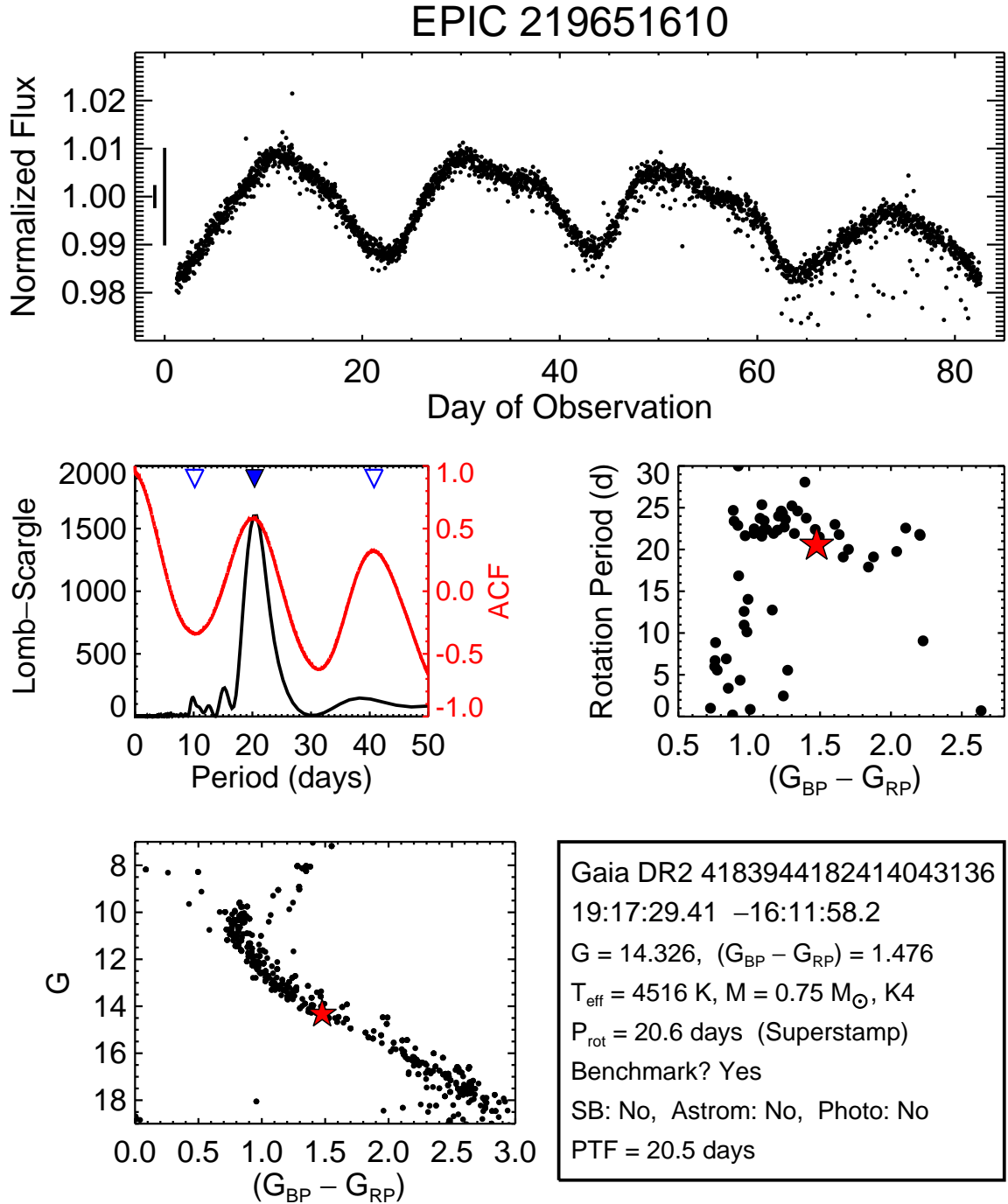}
  \caption{Analysis of \textit{K2} data 
  for EPIC~219651610 (\textit{Gaia}~DR2 4183944182414043136,
  CWW~108, NOMAD~0738-0797617), 
  a mid-K dwarf 
  ($\teff = 4516$~K, 
  $M_\star = 0.75$~\msun)
  member of Ruprecht~147.
  \textit{Top---}\textit{K2} light curve from the superstamp.  Two vertical lines at left mark the photometric precision (``sigma\_LC'') and amplitude (``Rvar\_LC''), respectively.
  \textit{Middle left---}Periodicity is measured with 
  the Lomb--Scargle periodogram 
  (left axis, black line) and autocorrelation function 
  (right axis, red line). 
  At top, the solid blue triangle marks the adopted period, 
  and the open triangles mark the half and double values.
  \textit{Middle right---}\prot\ versus \gbr\
  for Ruprecht~147 (black points), along with the target star (red star).
  \textit{Bottom left---}The \textit{Gaia} DR2 CMD for Ruprecht~147 and the 
  target star (red star) is used to check for binary photometric excess.
  We also queried \textit{Gaia}~DR2 for objects within 12$''$ of the target ($\approx$3 pixels), and plot the apparent (cyan) and absolute 
  (blue) magnitudes of any neighbors (none found near this target). 
  This is useful for assessing whether blends could be responsible for the 
  apparent rotation signal seen in the light curve. 
    \textit{Bottom right---}This information panel includes useful data for the target.
  Similar plots for all candidate members with
  \textit{K2} data are available in the online journal (151 total).
   \label{f:k2}}
\end{center}\end{figure*}

\begin{figure*}\begin{center}
\includegraphics[trim=0.0cm 0cm 0.0cm 0cm, clip=True,  width=6.5in]{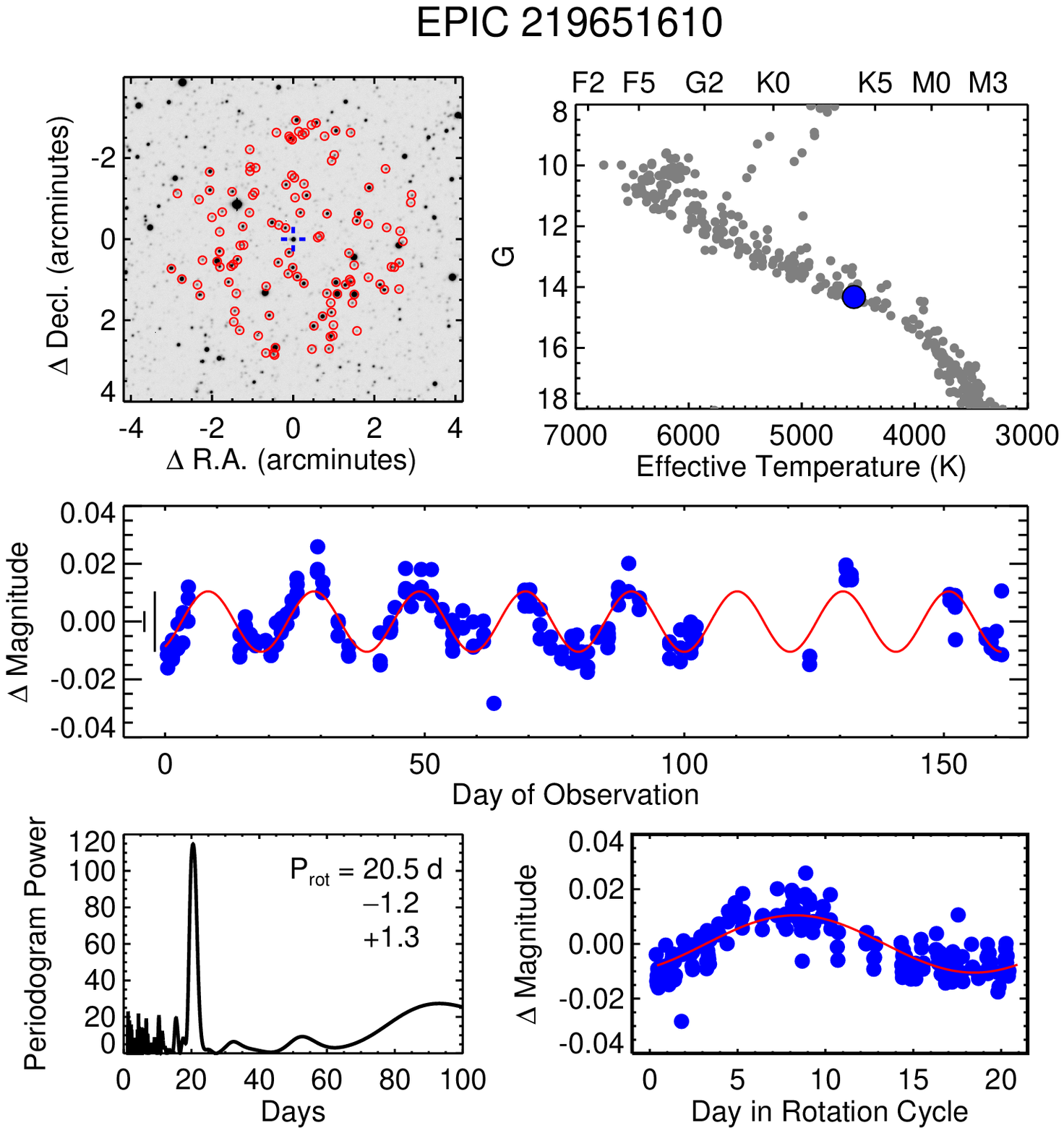}
  \caption{
Analysis of the PTF data for EPIC~219651610, the  star presented in Figure~\ref{f:k2}.  
  \textit{Top left---}PTF 
  image of a $8\amin \times 8\amin$
  region centered on the target, marked with blue cross-hairs. 
  Reference stars used to 
  calibrate the light curves 
  are highlighted with red circles
  ($\Delta G \lesssim 1.5$~mag).
  \textit{Top right---}Temperature 
  versus \textit{Gaia}~DR2 apparent $G$ magnitude for Ruprecht~147, 
  with the target star highlighted in blue.
  The \textit{Gaia} astrometry, 
  the RVs collected by our team, 
  and the proximity to the 
  single-star main sequence 
  collectively indicate that this star 
  is likely a single member.
  \textit{Middle---}The PTF 
  light curve, extracted with 
  simple aperture photometry, 
  and corrected using a systematics 
  light curve generated with the reference 
  stars shown in the image above.
  The photometric modulation due 
  to rotating spots is apparent 
  in this light curve. A sine curve fit to these data is overlaid in red to help illustrate the periodicity. 
  Two vertical lines at left mark the photometric precision (``sigma\_LC'') and amplitude (``Rvar\_LC''), respectively.
  \textit{Bottom left---}The 
  Lomb--Scargle periodogram for 
  the target shows a 20.5 day rotation period. 
  The half-width at half-power values for the main peak provide estimates of the period uncertainty ($\lesssim$10\% typically). 
  \textit{Bottom right---}The phase-folded 
  light curve.
  Similar plots for EPIC~219189038, 219234791, 
   219297228,
  219333882, 218984438,
  and 219665690 are available in the online journal.
   \label{f:ptf}}
\end{center}\end{figure*}


We monitored Ruprecht 147 from 2012 Apr 29 to 2012 Oct 07 as part of the PTF Open Cluster Survey \citep[][]{PTFpraesepe, Covey2016, Kraus2017, Agueros2018}. 
This survey used the robotic 48-in Oschin (P48) telescope at Palomar Observatory, CA. The P48 was equipped with the modified CFH12K mosaic camera: 11 CCDs, 92 megapixels, 1$\arcsec$ sampling, and a 7.26 deg$^2$ field-of-view \citep{Rahmer2008}. Under typical conditions (1$\farcs$1~seeing), it produced 2\arcsec~full-width half-maximum images with a 5$\sigma$ limiting $R_{\rm PTF} \approx21$ mag in 60~s \citep{Law2010}. 

Two slightly overlapping PTF fields, each 3.5$^\circ$~$\times$~2.3$^\circ$, covered the center of Ruprecht 147. The bulk of the cluster members later published in \citet{Curtis2013} fall in these two fields, centered at $\alpha$ = 19:07:43.4, $\delta$ = $-$16:52:30.0 and $\alpha$ = 19:21:58.8, $\delta$ = $-$16:52:30.0.
For most of the campaign, these fields were observed $\approx$three times a night, weather permitting. However, there were gaps in our coverage each month when PTF conducted its $g$-band and/or H$\alpha$ surveys. The result is a set of $R_{\rm PTF}$ light curves with $\approx$180 points unevenly spaced over the roughly five month baseline for our observations.    
We typically only kept data taken in the first $\approx$100 days 
due to long data gaps that followed. 
In the case shown in Figure~\ref{f:ptf}, 
this left us with 142 data points collected on 48 nights spread across 102 days.

We sought to improve the photometric precision for our targets' light curves by performing local differential photometry in the immediate vicinity of each target.
We downloaded the calibrated PTF images 
($8\amin \times 8\amin$ regions centered on each target)
from the NASA/IPAC 
Infrared Science Archive 
(IRSA).\footnote{\url{https://irsa.ipac.caltech.edu/applications/ptf/}}
We identified all sources within $\pm$1.5~mag of 
the target, and calculated simple aperture photometry for these stars on all images. We then subtracted off the median magnitude for each star and median combined the results for all stars to 
produce a light curve describing the systematic photometric zero-point. 
Finally, we subtracted this signal from the light curves 
for the target and all reference stars, 
and then computed LS periodograms for the full
sample
with periods ranging between 0.1 and 80 days.\footnote{Although we preferred ACF for the \textit{K2} light curves, that technique requires evenly sampled time series. In double-dip light curves where the LS returned the half-period harmonic, the true period was always represented at a somewhat weaker power. For the PTF analysis, we identified stars as rotators which had unambiguous, single-peaked LS periodograms so as to avoid the half-period harmonic issue; i.e., we trust our \prot\ measurements for stars with a single significant peak in the LS periodogram.} 
Some of the reference stars are themselves variables; 
phase-folded light curves for four examples are shown 
in Figure~\ref{f:ptf_variables} in Appendix~\ref{a:ptf} 
to illustrate the photometric precision and range of measurable \prot\ 
attainable with our procedure.

We applied this procedure to all stars on 
our list with $13 < G < 18$ mag (K to early M spectral types)
with $\Delta G < 0.5$ mag
and, where available, RVs consistent with being single-star members (RVs are available for 22 stars). 
Our list included 128 targets, 56 of which had usable data from PTF 
(62 stars were not observed in our two fields, and 10 others suffered other data quality problems
including proximity to bright neighbors or diffraction spikes).
Seven of our targets showed significant rotational modulation for which we report rotation periods, 
and nine others showed weak modulation. 

Figure~\ref{f:ptf} shows the results for EPIC~219651610, the mid-K dwarf observed with \textit{K2} and presented in Figure~\ref{f:k2} (similar figures for the other PTF rotators 
are available in the online journal).
Remarkably, we measured $\prot = 20.6$~days from \textit{K2} and 20.5~days from PTF,
despite the differences in photometric precision, cadence, and 
times of observation.
To estimate the \prot\ uncertainty, 
we calculate each half-width at half-maximum power of the primary peak in the periodogram to estimate upper and lower error bars, and find a typical value of 10\% for these stars ($\approx$2 days); see Appendix~\ref{a:eprot} for additional discussion.

\section{The Ruprecht 147 rotation catalog} 
\label{s:sample}

The Ruprecht 147 catalog, described in Table~\ref{t:prot}, 
includes identifiers from \textit{Gaia} DR2, 
the \textit{K2} Ecliptic Plane Input Catalog \citep[EPIC;][]{HuberEPIC}, 
2MASS, NOMAD, and the CWW numbers from \citet{Curtis2013}.
It also includes astrometry from \textit{Gaia} DR2 and proper motions from PPMXL for three stars lacking \textit{Gaia} data, 
photometry from \textit{Gaia} DR2 and 2MASS, RVs from \textit{Gaia} DR2 and our own database (see footnote~\ref{foot:rv}), 
stellar properties, and the results of our rotation period analyses.
We also provide information on membership and binarity, and list the $G$ magnitude and radial angular distance to the brightest neighbor within 12$''$ to assess possible contamination for \textit{K2} targets.

\subsection{Crafting a benchmark sample of rotators} \label{s:bench}

We measured preliminary periods for 72 stars, including 64 main-sequence dwarfs.
Seven of these \prot\ were measured using PTF, including four stars not observed by \textit{K2}.
However, this raw sample suffers from a variety of problems.

First, this list includes seven SB2s and seven short-period SB1s. These stars are susceptible to tidal interactions, 
which can alter the course of stellar angular momentum evolution
and keep stars rotating rapidly or even spin them back up depending on the circumstances.
For this reason, we remove such stars from our benchmark sample 
(14 stars; criteria were described in Section \ref{s:sb1}).

Second, we reject stars with excess luminosities from our sample because rotational modulation
from binary components can confuse the light curve analysis. 
In certain cases where the primaries are very inactive (e.g., late-F to early-G dwarfs), 
we are concerned that the rotational modulation apparent in some light curves might be solely 
attributable to more active lower-mass companions (e.g., EPIC 219404735 and EPIC 219442294; 
see also the discussion of EPIC 219661601 in Appendix~\ref{a:FK}; designated ``Benchmark'' = ``Yes-Prot\_Secondary?'' in Table~\ref{t:prot}). 
Photometric binaries are also problematic because the rotation period signals from both components 
can be visible in the light curve. We speculate that the spot modulation signals from both components of nearly equal mass binaries are interfering and confusing the periodicity analysis. Furthermore, even when the light curves present a clean, periodic pattern, 
it is impossible to reliably associate the period to the appropriate binary component. 
Therefore, it is imperative that we remove all such binary candidates 
from our sample, regardless of their light curve morphologies or rotation periods (13 stars total, including six of the SB2s and two of the short period SB1s; criteria were described in Section~\ref{s:pb}). 

Third, three stars show rotation periods that seem to be anomalously rapid for their mass and age. 
However, our \caiihk\ spectra (high resolution from MMT/Hectochelle and/or Magellan/MIKE, with high signal-to-noise ratios),
demonstrate that these stars have inactive chromospheres, 
thus invalidating the apparently rapid rotation seen in the \textit{K2} light curves 
For details, see Figure~\ref{f:ca} in Appendix~\ref{a:invalid}.

After removing such stars from our sample,
we are left with 39 dwarfs rotators. 
Of these, 26 satisfy our criteria for single-star membership; the remainder are wide binary candidates based on their astrometry 
(e.g., $\Delta \mu >$2~\mas\ and/or RUWE$>$1.4). 
The rejected stars are discussed in more detail in Appendix~\ref{a:valid} and their distribution in color--period space is shown in Figure~\ref{f:a:color-period_prelim}.

\begin{figure}[t]
\begin{center}
\includegraphics[trim=1.1cm 0.2cm 0.0cm 0.5cm, clip=True,  width=3.5in]{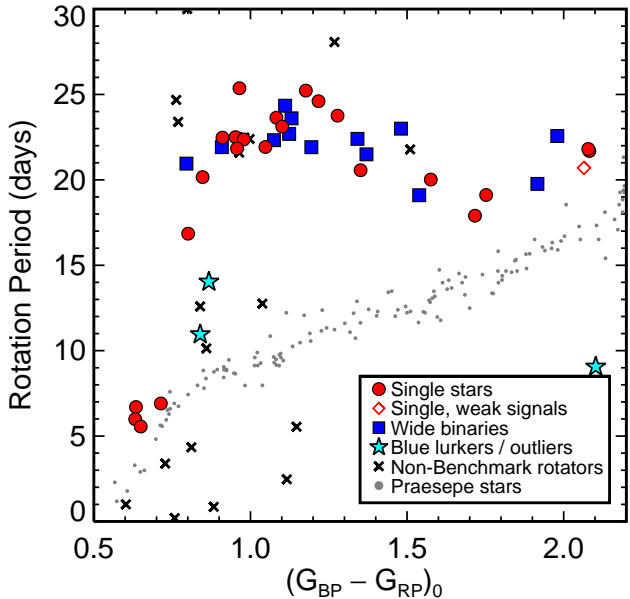}
\caption{Color--period distribution for benchmark stars in Ruprecht 147. 
The slowly rotating sequence for Ruprecht 147 appears remarkably flat relative to Praesepe's \citep[small gray points, 670 Myr;][]{Douglas2019} and other young clusters. 
Stars marked with ``$\times$'' symbols are removed from this sample because they 
have large photometric excesses or are short-period binaries (19 stars). 
Red circles mark single stars (according to photometry, astrometry, AO imaging where available, and RVs where available; 24 stars), 
while blue boxes mark stars ``effectively single'' wide binaries (single according to photometry and RVs where available, but have either proper motion deviations $\Delta \mu >$2~\mas\ and/or RUWE$>$1.4 indicating that they are likely wide binaries; 12 stars). 
One star with a weak signal is shown with an open symbol to indicate that its period is not counted as validated, 
and is caused by increased noise in its light curves relative to analogous cluster members.
Three single stars (marked with cyan five-point-star symbols) appear as rapid outliers relative to the main Ruprecht 147 trend. 
Two are solar twins, which we tentatively classify as candidate blue lurkers---blue stragglers embedded in the main sequence \citep{Leiner2019}. 
One is an M1 dwarf, which is rotating at twice the rate as Praesepe's slow sequence, suggesting to us that it might 
be a short period binary, a blue lurker, or is otherwise anomalous.
See Figure~\ref{f:a:color-period_prelim} in Appendix~\ref{a:valid} for details on the non-benchmark rotators and stars within invalidated periods.
\label{f:color-period}}
\end{center}\end{figure}

\subsection{The Ruprecht 147 \prot\ distribution} \label{s:protd}

Figure \ref{f:color-period} presents the color--period distribution for the benchmark rotators. 
In this figure, we differentiate likely-single stars from effectively-single stars using different symbols and colors (red circles versus blue squares). For this purpose, we define ``effectively-single'' as stars satisfying photometric and RV criteria, 
but which have astrometry suggesting that they are wide binaries. We expect that the primary stars in such binaries evolve as if they were single stars in isolation, and their light curves are unaffected by their companions.

Our benchmark rotator sample spans $0.62 < \gbr_0 < 2.09$, corresponding to $6350 > \teff > 3700$~K and $1.40 > M_\star > 0.55$~\msun. 
There are four warm, short period rotators, which are not expected to have spun down  significantly, as they have relatively thin convective envelopes and therefore weaker magnetic dynamos ($0.5 \lesssim \gbr_0 \lesssim 0.6$). 
Figure~\ref{f:color-period} also shows that most stars with $\gbr_0 > 0.8$ ($\teff < 5800$~K, $M_\star \lesssim 1$~\msun) 
congregate around $\prot \approx 22.5\pm1.6$~days, which appears to be the cluster's slowly rotating sequence. 
That sequence is relatively flat compared to rotation period sequences for younger clusters,
which tend to increase from relatively rapid G dwarfs to slower M dwarfs \citep[e.g., the figure shows that the single star rotator sequence for Praesepe ranges from $\approx$9 to 19 days over the same span in color;][]{Douglas2019}.

\subsubsection{Candidate Blue Lurkers} \label{s:bl}
Two rapid outliers appear to be single stars, according to all available astrometric, photometric, AO, and RV data. 
Furthermore, both stars, EPIC~219503117 (CWW~85, $\lrphk = -4.44$~dex) and  EPIC~219692101 (CWW~97, $\lrphk = -4.58$~dex), have anomalously high chromospheric emission, consistent with their rapid periods.
These stars are solar twins,
so we can compare them directly to the Sun, to solar twins in the field 
\citep{DiegoHK}, and analogous stars in other clusters (e.g., the fully-converged slow rotator sequence for Praesepe). 
According to all of these benchmarks, the behavior of these 
old, single, and rapidly rotating solar twins is anomalous. 

This is different behavior than found for 
EPIC~219388192
(CWW~89\,A), which \citet{BDposter2016} 
discovered harbors a warm, transiting brown dwarf and a distant M dwarf companion
\citep[see also][]{Nowak2017,Beatty2018}.
In that case, 
a 4~\kms\ discrepancy in its systemic RV and a NIR excess suggested the presence of a stellar companion, 
which was directly imaged with Keck/NIRC2. 
More critically, the \textit{K2} light curve revealed transits indicating a Jupiter-sized object, 
which the RV time series confirmed was a brown dwarf orbiting every 5.3 days. 
\citet{BDposter2016} speculated that the tidal interaction with the brown dwarf is responsible for that star's rapid rotation and overactive chromosphere.
In contrast, the high-precision RV time series for CWW~85 and CWW~97 from 
TRES and HARPS all but rule out the presence of tidally-interacting companions.

\citet{Leiner2019} found similar stars with anomalously rapid rotation 
in their survey of M67 (as well as rapidly rotating SB1s with long orbital periods where tides should be ineffective). 
They concluded that such stars are blue stragglers (i.e., stars that received a large influx of mass via accretion from or merger with a stellar companion). 
However, in contrast to the classic, higher mass blue stragglers that are found at warmer \teff\ beyond the main-sequence turnoff, 
these are embedded in the main sequence, and are referred to as ``blue lurkers'' by \citet{Leiner2019}.
Perhaps the two rapidly rotating and single solar twins we have identified in Ruprecht~147 are blue lurkers that were formed via mergers, leaving behind isolated blue stragglers with no other companions or remnants. Alternatively, perhaps they are indeed binaries with low mass companions and orbits oriented nearly in the plane of the sky.

Figure~\ref{f:color-period} shows a third rapid outlier---the M1 dwarf EPIC 219690421 ($\teff \approx 3740$~K, $M_\star \approx$0.54~\msun, $\gbr_0 = 2.10$), which is the faintest and reddest star in our rotation sample.
The \textit{K2} light curve shows a 9.1~day periodicity, 
and we detected an 8.8-day signal in the PTF light curve, albeit with a periodogram power below our quality threshold.
Analogous stars in Praesepe are rotating at 17.3~days.
Furthermore, the Praesepe sequence is tightly converged at this temperature, 
with only two outliers out of 23 stars within 100~K in the single-star sequence presented by \cite{Douglas2019}. 
This means EPIC~219690421 appears to rotate nearly twice as fast as analogous 670-Myr-old stars.
Based on the comparison with Praesepe, we do not consider this star to be a suitable benchmark for single-star rotation as it does not seem like it could be representative of 2.7-Gyr-old M1 dwarfs.
Unfortunately, we do not have any RV data to assess its binarity, 
nor an optical spectrum which we could use to diagnose enhanced chromospheric activity via \halpha\ emission.
It will be important to determine its membership and binarity with RV monitoring (and for the other stars currently lacking RVs for that matter) 
before we can hope to explain the cause for its rapid rotation 
(is it a tidally-interacting binary, a blue lurker, or have 0.55~\msun\ stars not fully converged yet?).
As this star has the lowest mass in our sample, ignoring it effectively refocuses this study on stars 
more massive than 0.55~\msun.

These three stars are designated as ``Benchmark'' = ``Yes-Rapid\_Outlier'' in Table~\ref{t:prot}.

\subsection{The role of binaries in shaping the color--period distribution}

Rapid outliers in color--period distributions are often short-period binaries
(see Figure~\ref{f:a:color-period_prelim} in Appendix~\ref{a:valid}), 
but not all binaries appear as outliers. 
Of the 39 rotators in our benchmark rotator sample, 
we classify 12 as wide binary candidates based on the \textit{Gaia} DR2 excess astrometric noise 
(i.e., RUWE $>$1.4) or large deviation in the proper motion ($\Delta \mu > 2$~\mas).
All of these candidate long period binaries are found on the slow sequence in Figure~\ref{f:color-period}. 
In addition, 
considering those stars we removed from our benchmark sample, 
two of the candidate short period binaries and two photometric binaries
are also found on the slow sequence. 

The components of long-period binaries should not be affected by 
gravitational tides (except maybe in rare cases where highly eccentric orbits result in close encounters), nor will the primary's photometric color necessarily be biased strongly by the secondary 
(assuming the primary hosts the rotation-period signal).
And any impact wide binaries might have on initial rotation rates, 
perhaps by prematurely dispersing circumstellar disks, 
will be largely erased through 2.7~Gyr of convergent spin-down.
For these reasons, we keep these twelve wide binary systems in our benchmark sample. 

We will continue to follow up on these binary, candidate binary, and likely single stars to better determine the properties of each system to study how binarity affects spin-down.

\subsection{Is there an undetected population of very slowly rotating K dwarf cluster members? No.}
\label{s:reservoir}


We reported periods for only 21 out of 64 benchmark stars with $0.50 \lesssim M_\star \lesssim 0.85$~\msun\ ($3700 \lesssim \teff < 5000$~K, $2.17 \gtrsim \gbr_0 \gtrsim 1.11$), or $\approx$33\% of the targets.
This includes 17 periods measured from \textit{K2} and four from PTF.
The periods for all range between 18 and 25 days, except for the rapid 9-day M1 dwarf noted earlier.
Given the relatively short duration of  Campaign~7, the presence of persistent systematics in the light curves, and expectations from standard gyrochronology models for much longer \prot\ (30-40~day), it is natural to wonder if we are missing a substantial number of longer period cool rotators. 
Unfortunately, only 19 of these benchmark targets were observed by \textit{K2}. 
We successfully measured \prot\ for 17 of these,
so the recovery rate is actually quite high at 89\%,\footnote{See Appendix~\ref{a:timing} and \ref{a:noise} for remarks on EPIC~219346771 and EPIC~219675090. The \prot\ for these stars are not immediately obvious from looking at the light curve; 
however, we measured their \prot\ by timing the arrivals of their minima, and consider our values to be accurate. 
Even if one rejects their \prot\ as inconclusive, the recovery rate for this sample would still be 79\%.}
and is on par with the rates for the \textit{K2} surveys of the Hyades and Praesepe clusters \citep[85-88\%;][]{Douglas2017, Douglas2019}, 
despite those clusters being only 25\% of the age of Ruprecht 147 ($\approx$700 Myr versus 2.7~Gyr).

Regarding the two non-detections, in Figure~\ref{f:a:amp_noise} in Appendix~\ref{a:noise} we examine the photometric noise and spot amplitudes for all targets, 
and we find that the light curve for EPIC~219616992 is 6.7$\times$ noisier than analogous targets, which is suppressing our sensitivity to the rotation signal. Still, we see evidence for a 20.7-day period in a smoothed version, which the LS analysis also picks up (this \prot\ is listed as a negative value in Table~\ref{t:prot} to distinguish it from stars with validated periods).
As for EPIC~219141523, the PDCSAP and EVEREST light curves show long-period variations that could be caused by rotating spots, but there is no obvious periodicity. The PTF light curve is noisier than those for stars with measured periods and we see no convincing period; the LS periodogram does shows a weak peak at 30 days. 

Unfortunately, Ruprecht 147 has not provided us with hundreds of rotators like the rich Pleiades and Praesepe clusters. The \textit{K2} surveys of the Pleiades and Praesepe have spoiled us, in a way, 
so that we expect large returns from these high-quality data. However, Ruprecht 147 is a very sparse cluster with a top-heavy mass function. It might be more fair to call it a cluster remnant, as it has clearly suffered from extreme dynamical evolution and mass loss
\citep{Curtis2016PhD, R147dissolve}.
Still, we are encouraged by the high rate of success for those 19 benchmark targets observed by \textit{K2}.

The two non-detection cases together with the rapid M1 dwarf are the three lowest mass stars in this particular sample.
Refocusing on those with $M_\star > 0.55$~\msun, we measured validated \prot\ for every benchmark target observed by \textit{K2}, 
and found that all share a common \prot\ to within a few days.

\section{Stellar Rotation at 2.7~Gyr with Ruprecht~147 and NGC~6819}                \label{s:25}

\begin{figure*}[t]\begin{center}
\includegraphics[trim=1.0cm 0cm 0.4cm 0cm, clip=True, width=3.4in]{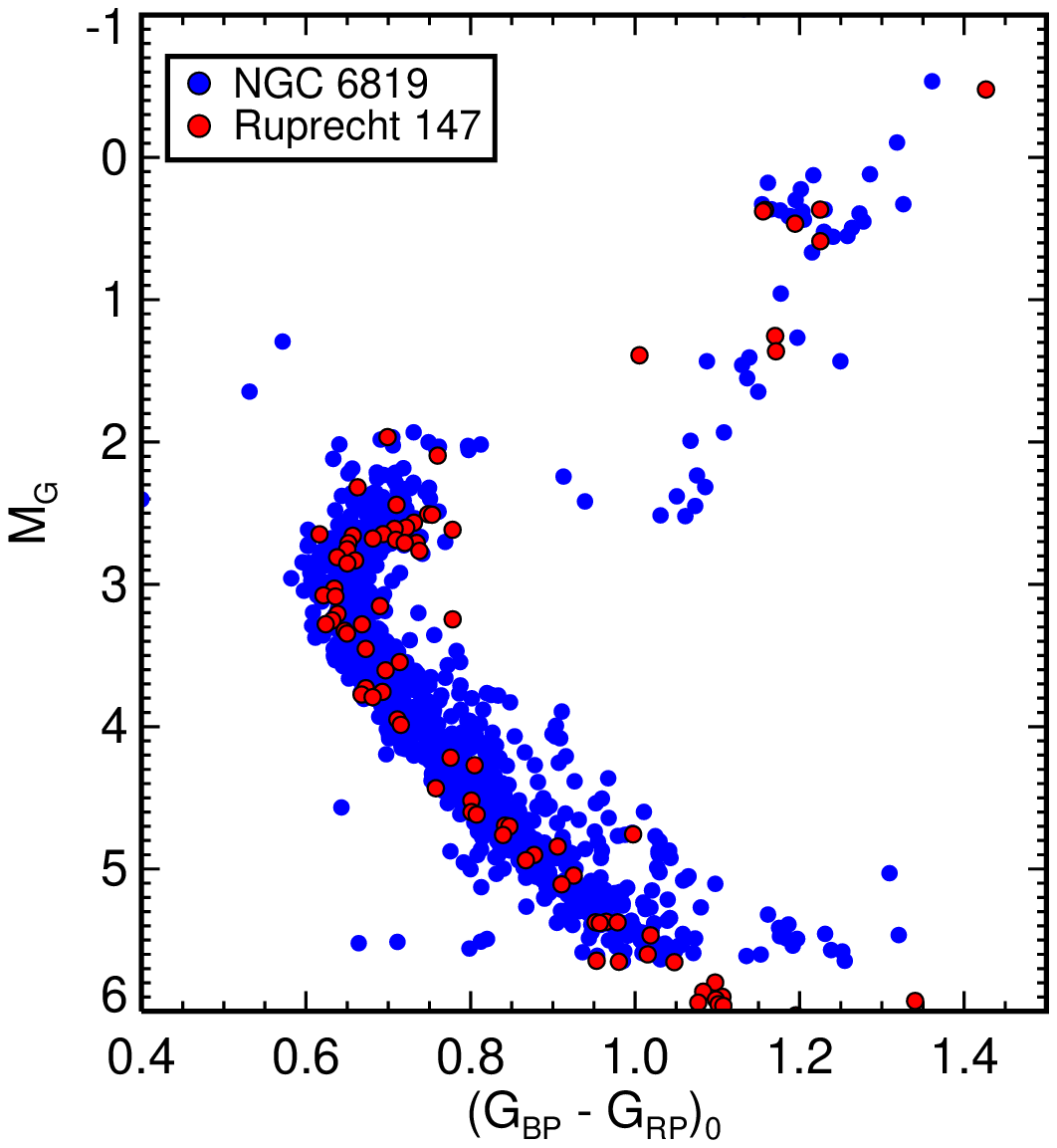}
\includegraphics[trim=1.0cm 0cm 0.4cm 0cm, clip=True, width=3.4in]{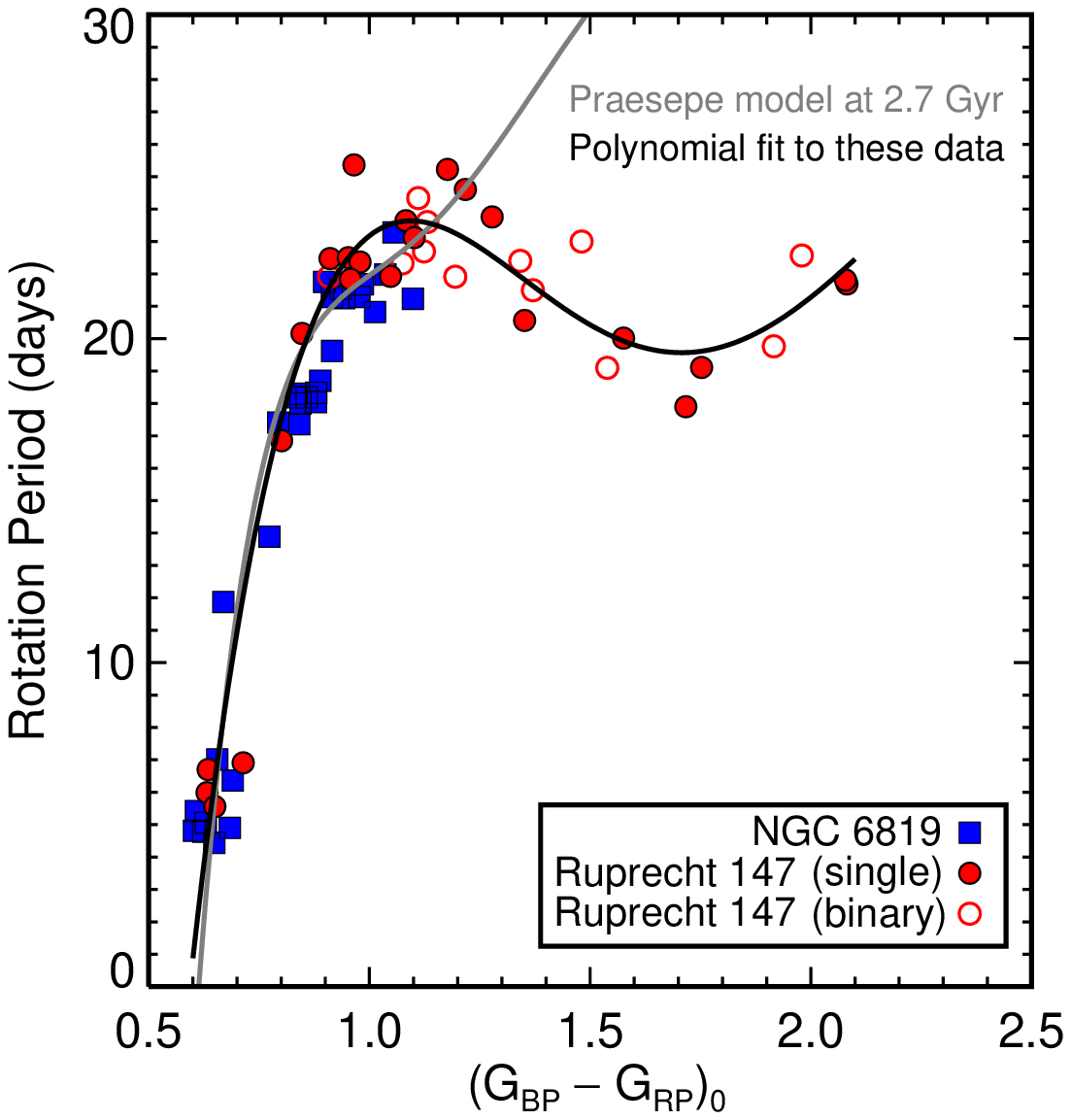}
  \caption{\textit{Left---}De-reddened \textit{Gaia} DR2 color versus absolute magnitude for the 2.5-Gyr-old NGC~6819 
  (blue points; members from \citeauthor{CG2018}~\citeyear{CG2018}, with spectroscopic binaries and RV non-members from 
  \citeauthor{Milliman2014}~\citeyear{Milliman2014} removed).
  and the 2.7-Gyr-old Ruprecht~147 (red points). 
  For Ruprecht~147, we applied $(m - M)_0 = 7.4$ and $A_V = 0.30$; 
  for NGC~6819, we calculated the values needed 
  to align each cluster's red clump stars (Table~\ref{t:clump}), 
  which should have approximately equal 
  absolute magnitudes in all photometric bands. 
  We find $\delta A_V$ = 0.14 and 
  $\delta (m - M)_0 = 4.57$, 
  corresponding to $A_V = 0.44$ and $(m - M)_0 = 11.97$ for NGC~6819.
  The main sequences and subgiant branches 
  also approximately align, 
  indicating that the clusters are
  approximately coeval. 
  \textit{Right---}Color--period diagram 
  for NGC~6819 \citep[blue squares;][]{Meibom2015} and 
  benchmark members of Ruprecht~147 (red circles), 
  divided into singles (filled) 
  and long-period binaries (open).
  The Ruprecht~147 sample extends 
  much redder than the NGC~6819 sample, 
  and appears remarkably flat compared 
  to expectations from standard 
  gyrochronology models, represented by 
  the gray line showing
  a polynomial fit to Praesepe that  
  is projected forward to 2.7~Gyr 
  with $n = 0.62$ 
  \citep[tuned with the Sun;][]{Douglas2019}.
  Also shown is a polynomial fit to these \prot\ data (black line); 
  while the 2.7-Gyr slow sequence is flat compared to past expectations, 
  this model emphasizes that the sequences in fact appears somewhat curved, 
  where the G dwarfs are spinning more slowly than the K dwarfs.
  The largest overlap between 
  the NGC~6819 and Ruprecht~147 rotation 
  sequences 
  occurs for 
  $0.91 < \gbr_0 < 1.09$
  ($5000 < \teff < 5500$~K): 
  the Ruprecht~147 stars in this range 
  are on average slower by $\approx$1~day, 
  corresponding to it being only 8\% older.
  This is consistent with the
  the 
  somewhat redder and fainter top 
  of the main sequence turnoff 
  in the left panel of this figure, 
  as well as the literature 
  ages for each cluster.
   \label{f:joint}}
\end{center}\end{figure*}

Ruprecht~147 is approximately coeval with the 2.5-Gyr-old open cluster NGC~6819, which was surveyed during the primary \textit{Kepler} mission
(a literature review of its fundamental properties is provided in Appendix~\ref{a:6819}).
\citet{Meibom2015} presented \prot\ for 30 NGC~6819 dwarfs with masses greater than $M_\star \gtrsim 0.85$~\msun. Unfortunately, because of the great distance to NGC 6819 ($\approx$2.5~kpc), lower-mass members remained inaccessible to \textit{Kepler}. As a result, our \prot\ sample for Ruprecht~147 extends to lower masses than does the \citet{Meibom2015} catalog, primarily because Ruprecht 147 stars are $\approx$77$\times$ brighter than analogous stars in NGC~6819 (see Figure~\ref{f:compare} in the appendix). 
However, the \textit{K2} light curves suffer increased systematics due to the increased pointing instability of the spacecraft.
This made it relatively more challenging to measure \prot\ for the inactive late-F and early G stars in Ruprecht~147 
compared to NGC~6819.

The result is that the NGC~6819 sample is primarily composed of F and G dwarfs, whereas the Ruprecht~147 sample is primarily mid-G to early-M dwarfs.
Fortunately, because the two clusters are approximately coeval, we can 
combine their data to form a $\approx$2.7~Gyr sample that covers a larger range in mass than is possible with just one of the two.

\subsection{Gyrochronology confirms that NGC~6819 and Ruprecht~147 are approximately coeval}

Before examining the rotation samples for Ruprecht~147 and NGC~6819, we must 
determine if their mean interstellar reddening values are determined consistently.
To do this, we make use of each cluster's red clump population, 
which should have nearly identical intrinsic luminosities. 
Differences in apparent photometric magnitudes can therefore be attributed 
to differences in distance and interstellar extinction and reddening.
Appendix~\ref{a:6819} describes our procedure:
using Ruprecht~147 as a reference, we find 
$(m - M)_0 = 11.97$ and $A_V = 0.44$ for NGC~6819.

In the left panel of Figure~\ref{f:joint}, we plot the \textit{Gaia} DR2 CMDs for NGC~6819 and Ruprecht~147 
with de-reddended color and absolute magnitudes. The red clumps align by design. 
The main sequences and subgiant branches also approximately align, 
supporting the conclusion from our literature review that the clusters are approximately coeval.
The presence of differential reddening across NGC~6819 \citep{Platais6819, Twarog6819} 
and the persistent uncertainty in its metallicity
preclude a more precise derivation of their relative ages from their CMD morphologies.
Eventually, the asteroseismic analysis of Ruprecht~147's evolved stars should enable 
an independent measurement of their relative ages; 
however, this will still require measuring a more precise metallicity of NGC~6819 relative to Ruprecht~147.

The right panel of Figure~\ref{f:joint} plots the 
color--period distributions for NGC~6819 and Ruprecht~147. 
The \prot\ samples for each cluster 
overlap most significantly between $5000 < \teff < 5500$~K: 
over this range, 
for 
NGC~6819 
$\prot = 21.3$$\pm$1.0~days (nine stars; median and standard deviation)
and for Ruprecht~147 
$\prot = 22.4$$\pm$1.2~days (nine stars).

Analogous stars in this \teff\ range in the 1-Gyr-old NGC~6811 lag behind the projection of the 670-Myr-old Praesepe rotation sequence to 1~Gyr using the Skumanich law 
\citep[see figure 5 in][]{Curtis2019}.
However, by 2.5~Gyr these stars have clearly resumed 
spinning down, and 
we hypothesize that they do so following a Skumanich-like law, meaning continuously and with a common braking index.
Therefore, the similarity in the average \prot\ for the 
$5000 < \teff < 5500$~K stars 
in NGC~6819 and Ruprecht~147 reinforces the 
conclusion from CMD analyses that the clusters 
are approximately coeval.

Applying the $n = 0.62$ braking index 
\citep[tuned by comparing solar-color stars
in Praesepe with the Sun;][]{Douglas2019}
using the Ruprecht~147 sequence as a reference, 
we calculate gyrochronology ages for the nine 
NGC~6819 stars in this \teff\ range 
and find 2.5$\pm$0.2~Gyr,
which is only 7\% younger than Ruprecht~147.
This calculation is relatively 
insensitive to the actual braking index 
because the \prot\ values are so similar between the clusters. 
For example, varying $n$ by $\pm$0.3 would modify the
gyrochronological age of NGC 6819 by $<$10\%.

\subsection{The joint 2.7 Gyr rotator sample}
In total, we have 35 benchmark rotators for Ruprecht~147
(23 of which are likely single stars, not counting the two blue lurkers and rapid M dwarf) 
and another 30 rotators from NGC~6819, 
for a total of 66 stars.
There is a notable lack of rotators with $6000 < \teff < 6200$~K, 
and there is a similar dearth in the \textit{Kepler} field (see the top right panel of Figure~\ref{f:clusters}).
This suggests that \prot\ is very challenging to measure for such 
old stars in this \teff\ range, 
as they have crossed into a very inactive regime.

While the NGC~6819 rotator sample was limited to $M_\star > 0.85$~\msun, 
the addition of the Ruprecht~147 rotators extends the joint 2.5-2.7~Gyr sample down to $M_\star \approx 0.55$~\msun, 
and includes 20 stars with masses below the NGC~6819 sample limit.

The joint sequence can be approximately described by a polynomial model, 
as shown in the right panel of Figure~\ref{f:joint}. The dispersion about the fit is 1.18 days, 
with a median percent deviation of 5\% and a maximum deviation of 12\%.
The formula is provided in Table~\ref{t:models} and details 
are provided in Appendix~\ref{a:cpd}.

\section{Discussion}    \label{s:dis}
\subsection{Stars do not spin down continuously: \\ the case for a temporary epoch of stalled braking}

Early empirical gyrochronology models posited that stars spin down continuously with a common braking index that is constant in time: $\prot \propto t^n$, 
with $n = 0.5$ according to \citet{skumanich1972}. 
In this framework, the mass dependence is then described by a separate function independent of age and derived from young open cluster color--period sequences
\citep[e.g.,][]{Barnes2003, barnes2007, Angus2019}. 
While the coefficients have been recalibrated using a variety of empirical data 
\citep[e.g.,][]{mamajek2008, MeibomM35, Angus2015}, 
this class of model has been unable to accurately describe the color--period distributions 
of all benchmark clusters.

Over the past decade, study after study found that the  \prot\ distributions for pairs of young clusters ($\leq$1~Gyr) 
do not align when projecting one population to the age of the other using the Skumanich law
\citep[e.g.,][and including the Pleiades, Blanco~1, M35, M34, the Hyades, M48, and NGC~6811]{MeibomM35, MeibomM34, Meibom2011, Cargile2014, BarnesM48}. But it was not clear in those studies
whether these discrepancies 
could be resolved with a color-dependent (but time-independent) braking index,
whereby K~dwarfs simply have a smaller $n$ than G~dwarfs, 
and they spin down continuously following that different power law. 

This is no longer a viable solution. \citet{Agueros2018} found  that \prot\ data for the 1.4-Gyr-old NGC~752 cluster
overlapped with the Praesepe sequence at $M_\star \approx 0.5$~\msun, 
despite being twice as old (700 Myr older). Then, when examining the color--period distribution for the 1-Gyr-old cluster NGC~6811, \citet{Curtis2019} found that the converged slow sequence departed from 
models at $M_\star < 0.95$~\msun\
and merged seamlessly with the Praesepe sequence at 
$M_\star \approx 0.7$~\msun, 
despite being 40\% older in age (300~Myr).

Furthermore, \citet{Curtis2019} found that a color-dependent braking law tuned with
Praesepe and NGC 6811 cannot reproduce the observed \textit{Kepler} \prot\ distribution. In such a scenario, the universe is not old enough for K dwarfs to spin all the way down to the 30-40~day periods measured in the field \citep{AmyKepler}.

These studies concluded that stars temporarily stop spinning down after converging 
on the slow sequence, 
and that stars remain stalled for an extended period of time before spin-down resumes. 
Based on where the Praesepe, NGC~6811, and NGC~752 sequences overlapped, 
the duration of this temporary epoch of stalled spin-down must increase toward lower stellar masses.\footnote{\citet{Douglas2019} found that solar-type Hyads rotate 0.4~days slower than analogs in Praesepe, indicating that the Hyades cluster is 60 Myr older than Praesepe. However, the 
lower-mass \prot\ sequences for each cluster precisely overlapped, 
which \citet{Douglas2019} interpreted as further evidence for 
the mass-dependent stalled braking phenomenon.}

To illustrate the impact of stalling on gyrochronal ages, consider the case of the wide binary 61~Cyg A and B. Using empirical gyrochronology formulas, we infer ages for the pair of 
2~Gyr \citep{barnes2007}, 
3.5~Gyr \citep{mamajek2008}, 
3.7~Gyr \citep{Barnes2010}, and 
2.9~Gyr \citep{Angus2019}. However, because 61 Cyg A and B are so bright, they have been extensively studied and have interferometric radii measurements, which \citet{Kervella2008} used to constrain their ages to 6$\pm$1~Gyr. Assuming this age is correct, the gyrochronology ages are all systematically too young by 40-70\%.
We suggest that this is due to the fact that these models assume, incorrectly, that stars spin down continuously. 
To repair such models, the mass-dependent stalling timescale must be determined.

\begin{deluxetable}{lcccccc}
\tablecaption{Data for old, nearby benchmark K dwarfs \label{t:kd}}
\tabletypesize{\scriptsize}
\tablewidth{0pt}
\tablehead{
\colhead{Name} & \colhead{Age} & 
\colhead{$\gbr_0$} & 
\colhead{\teff} & \colhead{\logg} &  \colhead{[Fe/H]} & 
\colhead{\prot} \\
\colhead{} & \colhead{(Gyr)} & 
\colhead{(mag)} & 
\colhead{(K)} & \colhead{(dex)} &  \colhead{(dex)} & 
\colhead{(days)} 
}
\startdata
\;$\alpha$~Cen\;B & $6\pm1$     & (1.0222) & 5178 & 4.56 & $+0.23$ & 37 \\
61~Cyg\;A       & $6\pm1$     & 1.4450 &  4374 & 4.63 & $-0.33$ & 35.3  \\
61~Cyg\;B       & $6\pm1$     & 1.6997 & 4044 & 4.67 & $-0.38$ & 37.8  \\
36~Oph\;A       & \nodata     & 1.0561 & 5100 & \nodata & $-0.27$ & 20.7 \\ 
36~Oph\;B       & \nodata     & 1.0470 & 5124 & \nodata & $-0.25$ & 21.1 \\ 
36~Oph\;C       & \nodata     & 1.4033 & 4428 & \nodata & $-0.20$ & 18.0 \\ 
\enddata
\tablecomments{
$\alpha$~Cen\;B: \teff, \logg, and [Fe/H] are taken from 
the Spectroscopic Properties Of Cool Stars catalog \citep[SPOCS;][]{valenti2005}. 
The \textit{Gaia} DR2 color is predicted from our color--temperature relation (see Appendix~\ref{a:temp}, Figure~\ref{f:color_temperature}).
The age is the mean calculated by \citet{Mamajek2014} 
from several studies including those of 
\citet{Thevenin2002}, \citet{Thoul2003}, \citet{Eggenberger2004}, \citet{Miglio2005}, \citet{Yildiz2007}, and \citet{Bazot2012}, 
and is consistent with the age found by \citet{Spada2019alphaCen}. 
The \prot\ was measured from X-ray and UV data \citep{DeWarf2010, Dumusque2012}. \\\\ 
61~Cyg\;A \& B: \teff, \logg, and [Fe/H] are taken from the \textit{Gaia} FGK Benchmark catalog \citep{Heiter2015}, the age is from \citet{Kervella2008}, 
and the \prot\ were measured from \caiihk\ monitoring \citep{Donahue1996, BoroSaikia2016}. 
\\\\
36~Oph\;A, B, C: \teff\ calculated using our \textit{Gaia} DR2 color--temperature relation (see Appendix~\ref{a:temp}, Figure~\ref{f:color_temperature}). The metallicities are averages of the values listed in SIMBAD. For a detailed review of the \prot\ values, 
see footnote 28 in \citet{barnes2007}, which drew on measurements reported by 
\citet{Donahue1996} and \citet{Baliunas1983}.}
\end{deluxetable}

\subsection{Building a sample of benchmark rotator data}

To this end, 
we build a sample of benchmark \prot\ data. To characterize rotational behavior at younger ages, we use the  data for the Pleiades \citep[120~Myr;][see also 
\citeauthor{Stauffer2016}~\citeyear{Stauffer2016}]{Rebull2016}, 
Praesepe \citep[670 Myr;][see also 
\citeauthor{Douglas2017}~\citeyear{Douglas2017}, \citeauthor{Rebull2017}~\citeyear{Rebull2017}]{Douglas2019}, 
NGC~6811 \citep[1.0~Gyr;][see also
\citeauthor{Meibom2011}~\citeyear{Meibom2011}]{Curtis2019}, and 
NGC~752 \citep[1.4~Gyr;][]{Agueros2018}.
Polynomial fits to these cluster sequences are provided in Table~\ref{t:models}, 
are described in Appendix~\ref{a:cpd}, and are plotted in Figures~\ref{f:cluster_prot} and \ref{f:cluster_prot_b}. 
The \prot\ catalog for these clusters is available for download in the online journal as data behind Figure~\ref{f:cluster_prot}.

We also include three old K~dwarfs in our analysis (properties summarized in Table~\ref{t:kd}):\footnote{The Living with a Red Dwarf program is also building a catalog of ages, rotation periods, and X-ray luminosities for nearby low-mass stars
\citep{Engle2018, livingreddwarf}. \citet{Guinan2019} provides their latest age--\prot\ relations for M dwarfs.}
$\alpha$~Cen\;B 
and the wide binary 61~Cyg\;A and B. 

\begin{figure*}\begin{center}
\includegraphics[trim=0.0cm 0cm 0.0cm 0cm,clip=True,width=3.5in]{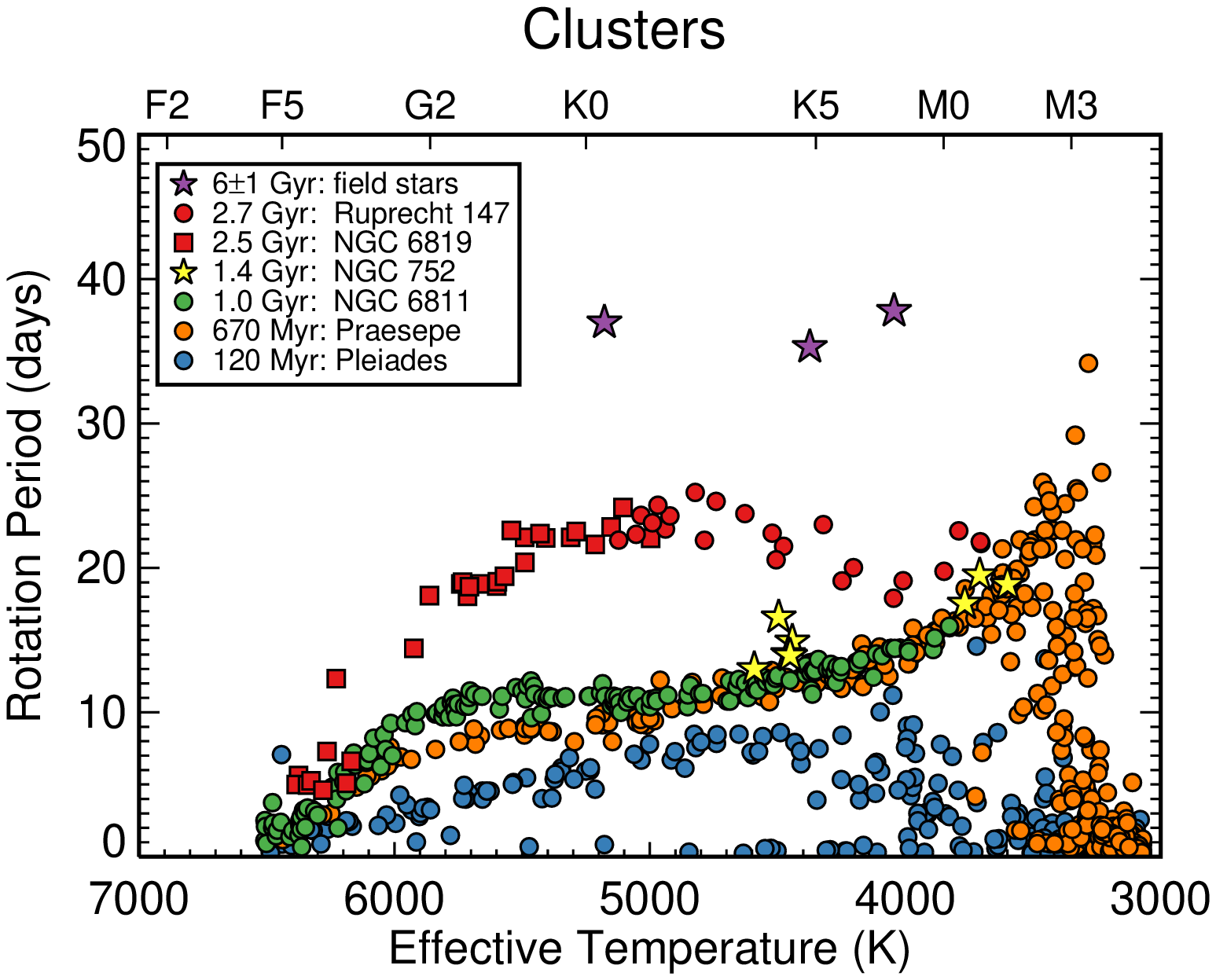}
\includegraphics[trim=0.0cm 0cm 0.0cm 0cm,clip=True,width=3.5in]{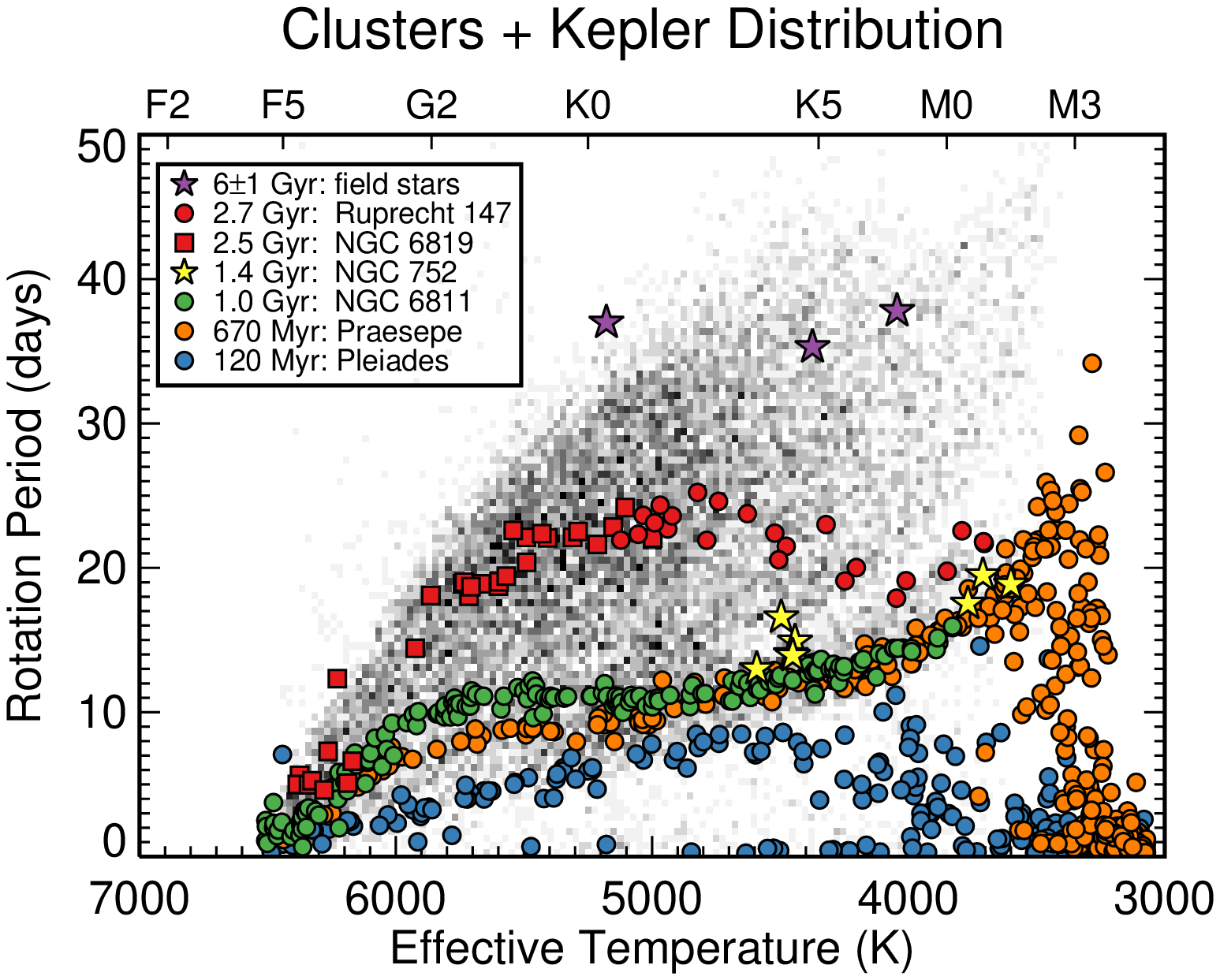}
\includegraphics[trim=0.0cm 0cm 0.0cm 0cm,clip=True,width=3.5in]{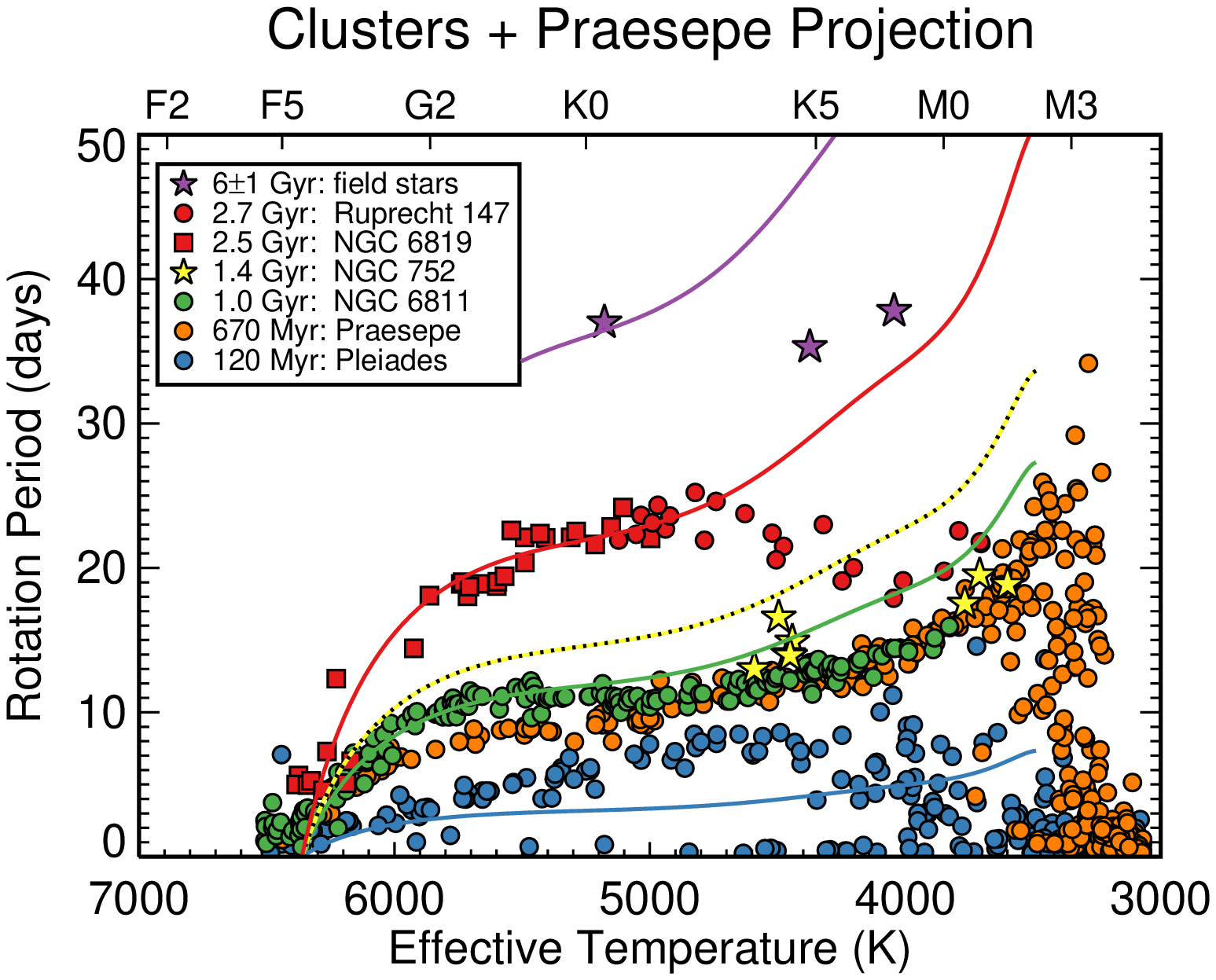}
\includegraphics[trim=0.0cm 0cm 0.0cm 0cm,clip=True,width=3.5in]{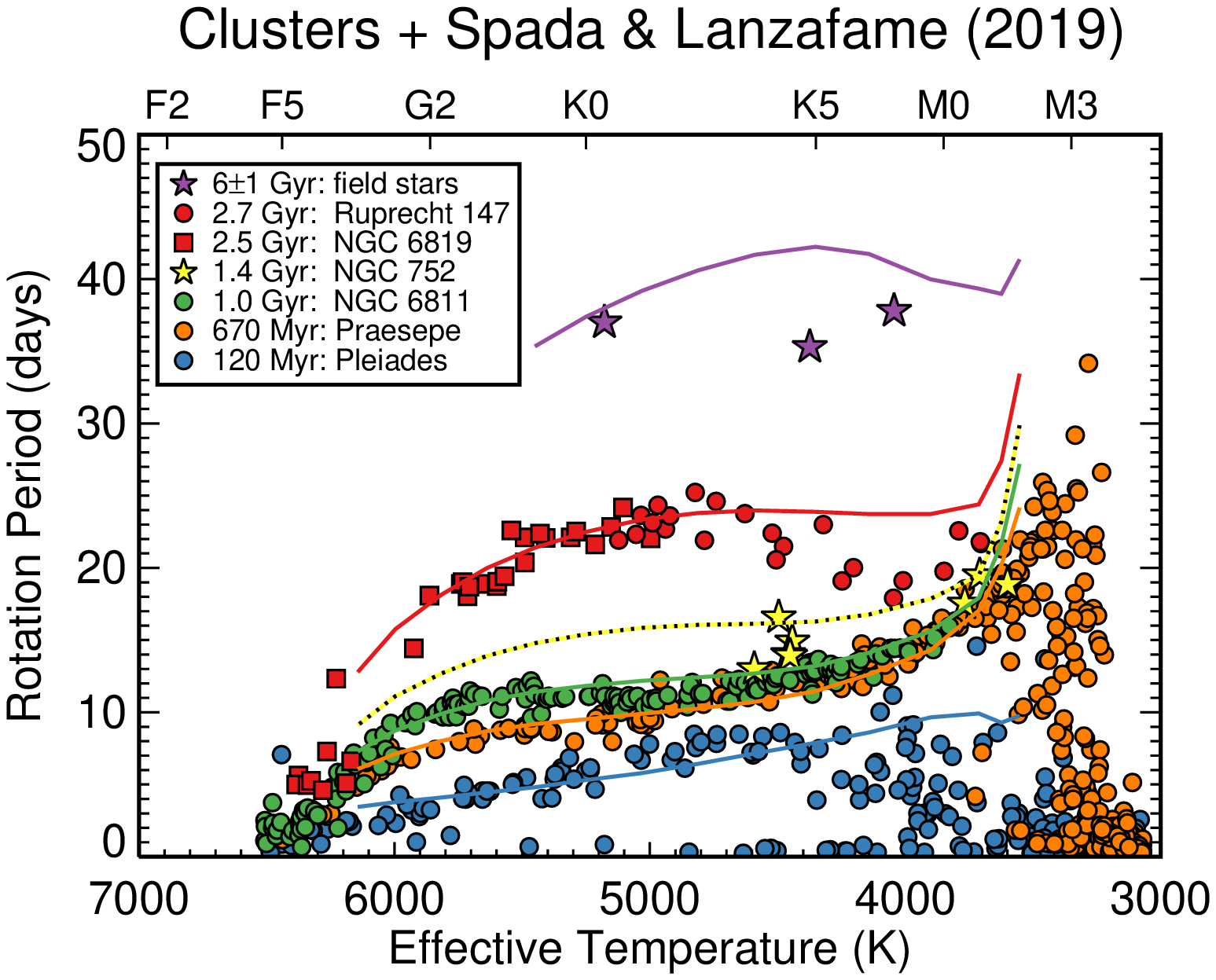}
  \caption{\textit{Top left---}\prot\ data for benchmark populations, including the Pleiades \citep[120 Myr;][]{Rebull2016}, 
    Praesepe \citep[670 Myr;][]{Douglas2017, Douglas2019},
    NGC~6811 \citep[1 Gyr;][]{Curtis2019},
    NGC~752 \citep[1.4~Gyr;][]{Agueros2018},
    NGC~6819 \citep[2.5 Gyr, and projected forward in time to 2.7~Gyr to better match the Ruprecht~147 sample;][]{Meibom2015},
    Ruprecht~147 (2.7~Gyr, this work), 
    and three old K~dwarfs: $\alpha$~Cen\,B, and 61~Cyg\,A and B ($6$$\pm$$1$~Gyr). 
    The Ruprecht~147 data show that the stars that were stalled between the ages of Praesepe and NGC~6811 have resumed spinning down 
    by 2.7~Gyr.
    Temperatures are computed from \textit{Gaia} DR2 $\gbr_0$ colors using our color--temperature relation presented in Appendix~\ref{a:temp}. Colors are dereddened using $E(G_\text{BP} - G_\text{RP}) = 0.415 \; A_V$, with cluster $A_V$ values provided in Appendix~\ref{a:cpd}.
    \textit{Top right---}Same data as the first panel, along with a binned subset of the \textit{Kepler} field distribution 
    \citep{AmyKepler}, focusing on single dwarfs within 1000~pc to minimize bias and smearing due to interstellar reddening. 
    The \textit{Kepler} colors are dereddened using the $E(B-V) = $ 0.04~mag\;kpc$^{-1}$ law, which approximates the median reddening pattern derived in Appendix~\ref{a:kepler}, and temperatures are calculated using our color--temperature relation. The data are binned by 25~K and 0.5 days.
    \textit{Bottom left---}Same data as the first panel, along with Praesepe-based gyrochrones (a polynomial fit to the Praesepe slow sequence, 
    scaled according to $(t_\star / t_{\rm Prae.})^{0.62}$). While this model works well for G dwarfs, for populations older than Praesepe, the models flare up to very long periods
    at temperatures cooler than $\teff \lesssim 5000$~K. 
    \textit{Bottom right---}Same data as the first panel, along with the 
    \citet{Spada2019} core--envelope coupling model evaluated at representative ages 
    for this sample (120~Myr, 700~Myr, 1.0~Gyr, 1.5~Gyr, 2.5~Gyr, and a projection of the solar age model to 6 Gyr 
    using $n = 0.5$).
    This model fits the joint 2.7~Gyr sample (NGC 6819 + Ruprecht~147).
   \label{f:clusters}}
\end{center}\end{figure*}

The star $\alpha$~Cen\,B appears as a long-period outlier in a relatively 
sparse region of the \prot\ distribution for the \textit{Kepler} field, 
shown in the top right panel of Figure~\ref{f:clusters}.
The upper envelope of this field distribution approximately corresponds to a line of constant Rossby number $Ro = 1.70$ using the \citet{Cranmer2011} convective turnover time formula
\citep[see also the discussion of a detection edge at $Ro_{\rm thresh} = 2.08$ in section 3.1 of][]{vanSaders2018}.
Whereas the \prot\ measured from \textit{Kepler} require brightness modulations in broadband visual light, 
the \prot\ for $\alpha$~Cen\,B was measured from variations in coronal and chromospheric emission observed in the UV, X-ray, and \caiihk. 
The \prot\ for 61~Cyg\;A and B were also measured from \caiihk\ modulation. 
With Rossby numbers of 1.73 and 1.80, the magnetic dynamos for these stars are relatively inactive as well. The \prot\ for similarly old stars in the field are likely to remain largely elusive in photometric time series surveys.


\subsection{The flat 2.7~Gyr color--period sequence: \\ a product of mass-dependent stalling}

The panels of Figure~\ref{f:clusters} show \prot\ data 
for the benchmark samples, including 
our new measurements for Ruprecht~147.
The 2.7~Gyr rotation period sequence appears remarkably flat, with most stars cooler than $\teff$~=~5700~K rotating within a narrow range of $\prot = 22$$\pm$2~days. This is in sharp contrast to the steep mass dependence seen in  younger clusters like the Pleiades and Praesepe, 
where G~dwarfs rotate more rapidly than K and early M~dwarfs
(e.g., \prot\ ranges between 6.7-18.6~days in Praesepe 
over the same \teff\ range).

However, it is reminiscent of the NGC~6811 sequence derived by \citet{Meibom2011};
the key difference is that the 2.7~Gyr flat sequence extends at least down to 0.55~\msun, 
whereas \citet{Curtis2019} demonstrated that at $M_\star \lesssim 0.8$~\msun, the NGC~6811 sequence begins to 
curve upward toward longer periods  as it merges seamlessly with the younger Praesepe sequence.

This extension of the flat sequence to lower masses is a consequence of the 
mass-dependence of the duration of the epoch of stalled braking. While the higher-mass stars spin more rapidly than lower-mass stars at the age 
of Praesepe, 
they resume spinning down earlier, and so are able to catch up to the K~dwarfs 
just as these resume spinning down. The process tends to flatten out the color--period sequence. 

While its appearance is flat compared to the steep younger sequences, 
the Ruprecht~147 sequence does seem to be subtly curved.
The sequence appears to dip downward from 
$\approx$23.7~days at spectral type K2 ($\gbr_0 \approx 1.09$, 0.85~\msun),
to slightly faster periods of $\approx$19.9~days at K7 ($\gbr_0 \approx 1.75$, 0.62~\msun), 
and then curves back up to $\approx$21.9~days at M1 ($\gbr_0 \approx 2.05$, 0.55~\msun).
However, this ``sagging'' curvature is only supported by five stars with $\prot < 21$~days. 

Recently, \citet{Angus2020} analyzed the \citet{AmyKepler} sample of $\sim$34,000 main-sequence dwarfs in the \textit{Kepler} field with measured rotation periods, combined with \textit{Gaia} data. 
The color--period distribution for these field stars were binned, and then color-coded according to $v_b$ velocity dispersion.\footnote{At the \textit{Kepler} field's low galactic latitude, the velocity in the direction of galactic latitude, $v_b$, approximates the vertical velocity perpendicular to the plane, $v_Z$ (often denoted $W$). The dispersion of $v_Z$ increases with age for a population of stars via dynamical heating. Therefore, the dispersion of $v_b$ for similarly-aged stars in the \textit{Kepler} field should likewise increase with age.}
Lines of constant velocity dispersion emerged, tracing out gyrochrones across the diagram. 
At young ages (i.e., low velocity dispersion), these gyrochrones increase from warm and rapid to cool and slow, similar to the pattern seen in young clusters like Praesepe.
However, the relationship between color and period flattens out at intermediate ages, 
similar to what we have found for Ruprecht 147.
At older ages, the relation inverts, where cooler stars spin more rapidly than their warmer coeval counterparts.
This result presents an independent verification of the flattening of middle-aged rotation sequences.

\subsection{When do stalled stars resume spinning down?} \label{s:resume}

With our new $\prot$ measurements for Ruprecht~147,
it is now clear that stalling is a temporary evolutionary stage. As expected based on the distribution of \prot\ 
measured for low-mass field stars, such as those in the \textit{Kepler} field 
in the top right panel of Figure~\ref{f:clusters}, low-mass stars do indeed eventually resume spinning down.

With our new data, we can empirically determine the age at which stars resume spinning down,
if we make a few simplifying assumptions:

\begin{enumerate}
    \item The lower envelope of the \textit{Kepler} \prot\ distribution represents the 
    \prot\ at which initially rapidly rotating stars converge 
    to join the slow sequence. 
    This process proceeds from higher to lower masses, which explains why 
    the Pleiades slow sequence is only converged down to $M_\star \approx 0.75$~\msun,
    whereas the Praesepe slow sequence extends down to near the fully convective boundary at $M_\star \approx 0.4$~\msun.
    For more details, see Appendix~\ref{a:lower}.
    
    \item Stars completely stall after converging on the slow 
    sequence (i.e., $n = 0$). 
    
    \item Once stars resume spinning down, they follow a Skumanich-like law with $\prot \propto t^n$, 
    where $n = 0.62$, irrespective of stellar mass.
    We examine the post-stalling braking index in Appendix~\ref{a:brake}.
\end{enumerate}

\begin{figure*}\begin{center}
\includegraphics[trim=0.0cm 0cm 0.0cm 0cm,clip=True,width=5.0in]{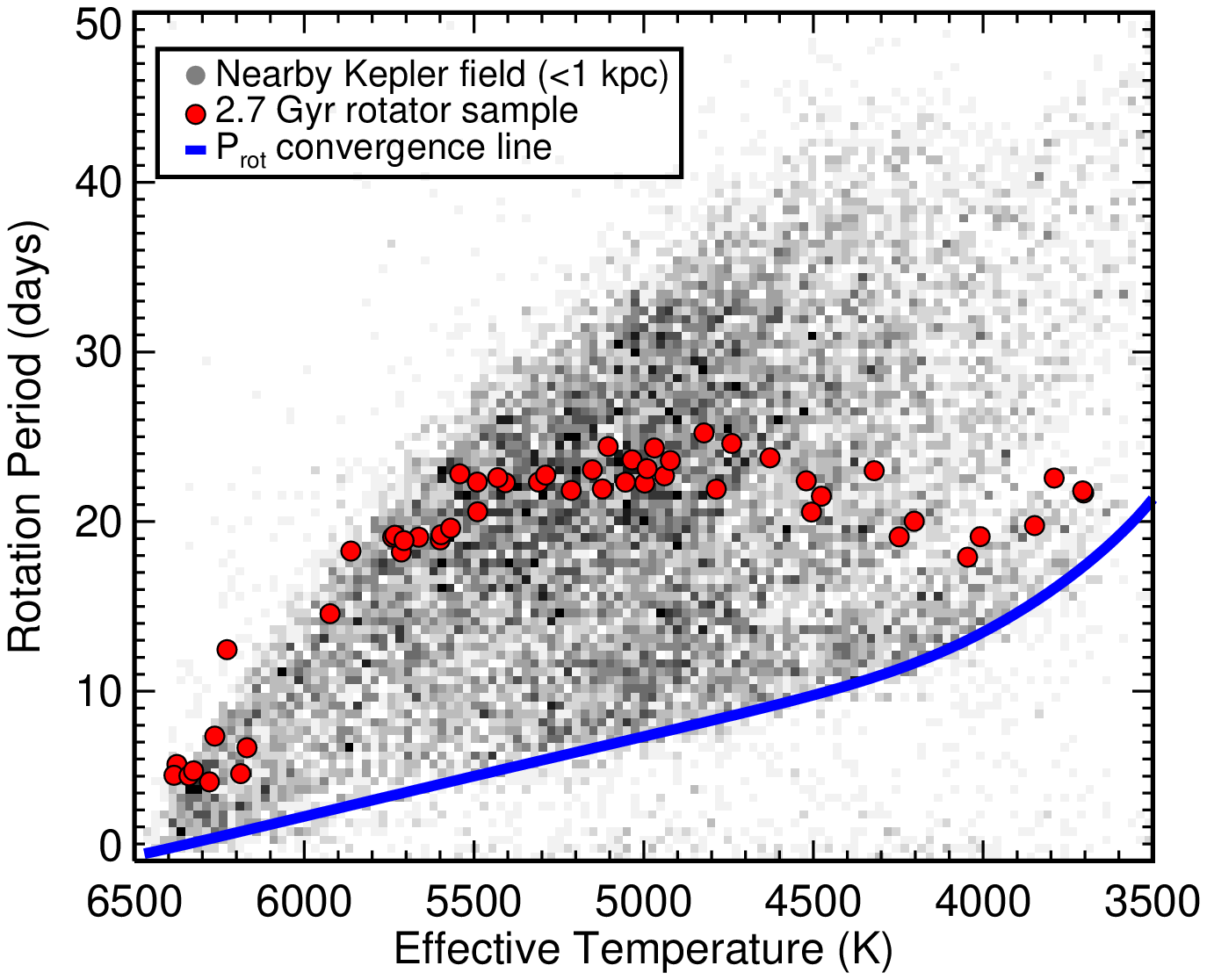}
\includegraphics[trim=0.0cm 0cm 0.0cm 0cm,clip=True,width=3.5in]{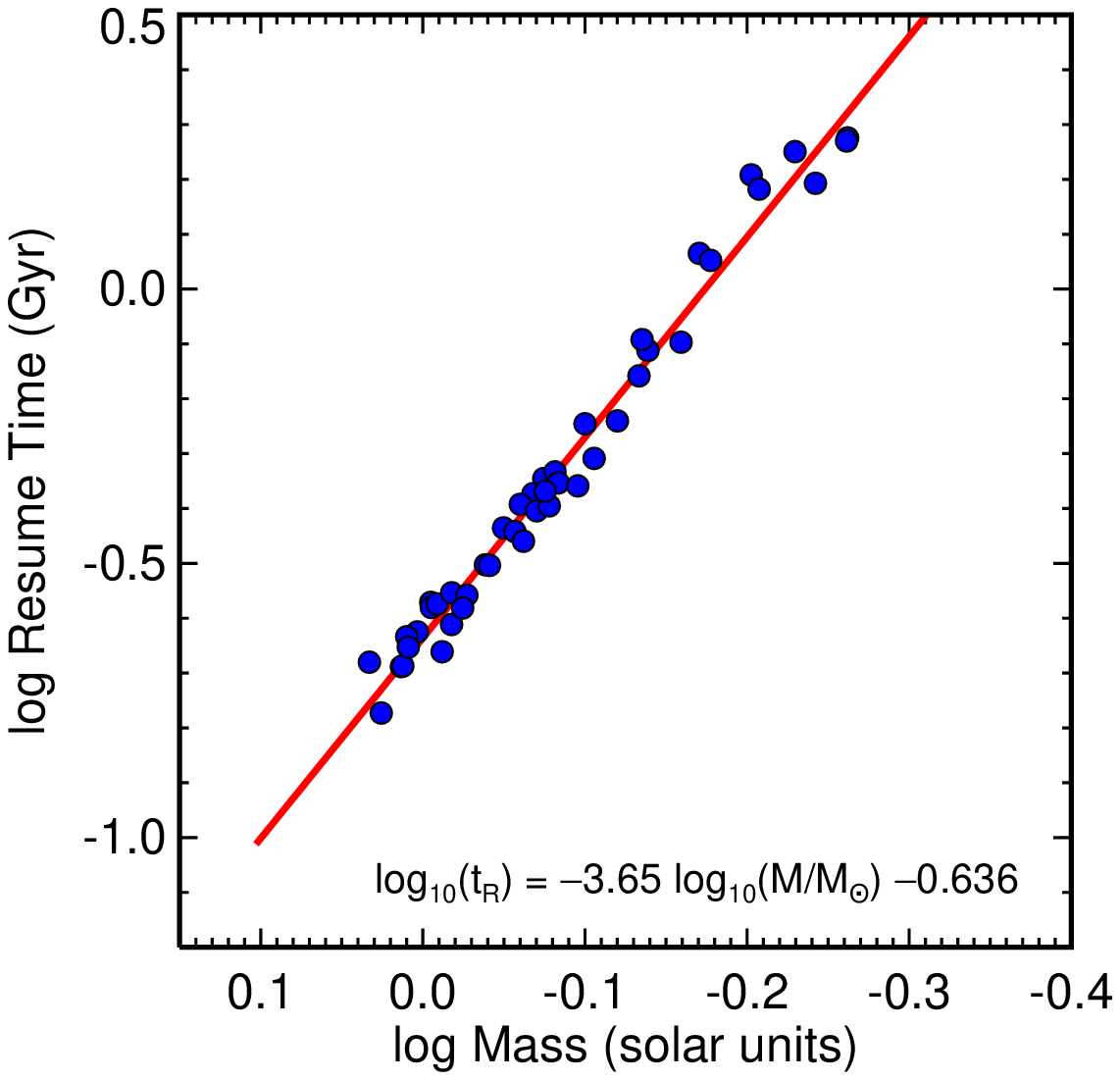}
\includegraphics[trim=0.0cm 0cm 0.0cm 0cm,clip=True,width=3.5in]{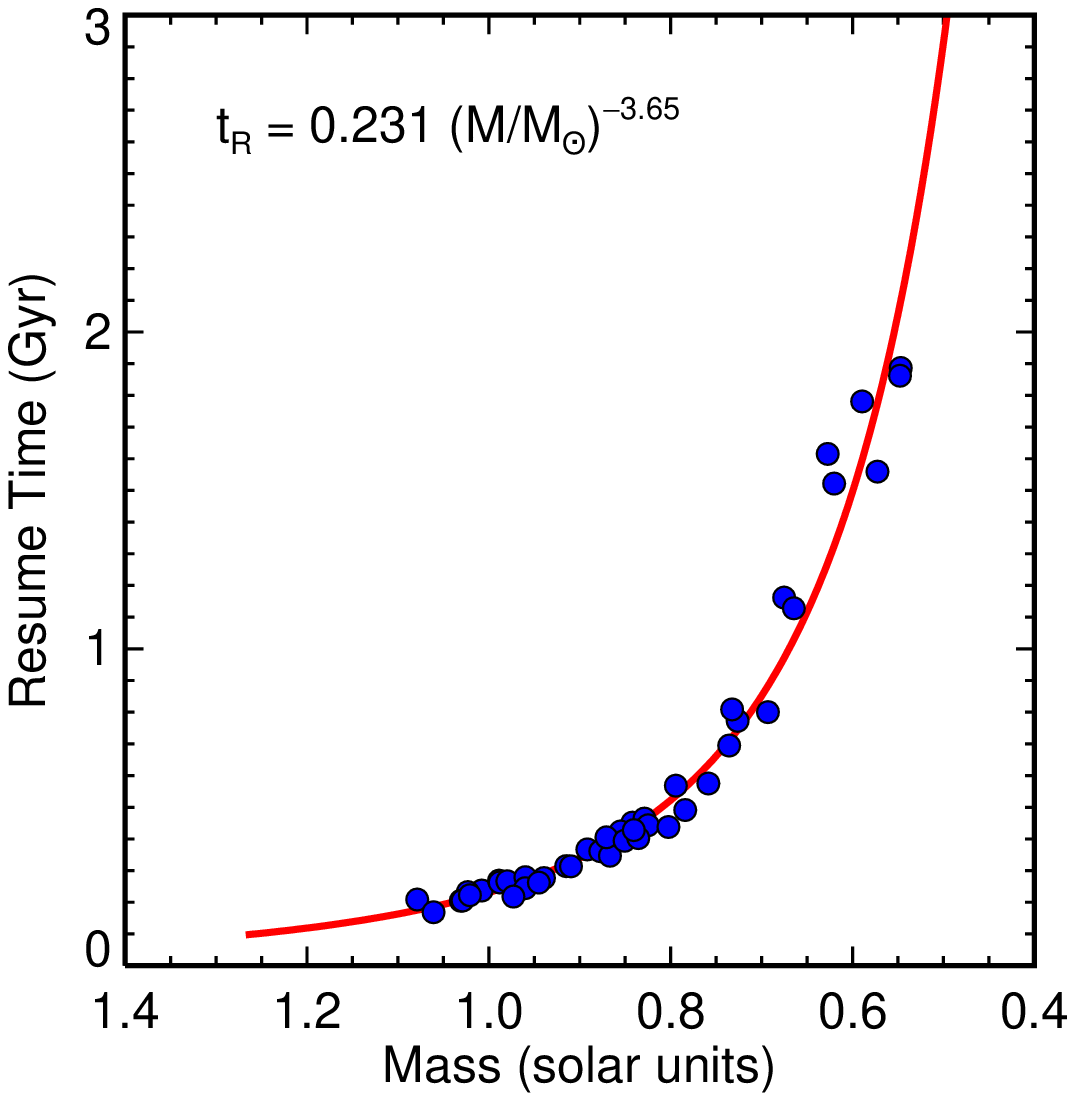}
  \caption{When do stalled stars resume spinning down? 
  \textit{Top---}The \textit{Kepler} \prot\ distribution 
  \citep{AmyKepler} has a border along the lower envelope, 
  which we posit represents the rotation period at which 
  stars converge onto the slow sequence (following Figure~\ref{f:clusters}, temperatures are computed from \textit{Gaia} DR2 colors using our color--temperature relation presented in Appendix~\ref{a:temp}, after de-reddening with our $E(B-V) = 0.04$~mag\;kpc$^{-1}$ relation derived in Appendix~\ref{a:kepler}; data are similarly binned by 25~K and 0.5~day).
  We fit a representation of this border with a 6\ith\ order 
  polynomial (blue line). 
  The representation was created by 
  combining a fit to the Pleiades sequence for more massive stars
  ($M_\star > 0.95$~\msun, modeled with a line of constant 
  Rossby number, where $R_o = 0.29$ using the \citet{Cranmer2011} convective turnover timescale), 
  a polynomial fit to the Praesepe sequence for lower mass stars 
  ($M_\star < 0.59$~\msun), 
  and points marked by hand for the intervening range.
  \textit{Bottom---}The age at which stars resume spinning down versus stellar mass (\msun), 
  plotted in logarithmic (\textit{left}) and linear (\textit{right}) scales.
  The blue points are the individual measurements of the resume time, $t_R$ (Equation~\ref{e:res}), calculated by dividing 
  the convergence line formula by the \prot\ data for the joint 2.7~Gyr sample (NGC~6819 + Ruprecht~147)
  raised to the $1/n$ power then scaled by the 2.7~Gyr age for the old-age reference sample (i.e., the gyrochronal age for the convergence line, which itself is not a gyrochrone). Here, we adopted $n = 0.62$ \citep{Douglas2019}. 
   \label{f:resume}}
\end{center}\end{figure*}

To calculate the age at which stars resume 
spinning down, $t_{\rm R}$, 
we fit the ratio of the lower envelope to the 
joint 2.7~Gyr sequence, 
raised to the $1/n$ power where $n = 0.62$, 
and scaled by the 2.7~Gyr age of Ruprecht~147. 
In other words, we calculate how long it takes to ``rewind'' the 2.7~Gyr sample 
back to the \prot\ convergence line, represented by the lower envelope of the 
\textit{Kepler} distribution. 
The bottom panels of Figure~\ref{f:resume} 
plot the resulting $t_{\rm R}$ in logarithmic and linear scales, 
which show a power law relationship with stellar mass.
We fit this relationship with the following function:
\begin{equation} \label{e:res}
    t_{\rm R} = t_{\rm R, \odot} \, (M_\star/\msun)^\alpha,
\end{equation}
\noindent and find $t_{\rm R, \odot} = 231 \pm 10$~Myr and $\alpha = -3.65 \pm 0.20$.\footnote{This approach was inspired by the \citet{Spada2019} 
calculation of the core--envelope timescale.}$^,\,$\footnote{The uncertainties are estimated by the bootstrap method, where Gaussian errors of 0.02~\msun\ for mass and 10\% for \prot\ are applied to 
simulated cluster samples.}$^,\,$\footnote{Fitting the resume time as a function of \teff\ and $\gbr_0$, we find $t_{\rm R} = 202 \, (\teff/ 5770~K)^{-5.11}$ and $t_{\rm R} = 328 \, (\gbr_0)^{2.47}$ in Myr.}
The scatter about this fit in Figure~\ref{f:resume} is clearly low, indicating that the statistical uncertainty is negligible. 
However, the true uncertainty is difficult to estimate because it is determined by the validity of our assumptions, not the quality or quantity of data at hand. We require \prot\ sequences for other intermediate-age clusters to constrain the post-stalling braking index, and we need \prot\ distributions for younger clusters to determine when stars converge on the slow sequence as a function of mass.

This formula has a strong mass dependence, 
where, for example, 
a $M_\star = 0.55$~\msun\ star will not resume spinning down until it is 2.0~Gyr old ($\teff \approx 3700$~K, $\gbr_0 \approx 2.1$, M1V).
Since analogous stars appear to be converged on the slow 
sequence by 670~Myr, the age of Praesepe, these stars apparently are stalled for $\geq$1.3~Gyr.

In this paper so far, we have advanced gyrochronology by illustrating two things: 
(1) stalled stars do in fact resume spinning down. That they do is already made obvious by the existence of slower stars in the \textit{Kepler} field, 
as pointed out in our study of NGC~6811 \citep{Curtis2019},
but it is encouraging to see this in an older cluster, and 
(2) cluster color--period gyrochrone sequences flatten out over time. 
However, we are not yet attempting to repair gyrochronology. There is still far too much work ahead of us, and it is premature to recalibrate or reformulate a rotation--age relationship until we have filled in  other critical gaps in time and mass with light curve data already available. The stalling timescale is not meant to be used for any reason other than to illustrate the magnitude of the problem. Until we can empirically determine the post-stalling braking index, and any dependence on mass, age, or metallicity, this resume time is fundamentally limited by our assumed braking index and the idea behind using the lower envelope of the \textit{Kepler} field as a reference point for where stars stall.

\subsection{Why do stars temporarily stall? Evaluating the core--envelope coupling model}

A purely empirical, data-driven model for spin-down (i.e., a gyrochronology relation) can be agnostic to the underlying physical causes for angular momentum evolution and still yield accurate and precise ages. 
In contrast, semi-physical models rely on prescriptions for the magnetic braking torque, which scales with various physical stellar properties, and includes some parameters that are empirically tuned (e.g., with cluster and solar data).
One approach assumes that stars rotate as solid bodies \citep[e.g.,][]{Matt2015, vanSaders2013}. In this case, one way to account for a short-term stalling epoch might be to temporarily reduce the magnetic braking efficiency. 

Another class of semi-physical model describes the stellar interior with a two-zone approximation, where the convective envelope is allowed to rotate at a different rate than the core. In these models, angular momentum is assumed to be exchanged between the envelope and core on some characteristic timescale \citep{MacGregor1991}. According to this scenario, when the age of the star is comparable to this core--envelope coupling timescale, the rotational braking of the envelope (which includes the visible surface) is temporarily stalled because the spin-down torque from the stellar wind is counteracted by a spin-up torque from the core.  
This does not necessarily cause stars to fully stall. 
We assumed $n = 0$ to estimate the resume time; 
however, core--envelope coupling can 
result in a reduced, but non-zero $n$, 
where the angular momentum received from the interior is not perfectly 
balanced against that lost via magnetic braking.

Unfortunately, the timescale for this coupling has not been modeled purely from theory, and instead requires empirical constraints from open clusters, 
which until now have not reached to low enough masses at old enough ages. 
For this reason, 
prior core--envelope models failed to predict the full impact of this phenomenon \citep[e.g.,][]{Gallet2013,Gallet2015,Lanzafame2015}.

\citet{Spada2019} recently recalibrated their model using our new \prot\ data for NGC~6811 \citep{Curtis2019}, 
and reported that their model can now account for most of the 
apparent stalling effect seen between Praesepe and NGC~6811.
In the appendix of their paper, 
these authors tabulated rotational isochrones,
and we can now test how well they perform. 

The bottom right panel of Figure~\ref{f:clusters} shows 
our benchmark sample along with the \citet{Spada2019} models.
According to their model, 
the strong mass dependence of the core--envelope coupling timescale 
introduces a change in slope from steep to flat in the color--period diagram. 
With respect to the Ruprecht~147 sequence,
this model is clearly superior to the 
Praesepe projection model (bottom left panel of the same figure). 
It also performs better for the NGC~752 stars 
and the old, nearby K dwarfs. 

\begin{figure}[b]
\begin{center}
\includegraphics[trim=1.1cm 0.2cm 0.0cm 0.5cm, clip=True,  width=3.5in]{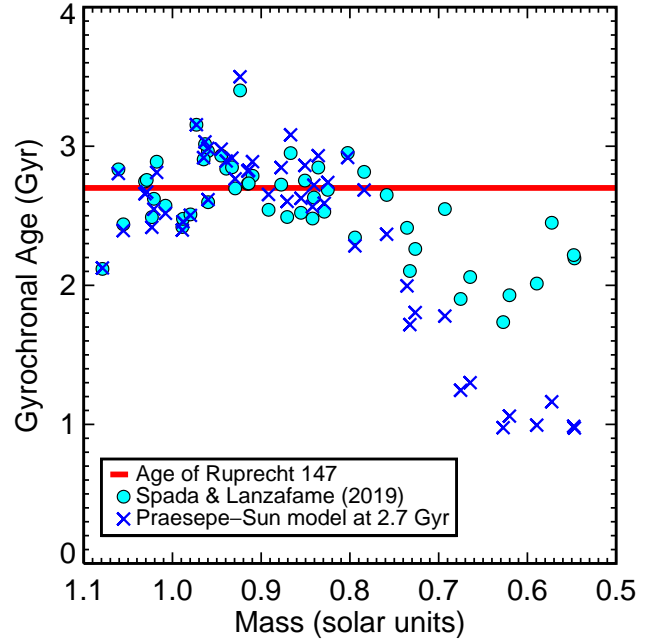}
\caption{Gyrochronal ages vs.~stellar mass for benchmark stars in Ruprecht 147 and NGC 6819 calculated using two models. 
The cyan points show ages calculated with the \citet{Spada2019} core--envelope coupling model, which was empirically tuned with data from the Pleiades, Praesepe, and NGC 6811 clusters.
The blue crosses show ages calculated with the Praesepe polynomial model projected forward in time from 670~Myr to 2.7~Gyr using the $n = 0.62$ braking law. The red line shows the 2.7 Gyr age of Ruprecht~147. 
Stars with $M_\star < 0.70$~\msun\ show the largest discrepancy with both models, with median gyrochronal ages of 1.1 and 2.2~Gyr for the Praesepe and \citet{Spada2019} models, respectively.  
The \citet{Spada2019} model significantly reduces the relative error from 60\% down to 20\%. As the core--envelope coupling timescale for this model is empirically tuned, incorporating data for NGC 752 and Ruprecht 147 into the calibration might improve this further.
\label{f:spada}}
\end{center}\end{figure}

The \citet{Spada2019} model does appear to mildly overshoot the mid-K dwarfs in NGC~752 (by 1.6 days) and Ruprecht~147 (by 2.7 days), 
suggesting that it needs refining with the data for these older clusters.  
In Figure~\ref{f:spada}, we calculate gyrochronal ages for the joint sample of benchmark rotators for NGC 6819 and Ruprecht 147 using this \citet{Spada2019} model,\footnote{We used the 2.5~Gyr model tabulated in \citet{Spada2019} as a function of stellar mass, and interpolated it for each star in the sample using masses estimated using the \citet{kraus2007} \teff--$M_\star$ table together with our relation for converting \textit{Gaia} DR2 $\gbr_0$ to \teff. We then applied our $n = 0.62$ braking index to calculate ages: $t = ({\rm observed} \, \prot / {\rm model} \, \prot )^{1/n} \times 2.5$~Gyr.}
and our polynomial fit to the Praesepe sequence projected forward in time using our solar-calibrated $n = 0.62$ braking index. 
Both models work fine at this age for $M_\star > 0.8$~\msun\ stars; 
however, each underestimates the true 2.7~Gyr age for $M_\star < 0.7$~\msun\ stars ($\teff \lesssim$4350~K, $\gbr_0 >$1.46, $>$K5V), 
and we find median ages of 2.2~Gyr and 1.1~Gyr from each model respectively. In other words, the large 60\% age bias that comes from ignoring stalling is significantly reduced to only 20\% with the empirically-tuned core--envelope coupling model. 

The \citet{Spada2019} model does not predict full stalling ($n = 0$) behavior, 
which would appear as truly horizontal lines in figure 4 of their paper. But \citet{Curtis2019} found that the Praesepe and NGC~6811 sequences precisely overlapped. 
To further constrain the exact $n$ during and after stalling, an expanded rotator sample for other immediate-aged clusters is needed, 
along with precise knowledge of the interstellar reddening toward each cluster,
and an understanding of the impact of metallicity on spin-down \citep{Amard2020}.

\subsection{Recommendation for age-dating old K~dwarfs: \\ the case of 36~Ophiuchi}
\label{s:gyro}

We believe it is premature to offer a full recalibration of gyrochronology at this stage.
First, we have additional clusters ready to analyze. 
Furthermore, it has only recently become possible to incorporate en masse the large sample of field stars with measured rotation periods into a calibration through the application of kinematic age-dating \citep{Angus2020}.
Finally, wide binaries offer great potential for studying stalling, 
expanding the parameter space covered by the benchmark sample to lower masses and older ages, 
and testing the impact of metallicity on spin-down; 
however, this potential has not yet been realized as there are only a handful of well-characterized systems (e.g, $\alpha$~Cen, 61~Cyg) and only a few preliminary investigations published using \textit{Kepler} targets \citep{Janes2017, Janes_CS20}.

\begin{figure}
\begin{center}
\includegraphics[trim=1.1cm 0.2cm 0.0cm 0.5cm, clip=True,  width=3.5in]{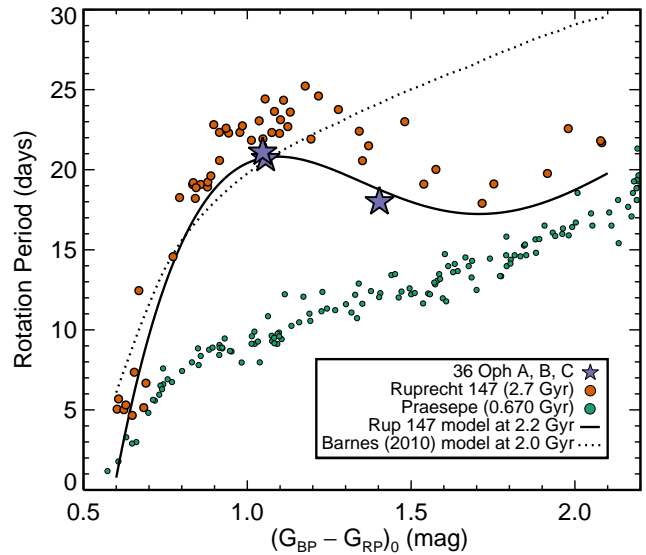}
\caption{Age-dating the triple star system 36~Ophiuchi. 
The purple five-point star symbols mark 36~Oph\;B, A, and C ordered from blue to red in \gbr; see Table~\ref{t:kd} for details. 
Classic gyrochronology relations yield ages for C that are half the value of A and B; 
e.g., 590~Myr vs 1.43 Myr from \citet{barnes2007}, 1.3~Gyr vs 2.4~Gyr with our Praesepe-based model, or 1.25~Gyr vs 2.05~Gyr from \citet{Barnes2010} shown as a dotted line in this figure.
Instead, for stars that have resumed spinning down (i.e., slower than the Praesepe sequence, shown with green points), 
we suggest an alternative approach: adopt the Ruprecht~147 color--period sequence as an empirical 2.7 Gyr gyrochrone (orange points), and project it forward or backward in time using a $t^n$ braking law (for now, we prefer $n = 0.62$). 
The 36~Oph periods almost overlap the Ruprecht~147 data, indicating that they share a similar age.
Spinning them down to best match the Ruprecht 147 sequence yields $t =$~2.2$\pm$0.1~Gyr for 36~Oph. 
Gyrochronology can now finally yield a consistent age for the  triple system.
\label{f:36oph}}
\end{center}\end{figure}

Apart from the \citet{Spada2019} model, how should one approach age-dating old K dwarfs according to their observed rotation? 
If we apply the original tenet of gyrochronology, that stars spin down continuously with a constant braking index common to all stars, then all we need is a rotation period sequence for one cluster (to delineate the mass dependence) and a braking index. 
Out of necessity, to calculate the resume time we posited that this assumption is valid once spin-down resumes. If true, one could then use the Ruprecht~147 color--period sequence in combination with our solar-calibrated braking index ($n = 0.62$).

To illustrate this concept, lets consider the case of the triple star system, 36~Ophiuchi, 
which is composed of an inner binary of early-K dwarfs (A+B), and a wider mid-K dwarf companion (C). 
Their properties are summarized in Table~\ref{t:kd}.
The three stars have rotation periods measured by the Mount Wilson Survey \citep{Donahue1996, Baliunas1983}, and the values are quite similar:  20.69, 21.11, and 18.0 days. 
According to classic gyrochronology relations and theoretical predictions, 
36~Oph\,C should be rotating more slowly than A or B. 

Stated differently, 
\citet{barnes2007} found gyrochronal ages for A and B of 1.43 Gyr, but C was only $590 \pm 70$~Myr old. 
Later, \citet{mamajek2008} ignored the C component and calculated 1.9~Gyr for A and B; 
however, that model also yields an age for C that is half the value for A and B; i.e., the \citet{mamajek2008} model predicts \prot\ = 25.4 days for C, which is 40\% slower than its observed value. 
The \citet{Barnes2010} model similarly yields 2.1~Gyr for A and B and 1.3~Gyr for C. 
Now that we have a well-defined sequence for Praesepe, we can use its model and our solar-calibrated braking index to age-date this trio, and we find 2.36, 2.45, and 1.30~Gyr. While our absolute values are systematically older (analogous stars in Praesepe spin at 9.48, 9.45, and 11.91 days, so 36~Oph must be significantly older than that cluster),
this model still finds C to be half the age of A and B.
\citet{barnes2007} explained that A and B share a highly eccentric orbit; although their orbital period is quite long ($\sim$500~yr), their periastron is estimated to be only $\sim$6~au. 
\citet{barnes2007} speculated that gravitational tides had altered the rotational evolution for A and B, leaving C to reveal the true age of the system.

However, we can now offer an alternative interpretation in light of the mass-dependent stalling phenomenon.
While we expect each star was temporarily stalled in the past, A and B should have resumed spinning down before C, allowing them to ``catch up'' and even surpass C in rotational period.
Instead of projecting the Praesepe model forward in time, if we rewind the Ruprecht 147 model backward 
using the same $n = 0.62$ time dependence, we find ages of 
2.24, 2.31, and 2.14~Gyr.\footnote{We also estimated ages using the \citet{Spada2019} model and found 2.24, 2.31, and 1.78 Gyr. To do this, we calculated a mass-dependent braking index using the 2.0 and 2.5~Gyr models, and then applied that to the 2.0~Gyr model tabulated according to $(B-V)$. The age of C is only 22\% lower than the average of A and B, which is a remarkable improvement over classic models. Even better, our approach using the Ruprecht~147 sequence finds C is only 6\% younger than A and B.} 
In other words, one no longer needs to invoke tides to explain the rotation periods for this trio. 

We expect that \prot\ will eventually be measured for other middle-aged wide binaries, which will corroborate our interpretation of the \prot\ data for 36~Oph and 61~Cyg. 
Once we have such data, we can test our critical assumption for the post-stalling braking index, and then we will be better positioned to reformulate an empirical gyrochronology relation and constrain theoretical models.
In the meantime, for K dwarfs with rotation periods slower than the stalled sequence (the overlapping portions of Praesepe and NGC~6811), we hypothesize that such stars have resumed spinning down, and can be similarly age-dated relative to the Ruprecht~147 sequence. 
Our procedure is illustrated in Figure~\ref{f:36oph}.

\subsection{The \textit{Kepler} intermediate period gap is not caused by a lull in star formation} \label{s:gap}
Measurements of stellar rotation in the \textit{Kepler} field revealed a bimodal distribution among cool dwarfs ($\teff < 5000$~K) at intermediate periods \citep[15-25 days;][]{AmyMdwarfs, AmyKepler}.
Several hypotheses have been advanced to explain this feature, including a lull in the local star formation rate \citep{AmyMdwarfs, AmyKepler}. 
\citet{Davenport2017} argued that this occurred $\approx$600~Myr ago, 
based on the fact that a gyrochrone from \citet{MeibomM35} 
appears to trace the gap \citep[see also][]{Davenport2018}.
However, there are two problems with this scenario:

\begin{enumerate}
    \item The intermediate period gap does not trace a 600-Myr-old population. 
    \item \prot\ sequences for older clusters cross the gap.
\end{enumerate}

Praesepe provides an \textit{empirical} gyrochrone at 670~Myr \citep{Douglas2017, Rebull2017},
and it does not trace the gap, 
but instead follows the lower envelope of the 
\textit{Kepler} \prot\ distribution. 
Sequences for the Hyades \citep[730 Myr;][]{Douglas2019, Douglas2016} 
and even NGC~6811 \citep[1 Gyr;][]{Meibom2011, Curtis2019} 
likewise do not trace the gap, and similarly track the lower envelope of the 
\textit{Kepler} \prot\ distribution at $\teff < 4500$~K as they remain stalled. 
Clearly, the \citet{Meibom2009} gyrochrone, 
and indeed all empirical gyrochrones as we have discussed here and elsewhere 
\citep{Agueros2018, Curtis2019, Douglas2019}, 
fail to accurately describe stellar spin-down for cool dwarfs.
That the intermediate period gap approximately coincides with that model is a coincidence. 

According to the open cluster data, the intermediate period gap does not have one common age.
In fact, the \prot\ sequences for NGC 752 (1.4~Gyr) and Ruprecht~147 (2.7~Gyr) intersect
the gap.\footnote{Figure~\ref{f:a:gap} in Appendix~\ref{a:gap} examines a member of Ruprecht 147, EPIC~219489683, which appears to fall right in the gap.}
This disproves any scenario for the creation of the gap occurring at a single point in time.
Indeed, these cluster sequences cross 
the gap at different points in the diagram,
at decreasing masses with increasing ages, 
indicating that it is formed at different times for stars of different masses.\footnote{\citet{Timo2020} reported \prot\ for 
stars in the \textit{K2} campaign fields, which probe different regions of the local Galaxy than did \textit{Kepler}, and they still find evidence for a ``dearth region,'' although it is does not appear as pronounced in their data set. The detection of an intermediate period gap along a variety of 
sight lines through the Galaxy would also challenge the variable star formation rate hypothesis.}

Alternatively, \citet{Timo2019} cited the gap as evidence for a transition 
from spot- to faculae-dominated photospheres, along with the observation that stars 
near the gap have weaker photometric amplitudes than slower or faster analogs.
In this case, the gap is actually filled with stars; their rotation periods are simply 
difficult to measure from photometric time series (however, they should be measurable with \caiihk\ monitoring).
\citet{Timo2019} ascribed a single age of 800~Myr to this transition, irrespective of stellar mass, which the open cluster data clearly refute. 
However, the idea of a spot-to-faculae transition is not necessarily in conflict with the 
open cluster data, as long as it is acknowledged that this happens at different times for different masses.
We note that the gap also appears to approximately follow a line of constant Rossby number of  
$Ro \approx 0.5$ using the \citet{Cranmer2011} formula for the convective turnover time.
This might suggest that the gap is caused by an event in the evolution 
of the magnetic dynamo, not an event in time, 
as the magnetic activity of stars of different masses evolve at different rates according to their mass-dependent convective turnover times.

Finally, having found that stars do not spin down continuously, 
we encourage theoretical work on the possibility that the gap is created by a
temporary increase in the braking efficiency, 
causing stars to ``jump'' across the gap \citep[an idea originally proposed by][]{AmyMdwarfs}.
In this case, 
the gap is actually relatively empty (i.e., there are few stars rotating with \prot\ that place them in the gap, irrespective of our ability to measure them).
Regarding this idea, we would like to call attention to the five Ruprecht~147 stars in or just beneath the gap (i.e., more rapid and cooler/redder)---they appear to trace the lower edge of the gap, as if they are ``waiting in line'' to cross it (see Figures~\ref{f:resume} and \ref{f:a:gap}).

\section{Conclusions} \label{s:concl}

We used the 2.7-Gyr-old Ruprecht~147 open cluster as a benchmark for studying stellar rotation, 
with the eventual goal of repairing gyrochronology for low-mass stars.
First, we assembled a catalog of 440 candidate members, and identified binaries and non-members using astrometric and photometric data from \textit{Gaia}~DR2, RV time series, and high spatial resolution imaging obtained with Robo-AO.
Next, we measured periodicities in \textit{K2} light curves for 68 stars, 
then used the chromospheric emission index \lrphk\ to invalidate 
\prot\ for three stars. 
Separately, we used PTF to construct sparser light curves with lower precision 
for K and early M dwarfs, and measured \prot\ for seven stars, 
including three for which we also have measured \prot\ from \textit{K2}. 
In these cases, we found nearly identical \prot\ values, despite the differences 
in cadence, photometric precision, and times of observation.
In all, we measured \prot\ for 58 dwarfs, including 35 stars we classify as benchmark rotators, of which 23 are likely single dwarfs.

We paired our Ruprecht~147 data with \prot\ data for the approximately 
coeval cluster NGC~6819, which was surveyed during the primary \textit{Kepler} mission.
Before merging the data sets, we derived a relative interstellar reddening and distance modulus for NGC~6819 by comparing the apparent magnitudes of red clump giants 
in each cluster, and found $A_V = 0.44$ for NGC~6819, after accounting for our 
$A_V = 0.30$ value for Ruprecht~147.

\citet{Meibom2015} measured \prot\ for 30 NGC~6819 dwarfs with masses $M_\star \gtrsim 0.85$~\msun.
Our Ruprecht~147 sample extends this mass limit down to $M_\star \approx 0.55$~\msun,
and includes 20 stars with masses below the NGC 6819 sample limit.

Using the overlapping portions of each sample ($0.84 < M_\star < 0.96$~\msun; $5000 < \teff < 5500$~K), 
we found that the Ruprecht~147 stars rotate systematically more slowly, although 
only by 5\%. 
Applying the $n = 0.62$ braking law \citep{Douglas2019}, 
and adopting 2.7~Gyr for the age of Ruprecht~147, 
we calculate a gyrochronological age of 2.5~Gyr for NGC~6819, 
which is consistent with the literature. 
Considering the merged $\approx$2.7~Gyr sample, 
we now have rotation periods for 67 benchmark stars with $0.55 < M_\star < 1.3$~\msun, 54 of which 
are likely single.

With this merged sample, we can now study how stars spin down from their youth, 
represented by younger clusters like the Pleiades and Praesepe, 
up to 2.7~Gyr.
The Ruprecht 147 rotation period sequence appears remarkably flat, 
with most stars cooler than $\teff < 5700$~K contained to within 
$\prot = 22$$\pm$2~days.
This is in sharp contrast with the steep mass dependence seen in the younger clusters, 
where periods tend to get longer toward decreasing mass. 

Now that we have identified a temporary epoch of stalled braking
\citep{Agueros2018, Curtis2019, Douglas2019}, 
we suggest that this flat sequence is produced by the mass-dependent duration of the epoch of stalled braking. While the higher-mass stars spin more rapidly than lower-mass stars at the age of Praesepe, they resume spinning down earlier, and so  catch up to their lower-mass siblings just as these resume spinning down. 

The rotation period distribution for cool dwarfs observed by 
\textit{Kepler} shows a gap at intermediate periods, 
which has been interpreted as evidence for a lull in the local star formation rate \citep{AmyMdwarfs, AmyKepler, Davenport2017, Davenport2018}. 
However, the \prot\ sequences for NGC~752 (1.4 Gyr) and Ruprecht~147 intersect this gap, thereby refuting the idea that this gap has a single age or was 
formed by any single event in time. 
Instead, perhaps the gap is created by an event in 
the evolution of the magnetic dynamo, which proceeds at different rates for stars of different masses (and convective turnover times).

We also used our new data for Ruprecht 147 to determine when stars resume spinning down as a function of mass,
and we find that 0.55~\msun\ stars are stalled for $\approx$1.3~Gyr. 
The steep mass dependence also means that this phenomenon might present a big obstacle for age-dating even lower-mass stars with rotation. 
Unfortunately, we were unable to measure \prot\ 
for $\leq$0.55~\msun\ stars in Ruprecht~147 to test this hypothesis.

\citet{Spada2019} used our new \prot\ data for NGC~6811 \citep{Curtis2019}
to re-calibrate an angular momentum evolution model that 
incorporates core--envelope coupling, which these authors  claim 
can explain the stalled braking phenomenon. 
Indeed, the \citet{Spada2019} model more closely matches our new Ruprecht~147 \prot\ 
data than any other model we are aware of, 
including the purely empirical \citep[e.g.,][]{Barnes2003, barnes2007, Barnes2010, Angus2019} or 
the semi-physical varieties 
\citep[e.g.,][]{vanSaders2013, Matt2015, Gallet2015}.
There is still some tension with the data, which perhaps can be mitigated by incorporating these measurements into their calibration. In that case, however, we would require \prot\ data for other old stars to validate a re-tuned model.
In the interim, to age-date old K~dwarfs, 
we recommend using the Ruprecht~147 \prot\ sequence as an empirical 2.7~Gyr gyrochrone and projecting it in time using a $t^n$ ($n = 0.62$) braking law.

Finally, to derive empirically {\it how long} stars are stalled, we need to know when stars converge on the slow sequence as a function of mass. The difference in age between the Pleiades and Praesepe  is too large to work this out from the $\prot$ data for these two clusters alone, however. There has been progress in filling in the age gap 
with other clusters using 
ground-based data; 
for example, 
M34 \citep[200-250~Myr;][]{MeibomM34},
M48 \citep[400-450~Myr;][]{BarnesM48},
and M37 \citep[550~Myr;][]{Hartman2009}.
The situation will hopefully improve in the near future 
from additional ground-based surveys of young clusters, 
and from the high-cadence photometric imaging from
NASA's 
\textit{Transiting Exoplanet Survey Satellite}
\citep[\textit{TESS};][]{TESS}.


\acknowledgments

J.L.C. is supported by the National Science Foundation 
Astronomy and Astrophysics Postdoctoral Fellowship under award AST-1602662 and
the National Aeronautics and Space Administration under
grant NNX16AE64G issued through the \textit{K2} Guest Observer Program (GO 7035).
M.A.A. acknowledges support provided by the NSF through
grant AST-1255419.
S.T.D. acknowledges support provided by the NSF through grant AST-1701468.
S.P.M. is supported by the European Research Council under the European Union's Horizon 2020 research and innovation program (agreement No. 682393, AWESoMeStars).
S.H.S. acknowledges support by NASA Heliophysics LWS grant NNX16AB79G.
C.Z. is supported by a Dunlap Fellowship at the Dunlap Institute for Astronomy \& Astrophysics, funded through an endowment established by the Dunlap family and the University of Toronto.
Funding for the Stellar  Astrophysics Centre is provided by The Danish National Research Foundation (Grant agreement no.:  DNRF106)
The Center for Exoplanets and Habitable Worlds and the Penn State Extraterrestrial Intelligence Center are supported by the Pennsylvania State University and the Eberly College of Science.

We thank Sydney Barnes, Mark Giampapa, Jennifer van Saders, Travis Metcalfe, Eric Mamajek, Garrett Somers, Federico Spada, David Gruner, Rayna Rampalli,
and the attendees of the Thinkshop~16 meeting hosted at 
the Leibniz Institute for Astrophysics in Potsdam, Germany (AIP) in 2019,\footnote{\url{https://thinkshop.aip.de/16/cms/}}
for enlightening conversations on rotation, activity, and our Ruprecht~147 results. 
We acknowledge preliminary work on the PTF data by Leo J.~Liu 
(not used in this study).
We are grateful to the \textit{K2} Guest Observer office and Ball Aerospace for 
re-positioning the Campaign 7 field to accommodate Ruprecht 147;
the staff at the various observatories cited in this study; 
the Harvard--Smithsonian Center for Astrophysics 
telescope allocation committee for granting access to Magellan and TRES;
Iv\'{a}n Ram\'{i}rez for assistance acquiring and analyzing MIKE spectra;
John O'Meara and John Bochanski for assistance with MagE; 
Edward Villanueva and Dan Kelson for their assistance with MIKE data;
Jeff Valenti, Debra Fischer, and John M.~Brewer for assistance with SME;
Katja Poppenhaeger for supporting our petition to repoint \textit{K2} Campaign 7,
Fabienne Bastien for supporting our GO 7035 proposal,
and 
Jennifer van Saders and Florian Gallet for sharing 
rotational isochrones generated from their 
angular momentum evolution models \citep{vanSaders2013, Gallet2015}.

This paper includes data collected by the \textit{Kepler} and \textit{K2} missions, 
which are funded by the NASA Science Mission directorate.
We obtained these data from the Mikulski Archive for Space Telescopes (MAST). 
STScI is operated by the Association of Universities for Research in Astronomy, Inc., 
under NASA contract NAS5-26555. 
Support for MAST for non-HST data is provided by the NASA Office of Space Science via 
grant NNX09AF08G and by other grants and contracts.

This work has made use of data from the European Space Agency (ESA)
mission {\it Gaia},\footnote{\url{https://www.cosmos.esa.int/gaia}} processed by
the {\it Gaia} Data Processing and Analysis Consortium (DPAC).\footnote{\url{https://www.cosmos.esa.int/web/gaia/dpac/consortium}} Funding
for the DPAC has been provided by national institutions, in particular
the institutions participating in the {\it Gaia} Multilateral Agreement.

This work made use of the \url{https://gaia-kepler.fun} cross-match database, created by Megan Bedell.

This paper is based on observations obtained with the Samuel
Oschin Telescope as part of the Palomar Transient Factory
project, a scientific collaboration between the California Institute
of Technology, Columbia University, Las Cumbres Observatory,
the Lawrence Berkeley National Laboratory, the National
Energy Research Scientific Computing Center, the University
of Oxford, and the Weizmann Institute of Science.

The Robo-AO system was developed by collaborating partner institutions, the California Institute of Technology and the Inter-University Centre for Astronomy and Astrophysics, and with the support of the National Science Foundation under Grant Nos. AST-0906060, AST-0960343, and AST-1207891, the Mt.~Cuba Astronomical Foundation and by a gift from Samuel Oschin.

This work utilized SOLIS data obtained by the NSO Integrated Synoptic Program (NISP), 
managed by the National Solar Observatory, 
which is operated by the Association of Universities for Research in Astronomy (AURA), Inc., 
under a cooperative agreement with the National Science Foundation.

This research has made use of NASA's Astrophysics Data System, 
and the VizieR \citep{vizier} and SIMBAD \citep{simbad} databases, 
operated at CDS, Strasbourg, France.

%

\vspace{5mm}
\facilities{\textit{Gaia}, \textit{K2}, PTF, PO:1.5m (Robo-AO)}


\software{The IDL Astronomy User's Library \citep{IDLastro}, 
          \texttt{K2fov} \citep{K2fov}}




\appendix
\section{Polynomial models for stellar property relations and color--period sequences} 
\label{a:models}

\subsection{An empirical color--temperature relation}\label{a:temp}

\begin{figure}[h]
\begin{center}
\includegraphics[width=4.in]{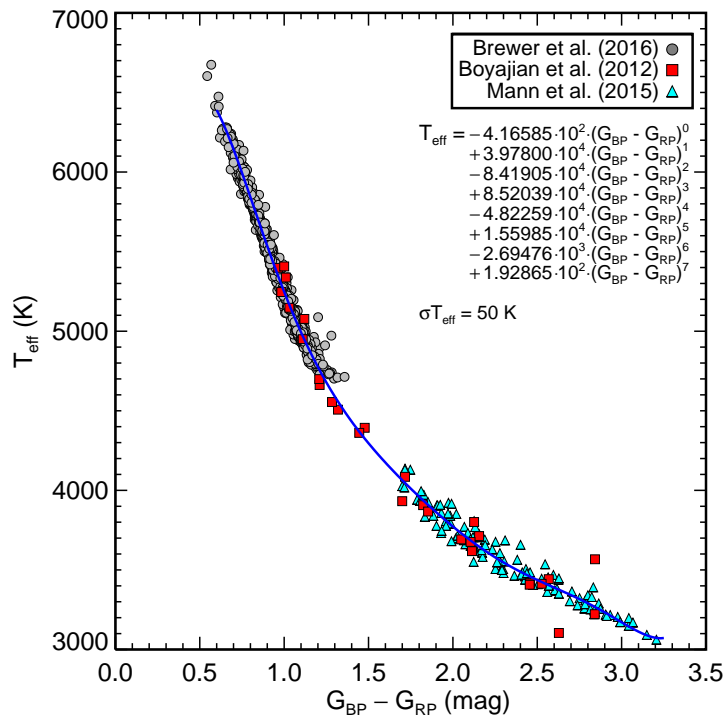}
\caption{An empirical color--temperature relation. This relation is constructed from benchmark stars drawn 
from \citet{Brewer2016}, \citet{Boyajian2012}, and \citet{Mann2015}. 
The scatter about the relationship implies a typical \teff\ precision of 50~K. 
The formula is provided in Table~\ref{t:models}.
\label{f:color_temperature}}
\end{center}\end{figure}

The \textit{Gaia} DR2 catalog included effective temperatures 
for $1.61\times10^8$ stars with $G<17$~mag and $3000 < \teff < 10,000$~K, 
determined from the broadband photometry \citep{Apsis2013, Apsis}. 
These \teff\ values can be heavily biased by interstellar reddening, which is unaccounted for in DR2.
For example, with $A_V = 0.3$, the \textit{Gaia} DR2 \teff\ values for solar-type stars in Ruprecht~147 are 
negatively biased by 300-400~K.
Fortunately, we know the amount of interstellar reddening for this and the other clusters we are studying, 
so we can de-redden the photometry and estimate accurate photometric temperatures with an empirical 
color--temperature relation of our own construction.
Figure~\ref{f:color_temperature} shows the empirical color--temperature relation 
we built from three benchmark samples: 
\begin{itemize}
    \item \citet{Brewer2016}: This catalog includes stellar properties and abundances calculated from HIRES spectra collected for the California Planet Survey, which were analyzed with SME \citep{sme, valenti2005} following the \citet{Brewer2015} procedure. We selected stars with high quality spectra 
    ($S/N > 70$) with $\teff > 4700$~K,
    low photometric errors ($\sigma G_{\rm BP}, G_{\rm RP} < 0.05; \sigma G < 0.025$), 
    fainter than $G > 3.7$~mag (to avoid any photometry problems for very bright stars), with $-0.5 < $[Fe/H]$<+0.5$~dex, and $d < 200$~pc to minimize the impact of interstellar reddening. 
    This includes 886 stars with $4702 < \teff < 6674$~K and $2.81 < \logg < 4.83$~dex. 
    \item \citet{Boyajian2012}: This catalog includes 30 stars with $3054 < \teff < 5407$~K and $-0.49 <$[Fe/H]$<+0.35$~dex, which were characterized with interferometry and spectral energy distribution analysis.
    \item \citet{Mann2015}: We selected 119 stars from this catalog with $3056 < \teff < 4131$~K and $-0.54 <$[Fe/H]$<+0.53$~dex, 
    after trimming 50 stars that were not cross-matched properly 
    (since the \teff\ range was adequately sampled by the properly-matched stars, we did not need to correct this). These stars were characterized with optical and  near-infrared spectroscopy.
\end{itemize}

The scatter about our color--temperature relation implies a typical \teff\ precision of 50~K. 
Furthermore, as noted by \citet{Rabus2019}, the \textit{Gaia} temperatures for 
stars cooler than $\teff < 4000$~K are inaccurate. Our relation,  
and the \citeauthor{Rabus2019} $M_G-\teff$ relation, more accurately represents the cool stars in our sample.

\subsection{The empirical Hyades main sequence}
\label{a:hyades}
We fit a polynomial to the \textit{Gaia} DR2 CMD for the Hyades cluster \citep{DR2HRD}, 
and we often use this as an empirical main sequence for selecting photometrically single stars in the field or in other clusters. For clusters, we typically use a broader cut to reject non-dwarf and large outlier stars, and then iteratively refit that cluster's CMD to make a model appropriate for selecting its single dwarfs.

\subsection{Polynomial fits to the converged, slow sequences in the color--period distributions for the Pleiades, Praesepe, NGC 6811, NGC 752, and NGC6819 + Ruprecht 147}
\label{a:cpd}
We fit polynomials to the de-reddened color--period distributions for various open clusters, 
focusing on the converged portions of their slow sequences. 
These are useful for qualitatively illustrating the empirical basis for gyrochronology and the stalling phenomenon, and for comparing field star periods to the benchmark cluster data in a way that is clean and simple, albeit approximate.

Prior to performing these fits, we used \textit{Gaia} DR2 astrometry, RVs, and CMDs to isolate likely-single stars, 
and also rejected binaries from the samples based on the identification of multiple rotation periods by the studies from which we collected the \prot\ values.
The data for these clusters are provided in the online journal in a table associated with Figure~\ref{f:cluster_prot}.
Our strategy for identifying benchmark single-star members for each of these samples is similar to the procedures used in this paper, as well as \citet{Douglas2019} and \citet{Curtis2019}.

Figures~\ref{f:cluster_prot} and \ref{f:cluster_prot_b} show these distributions in linear and logarithmic space, 
along with the models and their uncertainties. The coefficients for these models are provided in Table~\ref{t:models}. The uncertainties combine \prot\ measurement errors, apparent spread caused by differential rotation, and intrinsic dispersion in the true equatorial rotation periods (i.e., how tightly converged the sequences are). We opted for a single-valued rotation period deviation for each cluster, and found $\Delta \prot$ = 1~day for the Pleiades, Praesepe, and NGC 6811, and 2 days for NGC 752, and the joint sample of NGC 6819 and Ruprecht 147. A synthetic cluster sample can therefore be generated from these models with the application of Gaussian noise with $\sigma = \Delta \prot$.
For the Pleiades, its 1-day uncertainty translates to a larger relative dispersion compared to the more slowly rotating Praesepe and NGC 6811 sequences, and we suspect this reflects the intrinsic tightness of these intermediate-aged sequences compared to the Pleiades stars, which are still converging. For the older clusters, the 2-day dispersion likely reflects their larger \prot\ uncertainties (their longer rotation periods mean fewer cycles are captured in each light curve, inflating the uncertainty),
although differential rotation might also contribute.

\textit{Pleiades---}We adopt 120~Myr for the Pleiades, based on its lithium depletion boundary (125--130~Myr by \citeauthor{Stauffer1998}~\citeyear{Stauffer1998} and $115 \pm 5$~Myr by \citeauthor{Dahm2015}~\citeyear{Dahm2015}) and 
CMD isochrone ages (110-160~Myr from \citeauthor{Gossage2018}~\citeyear{Gossage2018} and  115-135~Myr from \citeauthor{Cummings2018young}~\citeyear{Cummings2018young}).
We adopted $A_V = 0.12$ for the interstellar extinction.
As \citet{PscEri} explained, the slow sequence is well-fit by a line of constant Rossby number, $Ro = 0.29$, 
    using the \citet{Cranmer2011} convective turnover time formula and our color--\teff\ relation. We also fit a quadratic polynomial to the sequence and provide both models in Table~\ref{t:models}.

\textit{Praesepe---}We adopt an age of 670~Myr and $A_V = 0.035$ for this cluster 
\citep[for more information, see table 1 and appendix A in][]{Douglas2019}. 
The polynomial fit to its single-star is taken from equation 1 in \citet{Douglas2019}.

\textit{NGC 6811---}We used 135 ``YY'' rotators from \citet{Curtis2019}, which satisfy single star membership criteria and are on the slow sequence; we adopted an age of 1.0~Gyr and $A_V = 0.15$ from that same work.

\textit{NGC~752---}\citet{Agueros2018} presented \prot\ for 12 members; however, 
only eight have \textit{Gaia} DR2 consistent with single-star membership. 
These rotators are clustered into two groupings in color, so the best we can do is fit a straight line between these clumps. We adopted an age of 1.4~Gyr, based on 
the 1.34~Gyr value from \citet{Agueros2018} and 1.45 Gyr value from \citet{Twarog2015}, 
and also applied $A_V = 0.1$ \citep{Twarog2015}.

\textit{NGC~6819 + Ruprecht~147---}The fundamental parameters for NGC 6819 and Ruprecht 147 were reviewed in Appendix~\ref{a:6819} and Section~\ref{s:147}, respectively. To construct this model, we divided the sequence into blue and red portions, 
split at $\gbr_0 = 0.85$. 
For the blue portion, we applied the Praesepe model
projected forward in time from 670~Myr to 2.7~Gyr using an $n = 0.62$ braking law. 
For the red portion, we fit a cubic polynomial to the $\prot > 18$~day rotators.  
We then refit this two-piece representation with 
a 5\ith\ order polynomial to produce a smooth relation, provided in Table~\ref{t:models}. This polynomial model is valid over the range $0.75 < \gbr_0 < 2.05$ and $3700 < \teff < 6000$~K. 
We opted for this two-piece approach because we assume that 
the Praesepe projection model remains valid at this age
based on the consistency found for the 1-Gyr-old NGC~6811 cluster, 
despite the lack of benchmark rotators in the $6000 < \teff < 6200$~K range in our joint 2.7~Gyr sample.

\begin{figure}\begin{center}
\includegraphics[trim=0.5cm 0.0cm 0.5cm 0cm, clip=True,  width=3.4in]{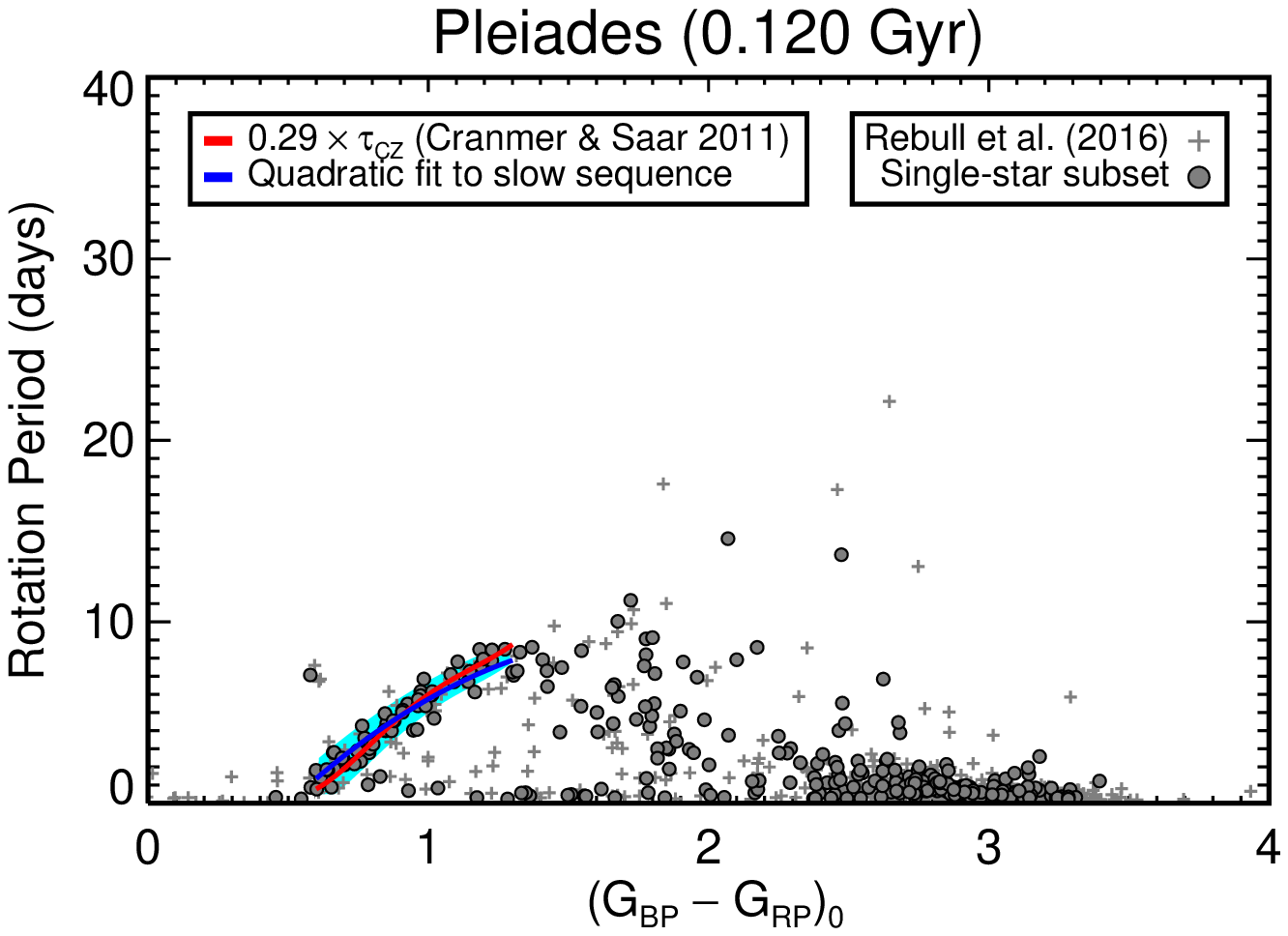}
\includegraphics[trim=0.5cm 0.0cm 0.5cm 0.0cm, clip=True,  width=3.4in]{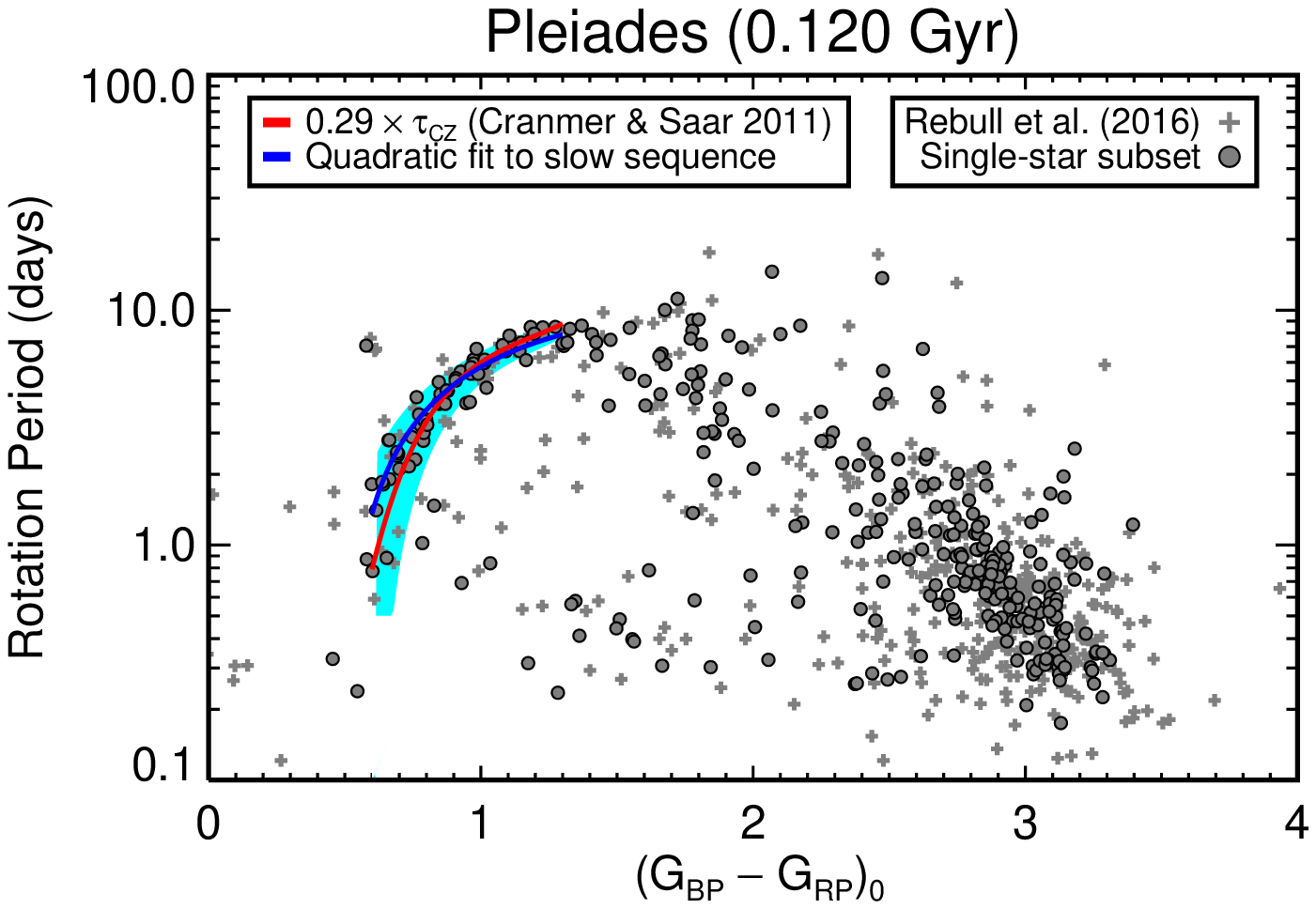}
\includegraphics[trim=0.5cm 0.0cm 0.5cm 0cm, clip=True,  width=3.4in]{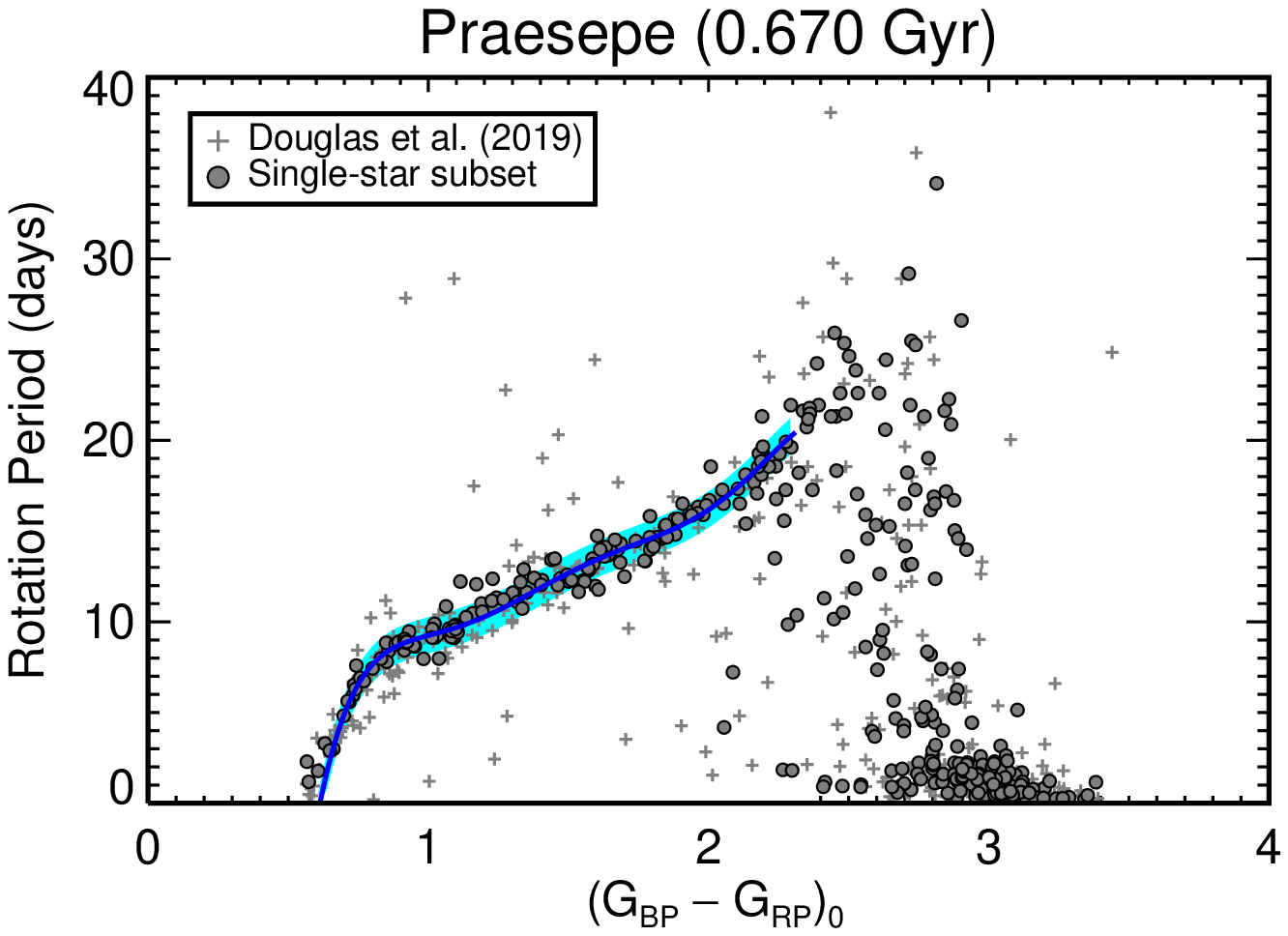}
\includegraphics[trim=0.5cm 0.0cm 0.5cm 0.0cm, clip=True,  width=3.4in]{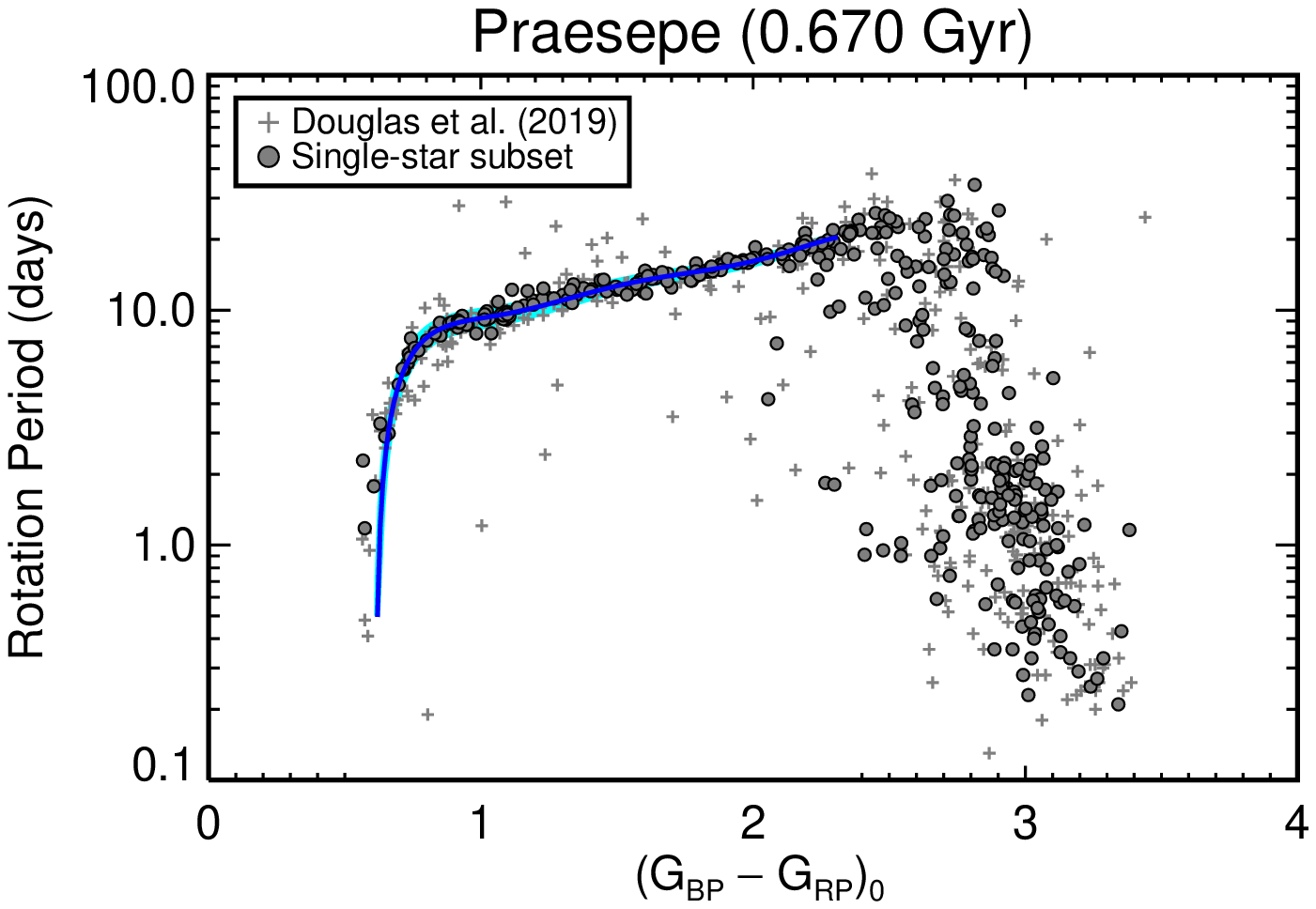}
\includegraphics[trim=0.5cm 0.0cm 0.5cm 0cm, clip=True,  width=3.4in]{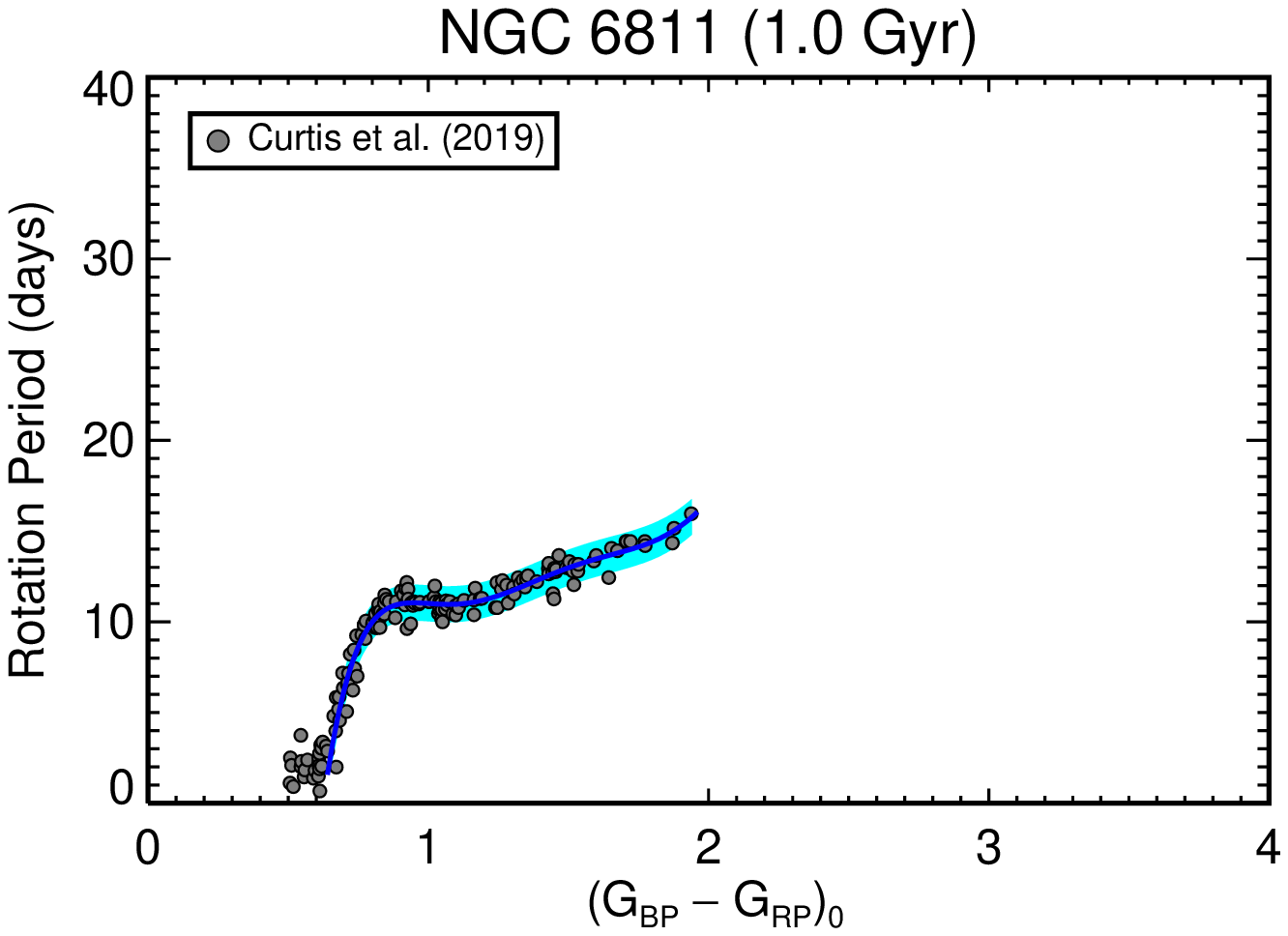}
\includegraphics[trim=0.5cm 0.0cm 0.5cm 0.0cm, clip=True,  width=3.4in]{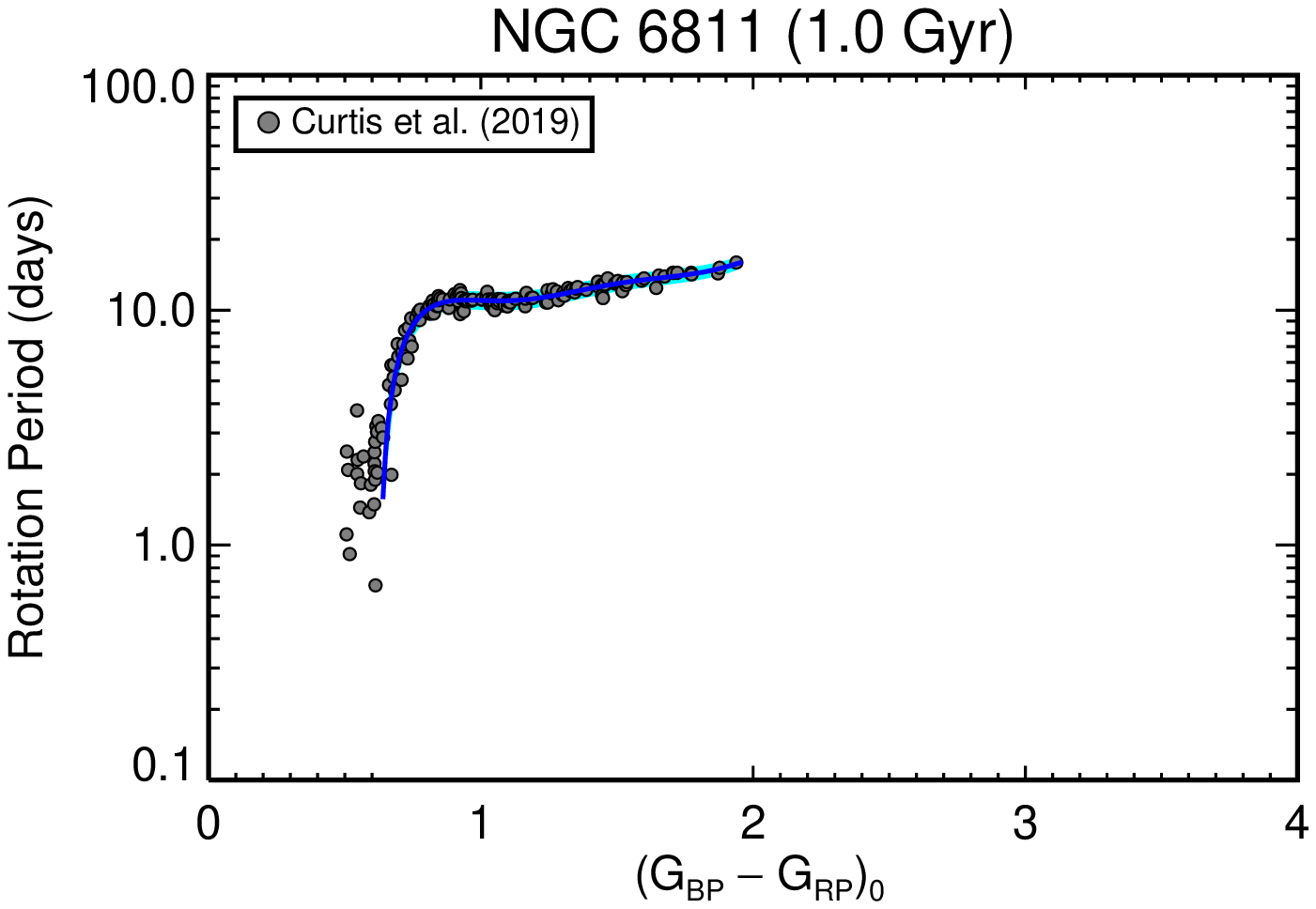}
  \caption{Color vs.~rotation period for 
  the Pleiades \citep[120 Myr;][]{Rebull2016},
  Praesepe \citep[670 Myr;][]{Douglas2019}.
  and   NGC~6811 \citep[1.0~Gyr;][]{Curtis2019}.
  (NGC~752 \citep[1.4~Gyr;][]{Agueros2018},
  and NGC~6819 \citep[2.5~Gyr;][]{Meibom2015} with Ruprecht~147 (2.7~Gyr; this work) are
  shown in Figure~\ref{f:cluster_prot_b}).
  Polynomial fits to the slowly rotating sequences are overlaid, and their coefficients are 
  provided in Table~\ref{t:models}.
  The cyan bands show an approximation
  of the dispersion for each sequence ($\Delta \prot$ = 1~day applied for Pleiades, Praesepe, and NGC~6811; 2 days applied for NGC~752, NGC~6819,  and Ruprecht~147). For younger clusters, the dispersion likely reflects the intrinsic tightness of the sequences, which are still converging. At older ages, the dispersion most likely reflect the \prot\ measurement uncertainties (although differential rotation might also contribute).
   \label{f:cluster_prot}}
\end{center}\end{figure}

\begin{figure}\begin{center}
\includegraphics[trim=0.5cm 0.0cm 0.5cm 0cm, clip=True,  width=3.4in]{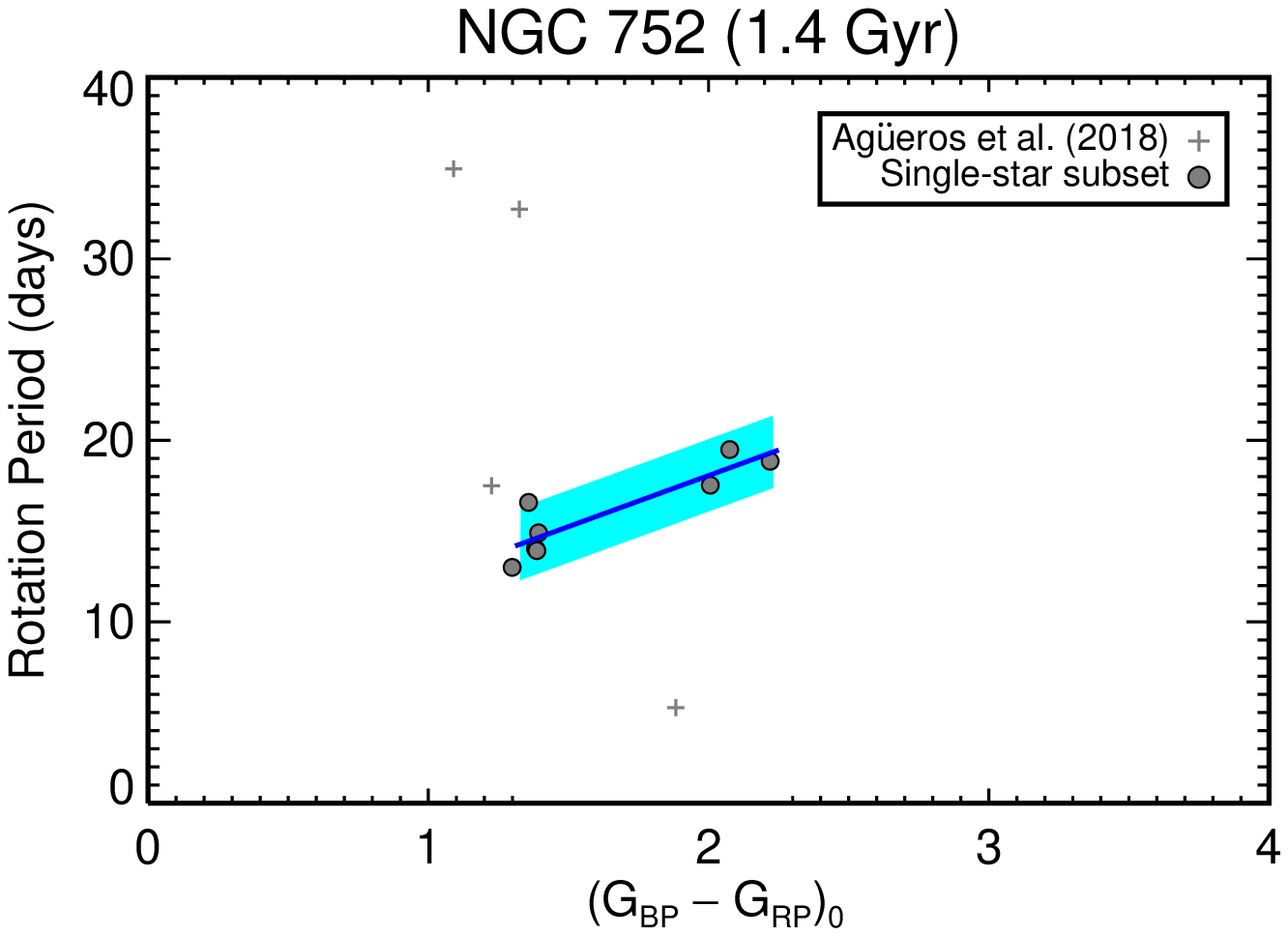}
\includegraphics[trim=0.5cm 0.0cm 0.5cm 0.0cm, clip=True,  width=3.4in]{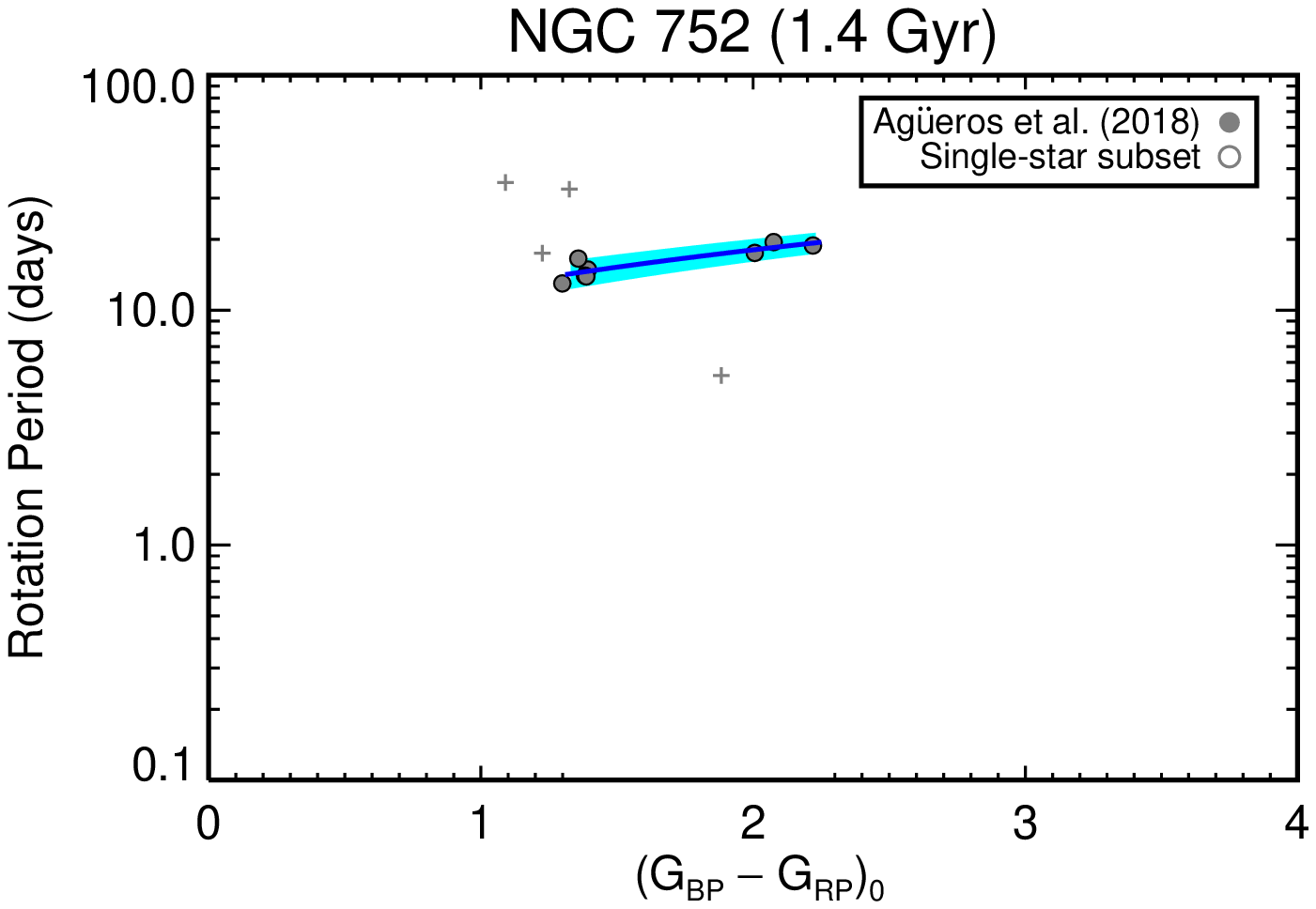}
\includegraphics[trim=0.5cm 0.0cm 0.5cm 0cm, clip=True,  width=3.4in]{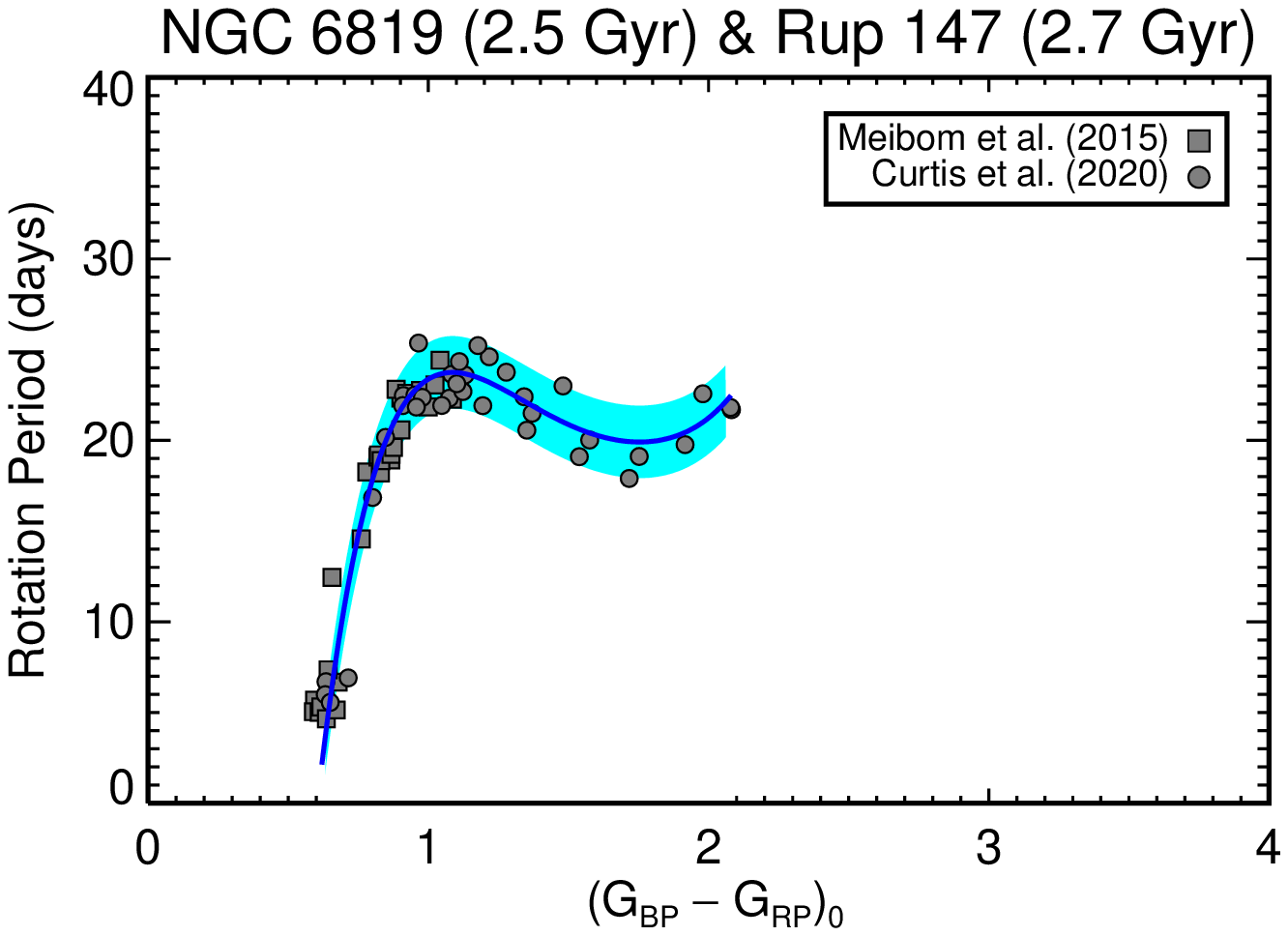}
\includegraphics[trim=0.5cm 0.0cm 0.5cm 0.0cm, clip=True,  width=3.4in]{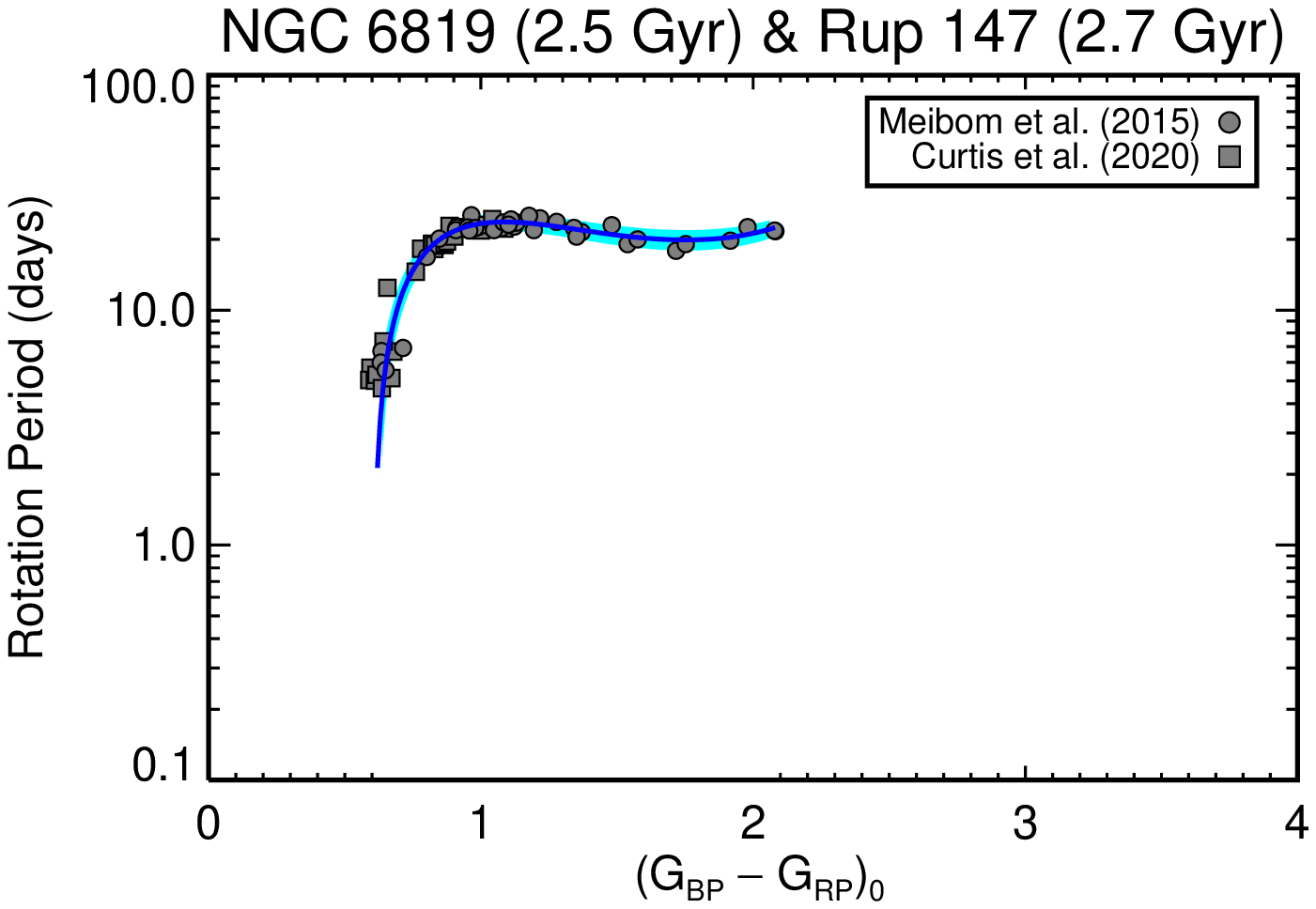}
  \caption{Continued from Figure~\ref{f:cluster_prot}---Color vs.~rotation period for 
    NGC~752 \citep[1.4~Gyr;][]{Agueros2018},
  and NGC~6819 \citep[2.5~Gyr;][]{Meibom2015} with Ruprecht~147 (2.7~Gyr; this work).
Polynomial fits to the slowly rotating sequences are overlaid, and their coefficients are 
  provided in Table~\ref{t:models}.
The cyan bands show an approximation
  of the dispersion for each sequence ($\Delta \prot$ = 1~day applied for Pleiades, Praesepe, and NGC~6811; 2 days applied for NGC~752, NGC~6819,  and Ruprecht~147). For younger clusters, the dispersion likely reflects the intrinsic tightness of the sequences, which are still converging. At older ages, the dispersion most likely reflect the \prot\ measurement uncertainties (although differential rotation might also contribute). 
   \label{f:cluster_prot_b}}
\end{center}\end{figure}

\subsection{The lower envelope of the \textit{Kepler} field \prot\ distribution}
\label{a:lower}
In Section~\ref{s:resume}, we posit that the lower envelope of the color--period distribution for 
\textit{Kepler} field stars measured by \citet{AmyKepler} represents the \prot\ at which initially rapidly rotating stars converge on the slow sequence. We then use this model together with the \prot\ data for NGC 6819 and Ruprecht 147 to estimate the age at which stalled stars resume spinning down. 

Regarding the stalled \prot\ sequence, we 
first created a representation of the 
lower envelope of the \textit{Kepler} \prot\ distribution, 
which we then fit with a sixth-order polynomial.
For this representation, 
we used the Pleiades sequence for $\gbr_0 < 0.93$, 
which we modeled with a line of constant Rossby number, 
$R_o = 0.29$, 
using the \teff-based 
convective turnover time formula from \citet{Cranmer2011}.
The density of the \textit{Kepler} \prot\ distribution 
drops significantly at $\gbr_0 > 2.1$ ($\teff < 3700$~K, $M_\star < 0.54$~\msun, M1). 
For these lower-mass stars, we adopt the Praesepe slow sequence, 
which appears to track the lower envelope 
at $\gbr_0 > 1.9$ ($M_\star < 0.59$~\msun), 
and which we know is stalled based on our analysis of the NGC~6811 cluster.
For the intervening color range, 
$0.93 < \gbr_0 < 1.90$, 
we marked points by eye, which we added to the representation.
The resulting polynomial is shown overlaid on the
trimmed \textit{Kepler} \prot\ distribution 
in the top panel of Figure~\ref{f:resume},\footnote{Appendix~\ref{a:kepler} describes our procedure for preparing the \textit{Kepler} \prot\ data 
from \citet{AmyKepler} for presentation in this study.}
and is provided in Table~\ref{t:models}.

\begin{deluxetable}{lcrrrrrrrr}
\tablecaption{Polynomial models for estimating stellar properties and for describing 
cluster rotation period sequences \label{t:models}}
\tabletypesize{\scriptsize}
\tablewidth{0pt}
\tablehead{
\colhead{Property, $Y$} & \colhead{$\gbr_0$} & 
\colhead{$c_0$} & \colhead{$c_1$} &  \colhead{$c_2$} &  \colhead{$c_3$} &  \colhead{$c_4$} &  \colhead{$c_5$} & \colhead{$c_6$} &  \colhead{$c_7$} 
}
\startdata
\multicolumn{10}{c}{\textbf{Stellar Properties}}\\
\multicolumn{10}{l}{\textit{Estimating effective temperature from the dereddened \textit{Gaia} DR2 color, $\gbr_0$}}\\
$\teff$ (K) & 0.55 to 3.25 & $-$416.585 & 39780.0 & $-$84190.5 & 85203.9 & $-$48225.9 & 15598.5 & $-$2694.76 & 192.865 \\\\
\multicolumn{10}{l}{\textit{The empirical Hyades main sequence: \gbr\ vs. $M_G$}}\\
$M_G$ (mag) & 0.2 to 4.2 & $-$0.0319809 & 4.08935 & 5.76321 & $-$6.98323 & 3.06721 & $-$0.589493 & 0.0417076 & $\cdots$ \\\\
\multicolumn{10}{c}{\textbf{Color--period sequences}}\\
\multicolumn{10}{l}{\textit{Kepler lower envelope}}\\
\prot\ (days) & 0.6 to 2.1 & 36.4756 & $-$202.718 & 414.752 & $-$395.161 & 197.800 & $-$50.0287 & 5.05738 & $\cdots$ \\\\
\multicolumn{10}{l}{\textit{Pleiades} (120 Myr)---Constant Rossby number, $Ro = 0.29$}\\
\prot\ (days) & 0.6 to 1.3 & 37.068 & $-$188.02 & 332.32 & $-$235.78 & 60.395 & $\cdots$ & $\cdots$ & $\cdots$ \\\\
\multicolumn{10}{l}{\textit{Pleiades} (120 Myr)---Quadratic fit}\\
\prot\ (days) & 0.6 to 1.3 & $-$8.467 & 19.64 & $-$5.438 &
 $\cdots$ & $\cdots$ & $\cdots$ & $\cdots$ & $\cdots$ \\\\
\multicolumn{10}{l}{\textit{Praesepe} (670 Myr)}\\
\prot\ (days) & 0.6 to 2.4 & $-$330.810 & 1462.48 & $-$2569.35 & 2347.13 & $-$1171.90 & 303.620 & $-$31.9227 & $\cdots$ \\\\
\multicolumn{10}{l}{\textit{NGC~6811} (1 Gyr)}\\
\prot\ (days) & 0.65 to 1.95 & $-$594.019 & 2671.90 & $-$4791.80 & 4462.64 & $-$2276.40 & 603.772 & $-$65.0830 & $\cdots$ \\\\
\multicolumn{10}{l}{\textit{NGC~752}: (1.4 Gyr)}\\
\prot\ (days) & 1.32 to 2.24 & 6.80 & 5.63 & $\cdots$ & $\cdots$ & $\cdots$ & $\cdots$ & $\cdots$ & $\cdots$ \\\\
\multicolumn{10}{l}{\textit{NGC~6819 + Ruprecht~147} (2.7 Gyr)}\\
\prot\ (days) & 0.62 to 2.07 & $-$271.783 & 932.879 & $-$1148.51 & 695.539 & $-$210.562 & 25.8119 & $\cdots$ & $\cdots$ \\
\enddata
\tablecomments{Model: $Y \approx \Sigma (c_i \, \gbr_0^i),$ where $Y$ is the modeled property. In each case, the degree of the polynomial was selected to minimize systematic patterns in the residuals, except for the NGC~725 relation which was effectively constrained by two points necessitating a linear model.}
\end{deluxetable}

\section{Super stamp light curves} \label{a:amc}
\begin{figure}[h]
\begin{center}
\includegraphics[trim=0.0cm 0.0cm 0.0cm 0cm, clip=True,  width=6.5in]{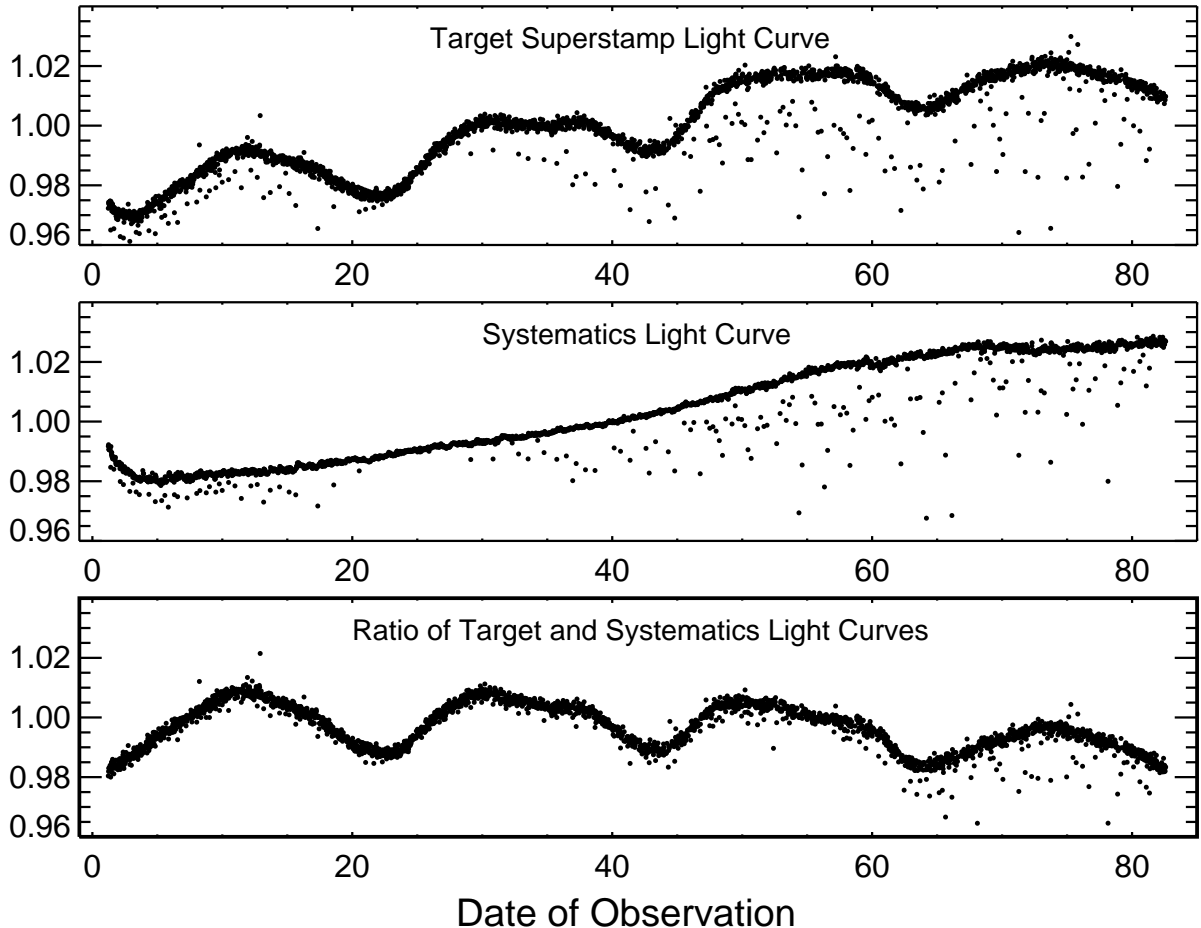}
\caption{Superstamp light curve preparation for 
EPIC~219651610, the star featured in 
Figures~\ref{f:k2} and \ref{f:ptf}.
(top) The target light curve, 
extracted from the superstamp 
with a 2-pixel-radius circular 
moving aperture. 
(middle) A systematics light curve 
produced by median-combining 
the normalized light curves 
for all superstamp targets 
in our sample. 
This shows that the superstamp target light curves share common 
systematic trends, 
including a prominent ramp-up 
in brightness over time.
(bottom) The ratio of the target 
and systematics light curves 
mostly removes this ramping 
and other features, 
isolating the periodic stellar rotation signal.
This procedure is largely for 
improving the aesthetics of 
our diagnostic plots, as the 
stellar rotation 
signal is prominent in the 
uncorrected version 
shown in the top panel. The superstamp light curves for individual targets and the systematics correction shown here are available for download in the online journal.
\label{f:lc_prep_amc}}
\end{center}\end{figure}

We provide light curve data for candidate cluster members 
located in the Ruprecht~147 superstamp, 
described in \citet{K2SUPERSTAMP}. 
The light curve procedure is described in \citet{Rebull2018}, 
and are extracted using 
1-, 2-, 3-, and 4-pixel circular moving apertures.
Position-dependent detrending was performed with the \texttt{K2SC} code \citep{K2SC}. 
We also calculated a systematics correction light curve for the 2-pixel apertures, 
shown in Figure~\ref{f:lc_prep_amc}, and the 3-pixel apertures.
The \prot\ values we measure from the systematics-corrected light curves are not negatively affected by this procedure; i.e., it does not introduce spurious signals, 
at least none with amplitudes comparable to the stellar rotation signals already 
prominent in the uncorrected version of the light curves.
However, this procedure does yield cleaner plots 
by removing a strong ramp-up in brightness that appears 
common to all of our superstamp targets.
The superstamp light curves and the systematic correction light curves are available for download from the online journal. 

\section{Gallery of selected variables in the vicinity of Ruprecht~147 rotators \\ observed with PTF} \label{a:ptf}

\begin{figure}[h]
\begin{center}
\includegraphics[width=3.in]{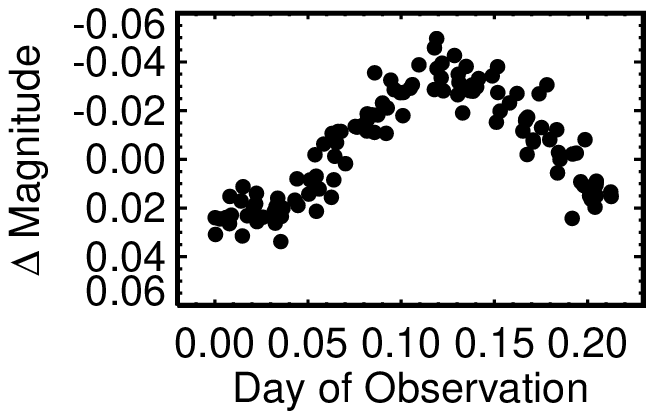}
\includegraphics[width=3.in]{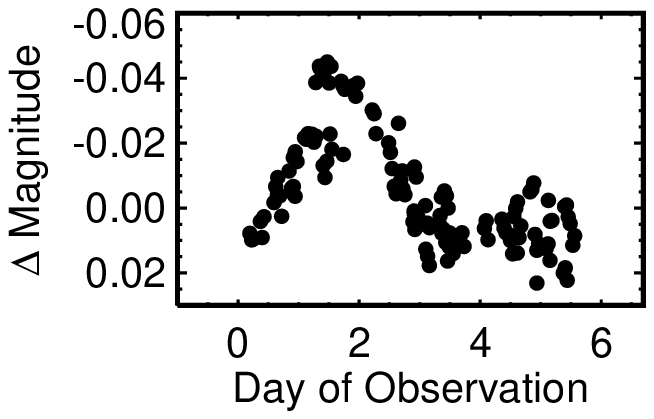}
\includegraphics[width=3.in]{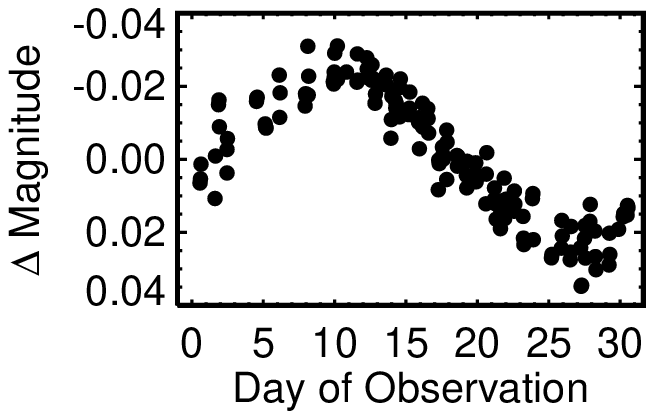}
\includegraphics[width=3.in]{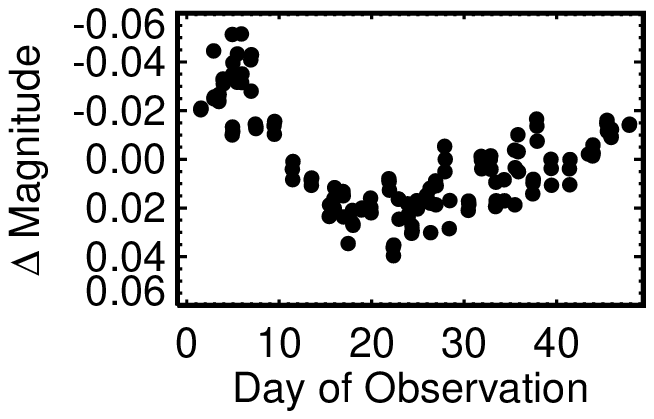}
\caption{Phase-folded PTF light curves for variable stars in the vicinity of Ruprecht~147 rotators, with periodicities ranging from 
5 hours to 50 day (shown), with additional longer-period variables noted. The cases shown here are drawn from the reference stars 
with strong peak powers in their Lomb--Scargle periodograms 
plotted in the panels of Figure~\ref{f:ptf}. 
(\textit{top left}) A neighbor to EPIC~219297228, Gaia DR2 4087800102544050688 is a distant G-type subgiant: $G = 15.13, \gbr = 1.017, \varpi = 0.409$~mas (2344~pc; RUWE = 1.84), with $\teff \approx 5650$-5810~K assuming $E(B-V) = 0.12$-0.16.
(\textit{top right}) A neighbor to EPIC~219234791, Gaia DR2 4087708155887598080 is a main-sequence K0 dwarf: $G = 14.10, \gbr = 1.21, \varpi = 2.63$~mas (377~pc; RUWE = 1.03), with $\teff \approx 5294$~K assuming $E(B-V) = 0.11$; the R147 value.
(\textit{bottom left}) A neighbor to EPIC~218984438, Gaia DR2 4084649348891870464 is a distant K giant: $G = 15.06, \gbr = 1.19, \varpi = 0.13$~mas (5575~pc; RUWE = 0.99).
(\textit{bottom right}) A neighbor to EPIC~219234791, Gaia DR2 4087710732868011904 is a distant K giant: $G = 14.46, \gbr = 1.42, \varpi = 0.18$~mas (4598~pc; RUWE = 1.03).
Distances are taken from \citet{Bailer_Jones2018}.
\label{f:ptf_variables}}
\end{center}\end{figure}

To illustrate the photometric precision of our PTF light curves, and our sensitivity to short and long 
periodicities, we present a gallery of four reference stars (non-members of Ruprecht~147) used in 
our photometric calibrations. Figure~\ref{f:ptf_variables} shows the phase-folded light curves for these 
stars, with periodicities ranging from 5 hours to 50 days. 

\section{\prot\ Validation Strategies, a Discussion of Outliers and Non-detections, and Notes on Other Stars} \label{a:valid}

Table~\ref{t:prot} provides notes on individual stars. This appendix provides supplemental material and expanded discussion to clarify those notes using case studies. References to the relevant appendices are provided in the table.

\subsection{The \prot\ distribution for the non-benchmark rotators of Ruprecht~147}
\begin{figure}
\begin{center}
\includegraphics[trim=0cm 0.0cm 0.0cm 0.0cm, clip=True,  width=3.5in]{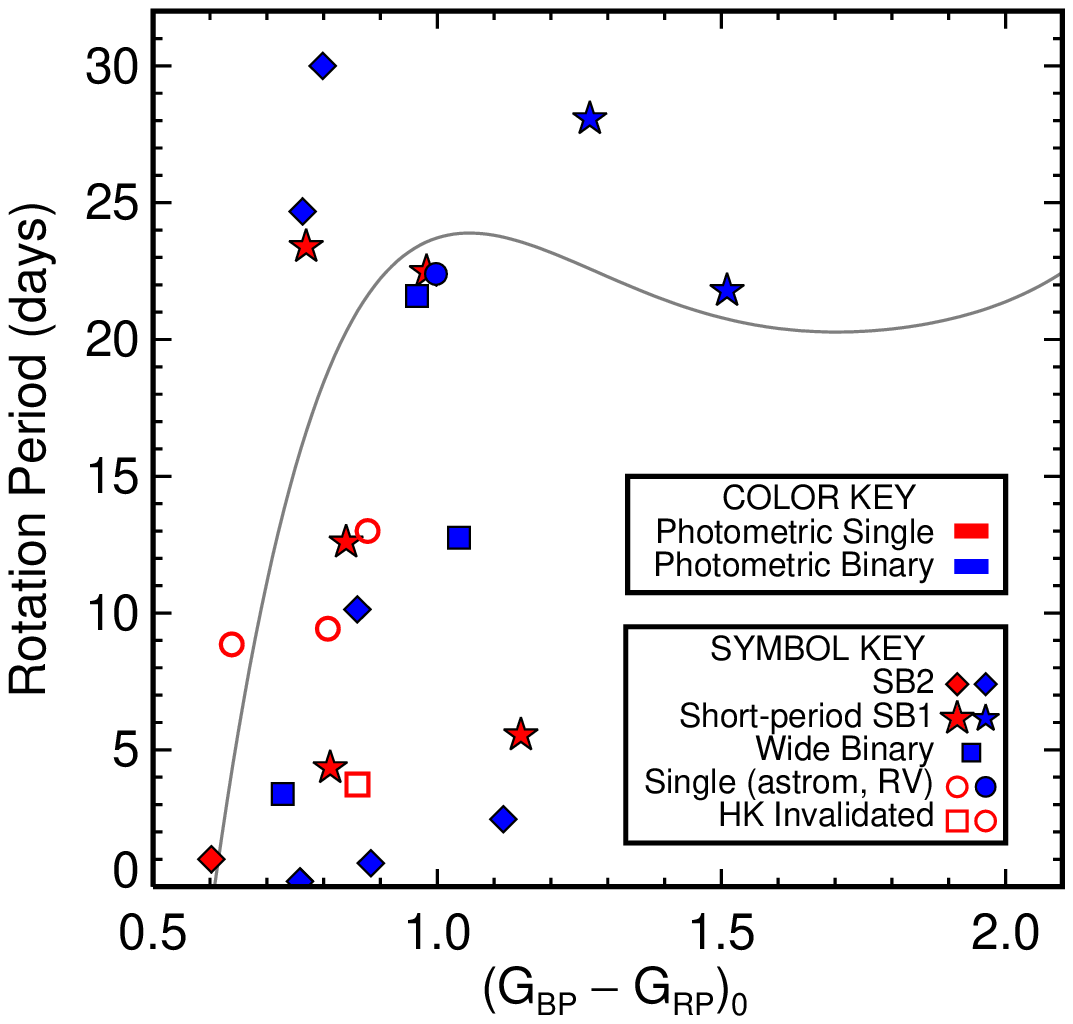}
\caption{Color--period distribution for non-benchmark stars in Ruprecht 147. The benchmark rotators of Ruprecht~147 are represented by the polynomial model (gray line, see Table~\ref{t:models} and Figures~\ref{f:color-period} and \ref{f:joint}).
Filtering out short-period binaries and photometric binaries rejects most \prot\ outliers. 
Stars that are classified as photometric binaries are shaded blue, and the photometrically single stars are shaded red.
Spectroscopic binaries with short orbital periods are shown with diamond symbols (SB2s) and five-point-star symbols (SB1s), and make up most of the $\lesssim5$~day outliers.
Astrometric binaries and long-period SB1s are marked with square symbols, but were only rejected as benchmarks if they also exhibited excess luminosities (i.e., no filled red squares are in this figure).
Stars that appear single according to astrometry and RVs, but which are photometric binaries are marked with circle symbols.
Three stars are plotted with open symbols to indicate that they have invalidated periods. In these cases, 
their \caiihk\ spectra reveal inactive chromospheres, which demonstrate that the rapid rotation apparent in their light curves cannot be hosted by these stars. 
\label{f:a:color-period_prelim}}
\end{center}\end{figure}

As explained in Section~\ref{s:bench}, we crafted a benchmark sample of rotators by rejecting photometric binaries 
and spectroscopic binaries that are known or suspected to have short orbital periods. 
We also rejected three stars based on discrepancies between the \prot\ apparent in  the light curve 
and the chromospheric activity level evident in their \caiihk\ spectra. 
The color--period diagram for these non-benchmark stars is presented in Figure~\ref{f:a:color-period_prelim}.
This figure demonstrates that most rapid outliers are short-period spectroscopic binaries. In these cases, 
tidal interactions are almost certainly responsible for their rapid rotation.
There are also a few long-period outliers at 28-30 days (EPIC 219442294 and EPIC 219634222); perhaps, the signals from both companions of each binary are interfering and causing spurious periodicities to dominate the periodogram analysis. 
In the remainder of this appendix section, we provide notes on additional stars not already included in the main paper. 

\subsection{Rapid solar analogs? Some yes, some no.}  \label{a:invalid}
At a given mass, rotation period  
tightly correlates with chromospheric emission. 
Considering a sample with a range of masses,
\citet{Noyes1984} found that 
dividing the rotation periods for stars 
by their convective turnover times (i.e., Rossby number, $R_o = \prot / \tau_{\rm CZ}$)
also yielded a tight relationship, referred to as the activity--Rossby relation
\citep[see also][]{mamajek2008}. 

We measured chromospheric activity via \caiihk\ line 
core emission for solar analog stars using 
high-resolution spectra with high signal-to-noise ratios 
obtained with the Magellan/MIKE and MMT/Hectochelle spectrographs
\citep[data collection and reduction described by][]{Curtis2016PhD}.
We measured the \lrphk\ activity index 
following \citet{Noyes1984} and \citet{wright2004}. 
\citet{PlanetR147} presents an example of this 
procedure applied to the planet host K2-231.
The results from our full \caiihk\ survey of FGK members will be presented in a separate study.

\begin{figure}
\begin{center}
\includegraphics[trim=0cm 0.0cm 0.0cm 0.0cm, clip=True,  width=3.5in]{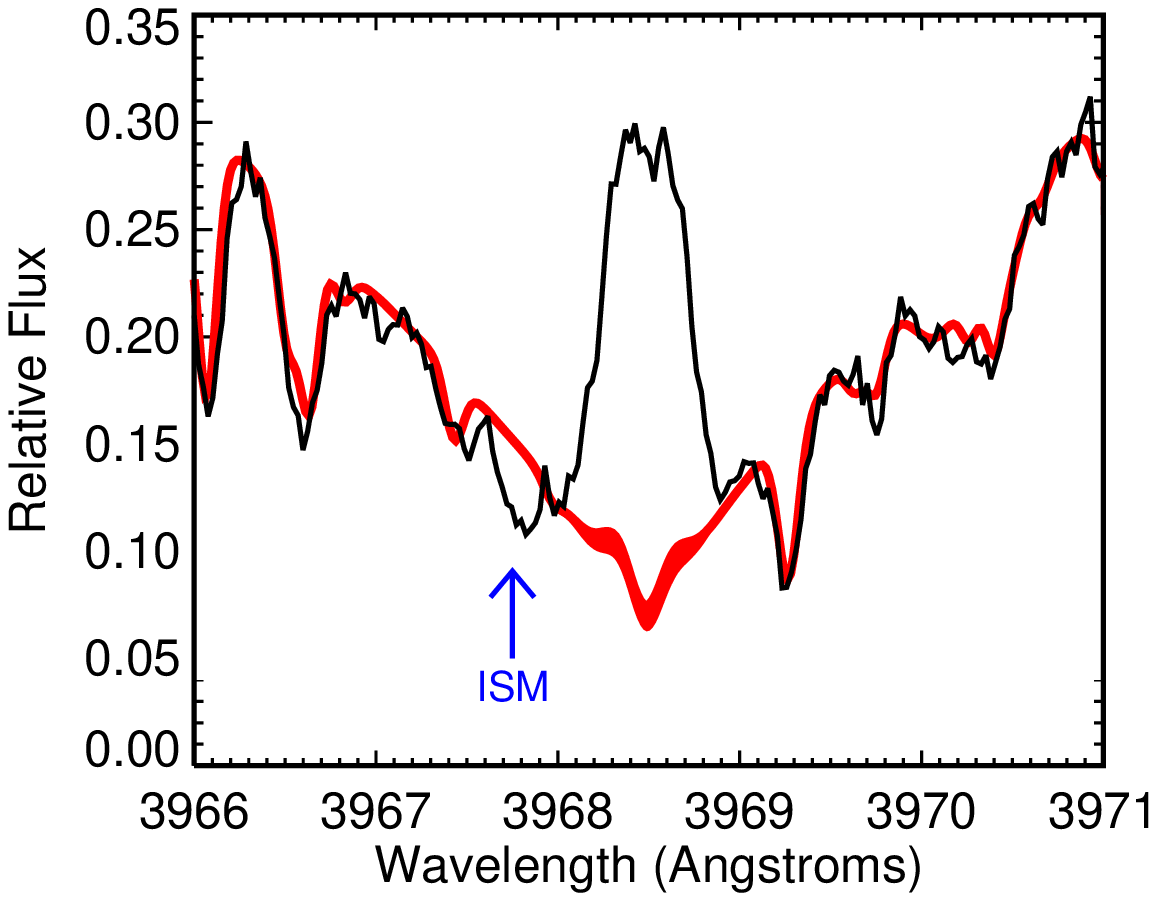}
\includegraphics[trim=0cm 0.0cm 0.0cm 0.0cm, clip=True,  width=3.5in]{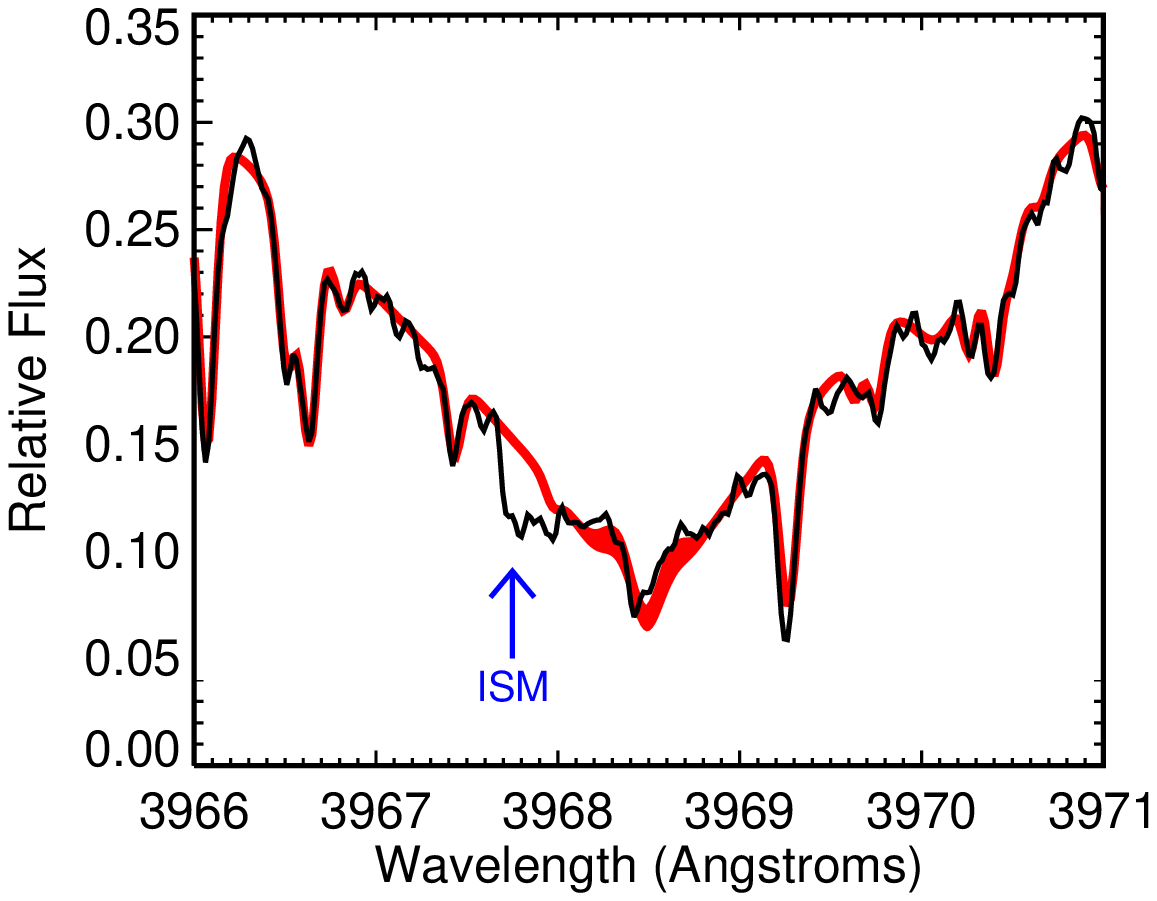}
\caption{\caii~H spectra for two solar twin members of Ruprecht 147 (black lines):
(\textit{left}) EPIC~219503117 (CWW~85, \teff\ = 5719~K, $\lrphk = -4.44$~dex, \prot\ =  11.0~days; spectrum from MMT/Hectochelle) 
and 
(\textit{right}) EPIC 219409830 (CWW~76, \teff\ = 5819~K, $\lrphk = -4.86$~dex, apparent \prot\ = 9.4~days; spectrum from Magellan/MIKE). The range of the contemporary solar cycle is represented with 
SOLIS/ISS spectra of the Sun taken on 2008 May 03 and 2014 July 01 (red shading). 
Note the interstellar absorption line blueward of the \caii~H line core (for more on ISM absorption and its impact on \caiihk\ emission metrics, see \citeauthor{Curtis2017}~\citeyear{Curtis2017}).
While the \textit{K2} light curves of each of these targets shows variability indicative of rapid $\sim$10-day rotation, 
only CWW~85 at left exhibits enhanced chromospheric emission at a level commensurate with this rotation. 
In contrast, the \caii\ emission for CWW~76 is consistent with the modern solar maximum, as is typical for the solar twins in this cluster and is expected for its age. 
The \textit{K2} light curve for the active star (CWW~85, EPIC~219503117) shows an eleven-day periodicity made clear by  asymmetries in the alternating dip patterns that reveals its true period. However, the \textit{K2} light curve for the inactive star (CWW~76, EPIC~219409830) shows a symmetric pattern with a period that is at odds with its inactive chromosphere. Doubling the apparent period would resolve the tension between its rotation and chromospheric emission. The photometric amplitude ($R_{\rm var}$) is also $5\times$ lower for this star relative to its active twin, and is inconsistent with analogous stars in the \textit{Kepler} field (median value of 95 reference stars is 8.3$\times$ higher); 
this likewise suggests that the true period is double the apparent value. 
Alternatively, the light curve for CWW~76 might be affected by a background star.
We searched \textit{Gaia} DR2 for neighbors within $20''$ with $G$ within 3 mag of our target and found one match: \textit{Gaia} DR2 4088004714786985344, at 7\farcs8 and 2.2~mag fainter than our target---a subgiant with a similar photometric color. If this scenario is correct then the background star would have a much larger photometric amplitude, 
which has been diluted by our brighter target.
These plots are produced following the same procedure used for K2-231 in figure~5 by \citet{PlanetR147}.
\label{f:ca}}
\end{center}\end{figure}

In Section~\ref{s:bl}, we identified two solar twins, which appear to be single according to all available 
astrometric, photometric, AO, and RV data. 
And yet, they are rotating rapidly relative to expectations from NGC~6819 and a gyrochronology model interpolating between Praesepe's color--period distribution and the Sun's period. 
Furthermore, both stars---EPIC~219503117 (CWW~85, $\lrphk = -4.44$~dex) and  EPIC~219692101 (CWW~97, $\lrphk = -4.58$~dex)---have anomalously high chromospheric emission, consistent with their rapid rotation periods.
Figure~\ref{f:ca} presents a portion of the MMT/Hectochelle spectrum for CWW~85 centered on the \caii~H line, 
which shows that the chromospheric activity for this star is greatly enhanced relative to the Sun, as is expected for its rapid rotation. 
However, their high activity levels are inconsistent with expectations for 2.7-Gyr-old solar twins based on the Sun's present range, the trend of declining activity with age seen in solar twins in the field \citep[e.g.,][]{DiegoHK}, 
and data for other benchmark clusters \citep{Giampapa2006, mamajek2008}.
We suggested that these overactive and rapid stars might be ``blue lurkers.'' Otherwise, perhaps they are merely binaries with short orbital periods and high mass ratios seen nearly face-on (so that they exhibit no detectable photometric excess or RV variability). 
Either way, they are clearly anomalous.


In the non-benchmark color--period distribution presented in Figure~\ref{f:a:color-period_prelim}, 
the symbols for most rotators are filled in, 
whereas three are represented by outlined symbols indicating 
that we rejected their apparent \prot\ from our catalog.
These three stars are also solar twins: 
EPIC 219800881 \citep[K2-231, CWW~93, \teff\ = 5695~K, \prot\ = 13.0~days, $\lrphk = -4.82$~dex][]{PlanetR147}, 
EPIC 219409830 (CWW~76, \teff\ = 5826~K, \prot\ = 9.43~days, $\lrphk = -4.86$~dex)
and EPIC 219256928 (CWW~88, \teff\ = 5641~K, \prot\ = 3.7~days, $\lrphk = -4.78$~dex).
All are photometrically single, the first two pass all single star criteria, 
and the third is an astrometric binary candidate and a known long-period SB1.
None are expected to be affected by tidal interactions, and yet their periods appear rapid.
However, the chromospheric emission we measured from MIKE spectra show these stars are inactive.
The right panel of Figure~\ref{f:ca} shows the case of EPIC~219409830, the activity of which is on par with solar maximum, 
as is expected for the age of Ruprecht 147.
According to the well-established relationship between chromospheric emission and stellar rotation 
\citep[e.g.,][]{Noyes1984, mamajek2008}, and between chromospheric emission and stellar age for solar twins 
\citep[e.g.,][]{DiegoHK}, 
we argue that there is no way that the periodic variability seen in the light curves indicates the 
true rotation periods for these stars. 
Thus, we reject these \prot\ measurements as invalid.

Regarding EPIC 219409830 (CWW 76, shown in the right panel of Figure~\ref{f:ca}), 
the \textit{Gaia} DR2 photometry and our MIKE spectroscopy indicate that it is only 80-100~K warmer than CWW~85 (the active star shown in the left panel of the same figure); i.e., they are nearly twins.
The LS periodograms for the \textit{K2} light curves for each indicate that they have similar rotation periods 
(9.4 and 11.0 days for CWW 76 and 85).
However, their \caiihk\ spectra definitively demonstrate that CWW~85 is magnetically active and CWW~76 is inactive; 
thus, their rotation periods cannot be similar. 
Doubling the apparent period of CWW~76 would increase its \prot\ to $\approx$19 days,
which is equal to the value expected from the activity--Rossby relation \citep{mamajek2008}.
Its photometric amplitude ($R_{\rm var}$) is also $5\times$ lower than CWW 85 (its active twin), 
and is inconsistent with analogous stars in the \textit{Kepler} field (the median value of 95 reference stars is 8.3$\times$ higher); 
this likewise suggests that the true period is double the apparent value. 
Alternatively, the light curve for CWW~76 might be affected by another star blended in the photometric aperture.
EPIC~219409830 appears to be single (11 HARPS RVs have a standard deviation of 40~\mps, and are consistent with RVs from Lick, 2012 MIKE, and \textit{Gaia}), so binarity is not likely the cause.
We searched \textit{Gaia} DR2 for neighbors within $20''$ with $G$ within 3 mag of our target and found one match: \textit{Gaia} DR2 4088004714786985344, at 7\farcs8 and 2.2~mag fainter than our target---a subgiant with a similar photometric color. If this scenario is correct then the background star would have a much larger photometric amplitude, 
which has been diluted by our brighter target.
We opt to remove this target from our benchmark sample at this time because we have not been able to 
conclusively determine its period.

\textit{EPIC 219256928---}This inactive star with an apparently rapid rotation period is an SB1 with an unknown orbital period. In this case, perhaps the light curve modulation comes from a fainter, more active secondary star. 

\textit{EPIC 219774323---}We report a period of 6.9 days for this star. However, the period we infer from its \lrphk\ using the activity--Rossby relation is approximately double that value. 
There is a weak hint of this in the light curve timing analysis, where the periods for the odd sets and even sets of dips are different; however, this difference is not significant enough to warrant our altering of the LS period.
We do note that our Praesepe-based gyrochrone predicts a period for this star of 12.9 days, which is only 7\% smaller than the double-period. We left the apparent \prot\ unchanged in the table. 

The \prot\ and \lrphk\ for EPIC~219601739, 219722212, and 219721519 are consistent. 
Results for all other members with \caiihk\ spectra will be presented in a separate study.

\subsection{Binaries with inactive primaries and low-mass secondaries} \label{a:FK}

Late-F and early-G dwarfs in Ruprecht~147 are expected to be inactive, according to the activity--Rossby--age relation 
\citep{mamajek2008}, and confirmed by our unpublished \caiihk\ survey.
In cases where such stars have lower-mass companions (e.g., a K dwarf), those secondaries are expected to be more active
due in part to their longer convective turnover times, 
and hence smaller Rossby numbers given the observed fact that the color--period sequence for the cluster is relatively flat.
In these situations, it is possible that the primary's brightness variations are negligible compared to the secondary's spot modulation signal, despite the difference in brightness that will dilute the amplitude for the secondary. 
Therefore, any apparent rotation period might be hosted by the secondary star.
We suspect this is the case for EPIC 219404735, EPIC 219442294, and EPIC 219661601.

\textit{EPIC 219404735}---This target is an SB2. Modeling the broadband photometry as a binary with a PARSEC isochrone, we find $M_1 = 1.14$ and $M_2 = 0.89$~\msun\ 
\citep[for details on the method, see section 3.1.3.1 in][]{Curtis2016PhD}. 
The \textit{K2} light curve shows a clear 24.7-day period, with a double-dip morphology typical for a K dwarf cluster member. Given the inferred masses for each component, we expect periods of 10.3 days and 23.6 days. We estimate $\tau_{\rm CZ} = 6.4$~days with 
the \citet{Cranmer2011} formula and the PARSEC \teff\ for the primary; 
the resulting Rossby number $Ro = 3.8$ would be extremely inactive. There are only nine stars in the \textit{Kepler} sample from \citet{AmyKepler} with larger Rossby numbers (prepared following the recipe in Appendix~\ref{a:kepler}). The more likely explanation is that the K0V secondary is responsible for the rotation period evident in the \textit{K2} light curve.

\textit{EPIC 219661601}---The \gbr\ for this star indicates \teff\ = 5856~K. 
It has a modest photometric excess ($\Delta G$ = 0.33 mag), but not large enough to satisfy our photometric binary criteria. With RUWE = 1.53, the star is a candidate wide binary. Our SME analysis of a MIKE spectrum indicates $\teff = 6045$~K 
(\citeauthor{Bragaglia2018}~\citeyear{Bragaglia2018} measured \teff\ = 5987~K from HARPS spectra).
Photometric modeling yields $M_1 = 1.107$~\msun\ (6054~K) and $M_2 = 0.721$~\msun\ (4477~K). 
This revised photometric temperature for the primary is consistent with our spectroscopic measurement.  
According to our gyrochrone model for NGC~6819 and Ruprecht~147, we expect such a star to have a 13.5-day period. 
Even if the 5856~K temperature is correct, this would make it a twin of 
EPIC~219601739 (5834~K), which has \prot\ = 16.7 days.
In other words, the apparent period of 20.5 days for EPIC 219661601 is too slow.
However, the \citet{mamajek2008} activity--Rossby--age relation predicts \lrphk\ = $-4.975$~dex for the primary, with a Rossby number $Ro \approx 2.1$, which means 
we should expect the primary to be quite inactive. 
This should make measuring the rotation period from white light brightness variations 
challenging. 
Instead, we suggest that the 20.5-day rotation signal is hosted by the K~dwarf secondary. According to the temperature found from our isochrone analysis, our Ruprecht~147  model expects a 21.6~day period, which is equal to the period measured from the light curve.

\subsection{K dwarfs with ambiguous rotation periods and the power of dip timing analysis}
 \label{a:timing}

\textit{EPIC 219333882---}This star ($\teff \approx 4950$~K, $M_\star \approx 0.83$~\msun, K2V) 
has a \textit{K2} light curve that we found challenging to interpret. 
There is a clear 11.6-day signal, which is favored by the LS periodogram. 
However, every third minimum is shallower than the other two, leading the ACF to favor a 34.8-day period.
Furthermore, the PTF light curve shows a 24.5-day periodicity. 
We have only one RV for this star, from MIKE, and it is consistent with being a single member.
The \textit{Gaia} proper motion and photometry also indicate that it is a single star; 
however, RUWE = 1.6, so it might actually be a wide binary.
There are four stars in \textit{Gaia} DR2 within 15$''$, 
but all are significantly fainter (4.1-7.3 mag in $G$), 
so background blending is unlikely a concern.

We examined the \textit{K2} light curve by timing the dips, and found an asymmetric spacing between the minima 
indicating that they come in pairs. The average spacing of the odd minima is $11.0 \pm 0.7$~day, and the spacing 
of the even minima is $12.3 \pm 0.6$~days, totalling to 23.3 days for a full rotation. 
If true, the first LS and ACF peaks would indicate the half-period harmonics. 
We suggest that this confusing pattern is caused by two active regions with particular growth and decay patterns 
that make it look like three peaks per rotation, 
whereas the PTF data happened to catch the star with only one dominant active region. 
This switching between one and two active regions is fairly common at this temperature and rotation period 
($\teff \approx 5000$~K, $\prot \approx 24$~days); figure 4 in \citet{Basri2018} indicates 
$N_{\rm single}/N_{\rm double} \approx 0.5$.
Alternatively, perhaps the confusion arises from interference from a faint companion. 
Although this star is RV- and photometrically-single, the astrometry indicates it is a wide binary. 
Perhaps, the variability of the faint secondary is interfering with the primary's rotational modulation pattern 
and producing this confusing signal. 
Based on this careful inspection of the \textit{K2} light curve minima timing, and supported by the clear signal in the PTF light curve, 
we conclude that the most likely rotation period for this star is 23.4 days, which places it on the slow sequence.

\textit{EPIC~219346771---}The \textit{K2} light curve for this star 
($\teff \approx 4247$~K, $M_\star \approx 0.72$~\msun, K5V) 
is also confusing at first glance, and appears to show two sets of periods: a long period of 27-30 days, with a second set of dips spaced every 8-10 days. However, timing successive pairs of dips shows that the odd pairs (first--second, third--fourth, fifth--sixth) are spaced at 8.1$\pm$0.7~days, and the even pairs (second--third, fourth--fifth, sixth--seventh) are spaced at 11.0$\pm$0.5~days. Combining these timings, we conclude that the period 
is likely 19.1~days, which would place it on the slow sequence as well. 
The \textit{K2} analysis panel for this star (in the Figure~\ref{f:k2} set online) includes vertical lines marking off the odd dips to aid the reader's interpretation of this timing analysis. 
Unfortunately, this star was not observed by PTF, so we cannot independently verify the timing period. 

\subsection{Unresolved background blends} \label{a:blends}
\textit{EPIC~219610822---}The \textit{K2} light curve for this star
($\teff \approx 5035$~K, $M_\star \approx 0.85$~\msun, K0V)
contains an EB with $P_{\rm orb} = 29.55$~d, 
with apparent eclipse depths of 0.1\% and 0.08\%. 
We suspect that the EB is in the distant background  
(based on TRES RVs that show the cluster member is not the host),
and is blended with the cluster member.
We also measured $\prot = 22.8$~d in the \textit{K2} light curve.
Given the large amplitude of the spot modulation signal, 
we assume that it is associated with the cluster member and 
not the background EB.

It is possible that some of the rotation signals we measure in our sample 
are caused by other background blends.
We can check \textit{Gaia} DR2 for bright neighbors in the \textit{K2} 
photometric apertures, and assess how such blends 
might bias our results.
The CMD panels in Figure Set~\ref{f:k2} show 
the full Ruprecht~147 catalog, together with 
the target star, and neighboring stars 
within 12$''$ that might be blended in the 
\textit{K2} photometric aperture. 
We plot both the apparent $G$ and 
absolute $M_G$ magnitudes (calculated using the parallax, and then adjusted to the R147 distance modulus 
by adding 7.4 mag) to illustrate that these blended stars 
tend to be significantly fainter than the target. 
Furthermore, their spectral types and luminosity 
classes are often inconsistent with the rotation 
signal apparent in the light curve.
For example, 
rapidly rotating giants are rare 
\citep{Pinsonneault2014}, 
although they can rotate as fast as 13-55~days \citep{Costa2015}.
Still, it is unlikely that each one of these background giants 
that happens to contaminate our targets also rotates rapidly.

\textit{EPIC~219297228---}This star has a bright neighbor (Gaia DR2 4087799651563238912), 
which is only 0.43 mag fainter, 0.1 mag redder in \gbr, 
and is separated by only 8\farcs25 ($\approx$two \textit{K2} pixels). 
This flux from this neighbor is certainly affecting the \textit{K2} observations. 
However, we can confidently attribute the 23-day rotation period signal 
to our target and not this interloper for two reasons. 
First, its absolute magnitude indicates it is a distant AGB star, 
whereas the light curve modulation looks like that of a middle-aged K dwarf. 
Second, the two stars are resolved in our PTF imaging, so we can 
cleanly extract light curves for each star.
According to our PTF light curve, our target has a 23.8 day period, 
confirming the \textit{K2} result.

\subsection{Photometric amplitudes, noise, and non-detections} \label{a:noise}

\begin{figure}
\begin{center}
\includegraphics[trim=0.0cm 0.0cm 0.0cm 0.0cm, clip=True,  width=6.5in]{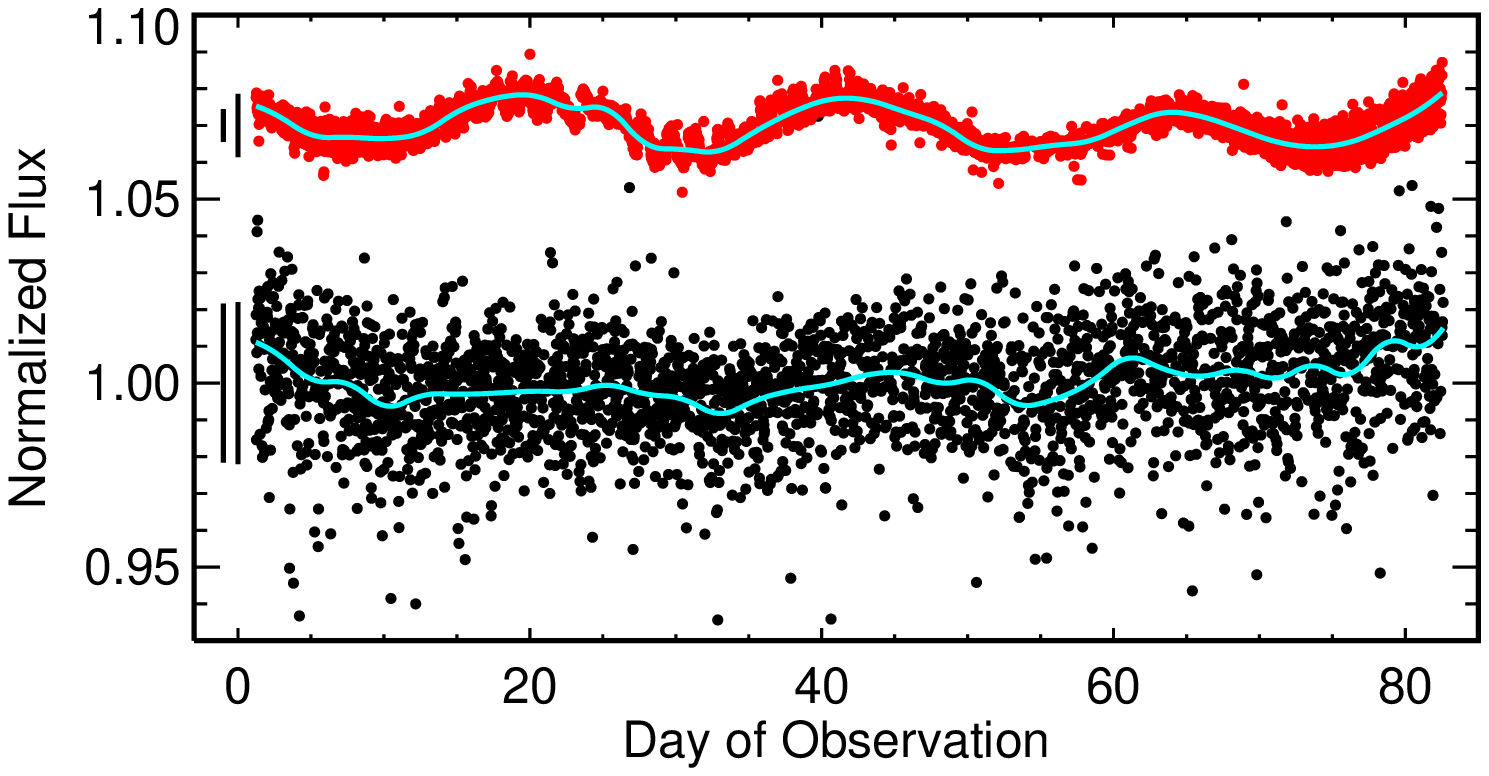}
\includegraphics[trim=0.0cm 0.0cm 0.0cm 0.0cm, clip=True,  width=3.0in]{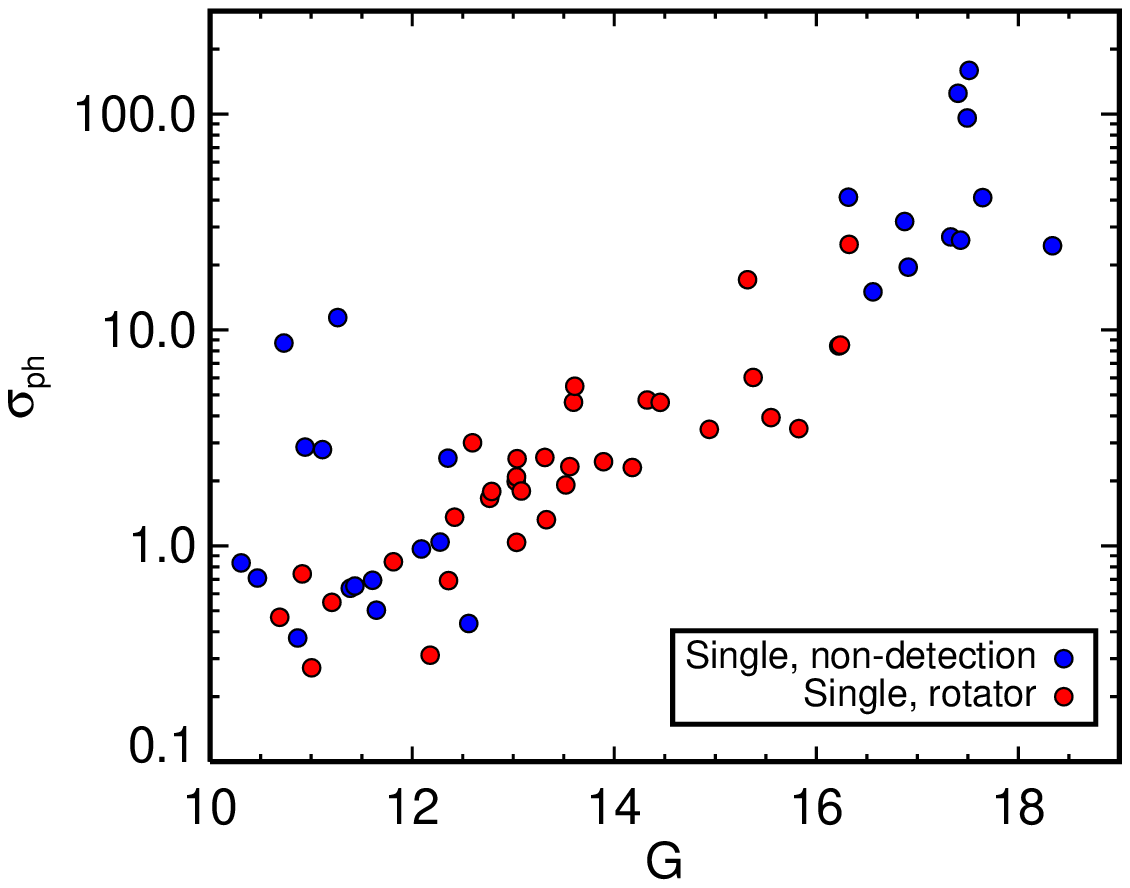}
\includegraphics[trim=0.0cm 0.0cm 0.0cm 0.0cm, clip=True,  width=3.0in]{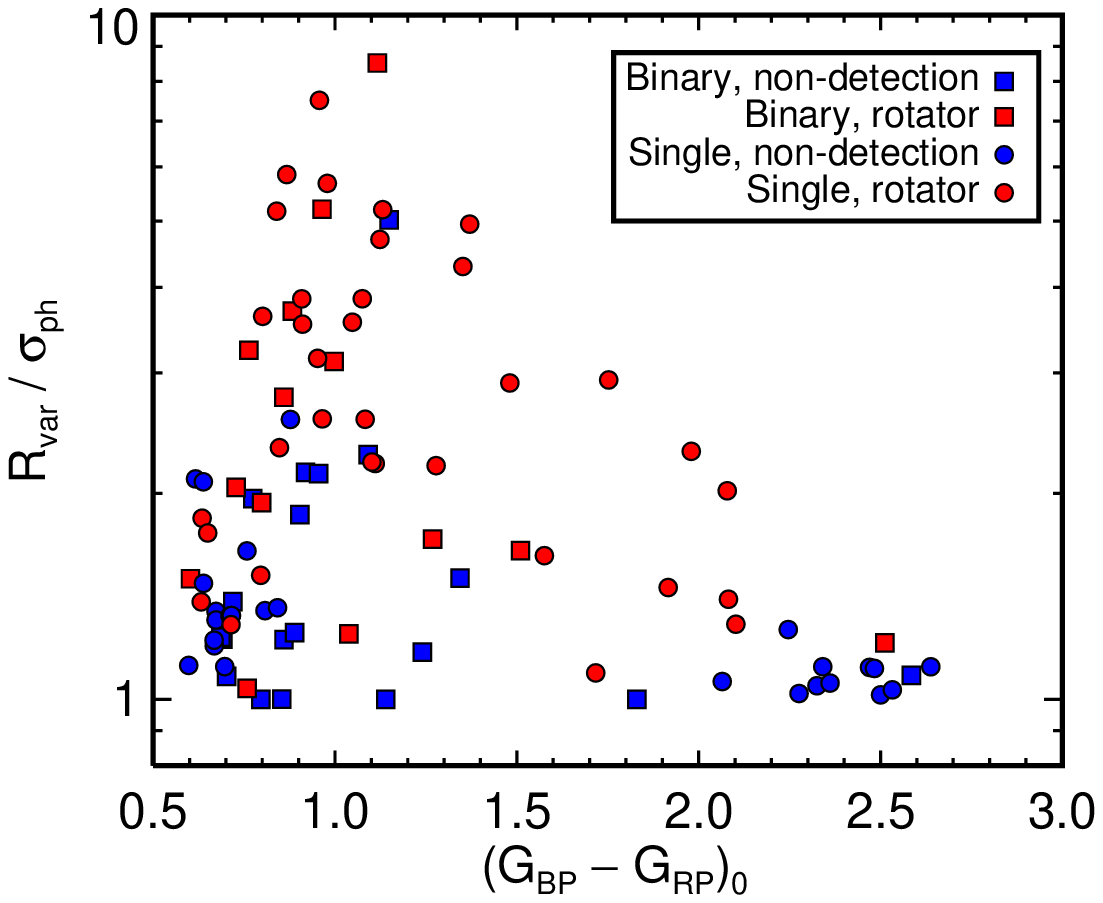}
\caption{(\textit{Top}) Light curves for two analogous stars illustrate how abnormally high photometric noise 
is suppressing our sensitivity to \prot\ in certain non-detection cases. 
These two stars are EPIC 219353203 (top, red points, \prot\ = 21.8 days) and EPIC 219616992 (bottom, black points, \prot\ is not clearly evident, but the Lomb--Scargle periodogram favors a 20.7-day period). 
The vertical lines at the left of the figure mark the photometric precision ($\sigma_{\rm ph})$ and 
amplitude ($R_{\rm var}$). In the top case, the photometric precision for this PDCSAP light curve is 
6.7$\times$ better than for the superstamp light curve for the bottom case.
(\textit{Bottom left}) The photometric precision for likely or effectively single stars is plotted as a function of apparent $G$ magnitude. The noise increases toward fainter magnitudes. We report \prot\ for stars shaded red, while the blue-shaded stars are non-detections. Between $13 < G < 16$, our success rate is high, 
with only three non-detections. Brighter (i.e., more massive) stars tend to be inactive by 2.7 Gyr, making their photometric amplitudes very weak, and we therefore report only few \prot\ detections. 
At $G > 16$, the typical noise increases by six times compared to the $14 < G < 16$ range, which likely is responsible for their non-detections.
(\textit{Bottom right}) We plot the ratio of the amplitude to the noise for an expanded sample that includes the stars from the left panel, and adds in the binaries (marked with squares). Stars with ratios $>$1 tend to have visually-validated \prot\ detections. Where this is not the case, it is often because the light curves exhibit morphologies that resist simple interpretation; i.e., they are binaries which show clear spot modulation patterns (and therefore have a high $R_{\rm var}/\sigma$ ratio), but the signals from both companions are likely interfering and confusing the periodogram analysis.
\label{f:a:amp_noise}}
\end{center}\end{figure}

The amplitude of the brightness fluctuations due to rotating spots can be assessed 
from the typical range of minima to maxima, called $R_{\rm var}$ \citep{Basri2010}.
We calculate this according to the interval between the 5\ith\ and 95\ith\ flux percentiles 
for the normalized light curves \citep{AmyMdwarfs}.
We also estimate the photometric precision using this interval technique performed on 
detrended light curves. 
For this, we fit a cubic basis spline to the light curves with break points spaced at 2.5 days, and then
divide out the model.
This removes spot modulation patterns and any remaining low-order systematics, 
and isolates the photometric noise, which we refer to as $\sigma_{\rm ph}$ and plot in the bottom left panel of Figure~\ref{f:a:amp_noise}. 
The ratio of $R_{\rm var} / \sigma_{\rm ph}$, plotted in the bottom right panel of the same figure, 
is $\approx$1 for most of our non-detections. 
Apparently in these cases the noise is too high relative to the spot signal.
Stars with clear rotation signals tend to have a higher ratio: the median value for 33 rotators is 2.3, and ranges from 1.2 to 7.1. 

There are other stars with ratios $>$1 for which we do not report a \prot. 
Many are binaries with large photometric excesses; 
in these cases, the rotation signal from the secondary is likely interfering with the primary's 
and confusing the \prot\ analysis.
Although multiple periods have been measured for binaries in the Pleiades \citep{Rebull2016} and 
Praesepe \citep{Douglas2017, Rebull2017}, 
those younger stars tend to spin much more rapidly than Ruprecht 147 stars, 
and often the second periodicity is fairly rapid. 
For example, 46 of the 58 multi-periodic stars in Praesepe analyzed by \citet{Douglas2017} have 
secondary $\prot < 5$~days.
This helps disentangle the two signals.
However, in our case where there are two stars likely spinning at $\sim$20~days, 
this is all but hopeless.

Restricting our list to only the main sequence dwarfs that are not classified 
as photometric binaries, SB2s, or short-period SB1s (i.e., our ``benchmark'' targets), 
there are three stars lacking \prot\ yet have ratios $>1$. 
EPIC 219800881 and EPIC 219256928 are addressed in Appendix~\ref{a:invalid}; briefly, 
their \caiihk\ spectra show them to be magnetically inactive, which invalidates 
the rapid periods apparent in their light curves.
Regarding EPIC 219698970, its superstamp light curve is strongly affected by a 
systematic, which probably biased $R_{\rm var}$ toward an artificially high value. 

\textit{EPIC~219616992---}This star ($\teff \approx 3765$~K, $M_\star \approx 0.54$~\msun, M1V) is located in the \textit{K2} superstamp.
The top panel of Figure~\ref{f:a:amp_noise} plots light curves for it and an analogous star (i.e., similar color and magnitude)---the light curve plotted in red at the top of the panel shows EPIC~219353203, which has a clear period of 21.8 days. The light curve plotted in black below that star shows EPIC~219616992---its period is not obvious in this light curve and we suspect the cause is the abnormally high photometric noise (6.7$\times$), which is suppressing our sensitivity to its period, given the spot amplitude expected from the analogous star. The figure shows a smoothed version of the noisy light curve (using a cubic basis spline used to calculate $\sigma_{\rm ph}$); there are subtle periodic dips spaced at 20.7~days. However, we require unambiguous visual validation of the \prot\ before accepting it into our benchmark rotator sample, and so this star remains a non-detection. 

\textit{EPIC~219675090---}A similar analysis of the \textit{K2} superstamp light curve for this star 
($\teff \approx 4055$~K, $M_\star \approx 0.64$~\msun, K5V) 
relative to an analog shows its light curve is also abnormally noisy---4.8$\times$ higher than stars of similar color and magnitude, which appears to be masking some periodic structure in the light curve  
that is weakly apparent in the second half of the light curve.
The \textit{K2} analysis panel in Figure Set~\ref{f:k2} shows the smoothed version of its light curve, 
which reveals a repeating sequence of narrow and deep dips followed by shallower and broader dips.
The timing of these dips indicates a period of 17.9 days, which we adopt as the validated \prot\ for this star.
If this is the true period, it too falls on the slow sequence for Ruprecht 147.

The bottom left panel of Figure~\ref{f:a:amp_noise} shows that the 
photometric noise increases by nearly six times at $G > 16$ (based on the median $\sigma_{\rm ph}$
for the 12 stars on our list with $14 < G < 16$ and the 17 with $G>16$). 
We selected analogous rotators from the \textit{Kepler} sample \citep{AmyKepler}: for the 721 stars with $\gbr > 2$, $\prot > 10$~days, the median amplitude is $R_{\rm var} \approx 7$~ppt, and 90\% range between 3-20 ppt. 
Our nine faintest stars lacking periods with $16.3 < G < 18.3$ have $\sigma_{\rm ph}$ = 
15, 19, 25, 27, 28, 32, 43, 126, 161 ppt.  
We conclude that the light curves for our $G > 16$ targets are too noisy to confidently detect stellar rotation.

\subsection{EPIC 219489683 is in the \textit{Kepler} intermediate period gap} \label{a:gap}

Section~\ref{s:gap} discussed how the distribution of color vs.~rotation period for the \textit{Kepler} field derived by 
\citet{AmyKepler} shows a bimodality, where there is a narrow gap that approximately traces a line of constant Rossby number. We explained that the Ruprecht 147 rotation period sequence crosses the gap, and concluded that this challenges the scenario where the gap is formed by a temporary lull in star formation rate (i.e., the gap does not have a single-valued age, so cannot have been created by a single event in time). 

\begin{figure}[h]
\begin{center}
\includegraphics[trim=0.0cm 0.0cm 0.0cm 0.0cm, clip=True,  width=3.5in]{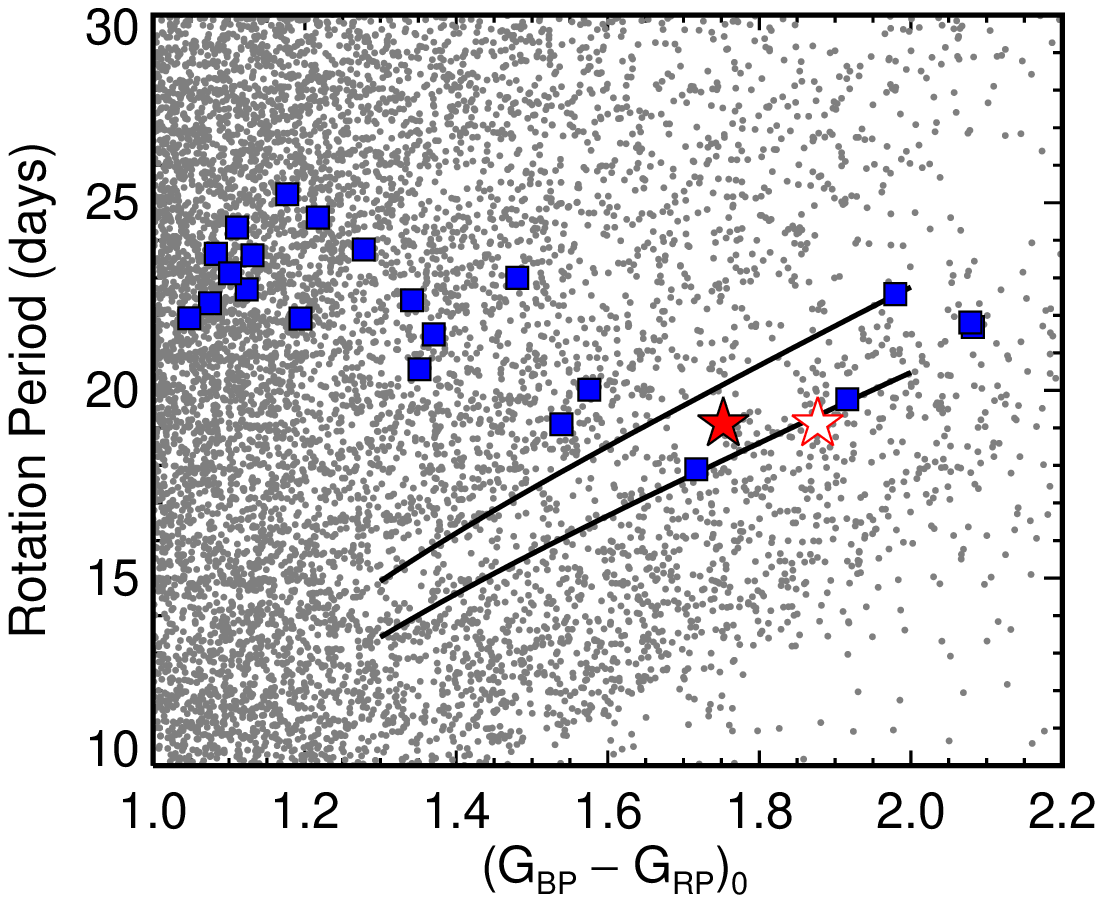}
\includegraphics[trim=0.0cm 0.0cm 0.0cm 0.0cm, clip=True,  width=3.5in]{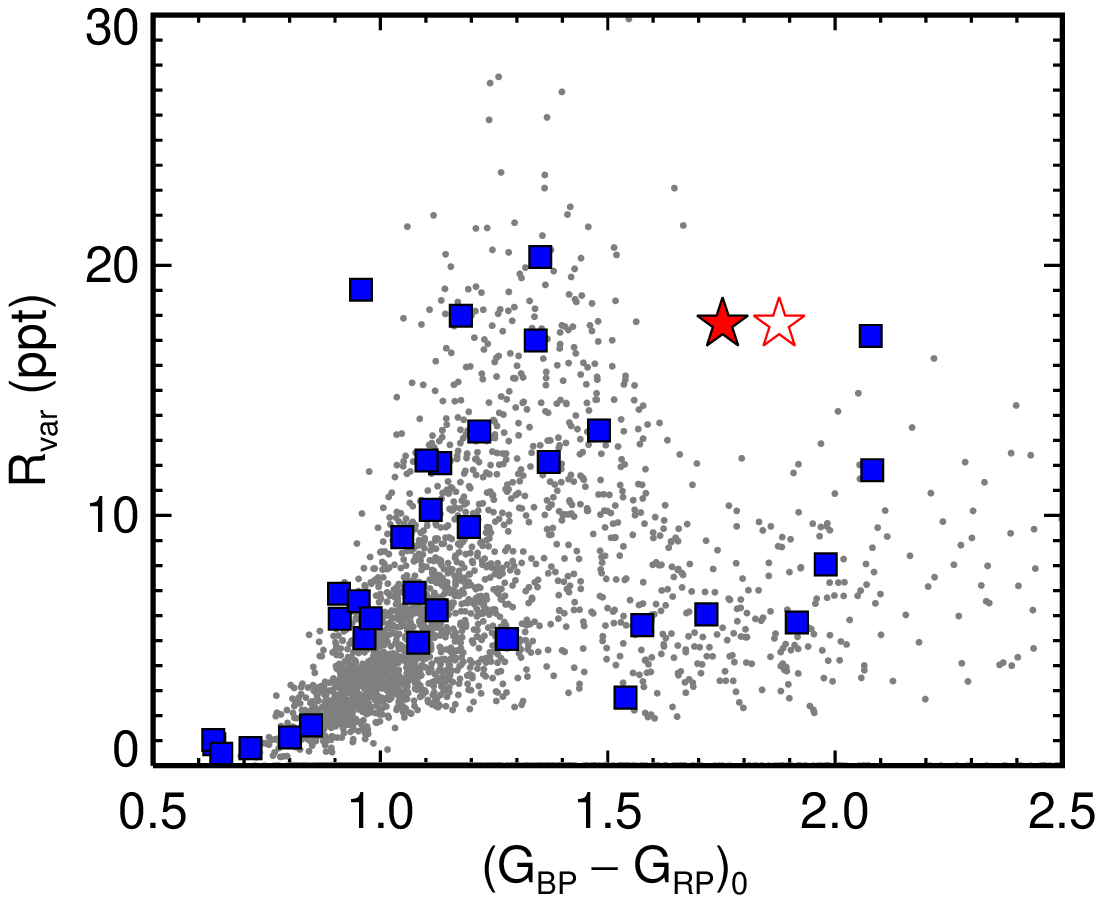}
\caption{EPIC 219489683 falls in the intermediate period gap. (\textit{Left}) Color versus period for \textit{Kepler} stars \citep[gray points;][]{AmyKepler}. 
The \textit{Kepler} colors have been de-reddened using the $E(B-V) = $ 0.04~mag\;kpc$^{-1}$ law, which approximates the median reddening pattern derived in Appendix~\ref{a:kepler}, and the following extinction coefficients: $E(G_\text{BP} - G_\text{RP}) = 0.415 \; A_V = 0.134 \; E(B-V).$
The black lines trace $Ro = $0.45 and 0.50 using \citet{Cranmer2011} convective turnover times, and approximately trace the intermediate period gap.
The benchmark rotators for Ruprecht 147 (blue squares, dereddened using $A_V = 0.30$~mag) intersect this gap, and 
EPIC 219489683 (marked with a filled red star with $A_V = 0.3$ applied) falls right in the middle of the gap---it is also marked with an open red star with no reddening correction applied. This illustrates how essential accurate and precise reddening corrections are for analyzing such features in color--period distributions.
(\textit{Right}) Color versus $R_{\rm var}$ for \textit{Kepler} stars with periods within 10\% of the Ruprecht 147 sequence model. The benchmark rotators for Ruprecht~147 approximately follow the same distribution, with one notable exception: EPIC 219489683 has an abnormally large $R_{\rm var}$ for its color and period. The light curve for the outlying blue square at ($\gbr_0 \approx 0.96$, $R_{\rm var} \approx 19$ ppt) shows significant spot evolution; the amplitude for the later 2/3 of the light curve is more consistent with the \textit{Kepler} and Ruprecht~147 distributions.
\label{f:a:gap}}
\end{center}\end{figure}

One member of Ruprecht~147, EPIC 219489683, has a color and period that appears to place it in the gap, shown in the left panel of Figure~\ref{f:a:gap}. However, interstellar reddening smears out the gap, and shifts it relative to the Ruprecht 147 sequence which has a known reddening value. If no reddening correction is applied to Ruprecht 147, then the star would fall on the edge of the gap. The stars in the \textit{Kepler} field are likely reddened somewhat less than Ruprecht 147 (see Appendix~\ref{a:kepler}), so the true location of EPIC 219489683 in this distribution is probably somewhere between the $A_V = 0.0$ and $A_V = 0.3$ cases.

\citet{Timo2019} explained that the gap coincided with a decrease in amplitude of the spot modulation signal, 
which they suggested is caused by a net cancellation of dimming from spots and brightening from plage. 
The right panel of Figure~\ref{f:a:gap} plots the $\gbr_0$ vs.~$R_{\rm var}$ for a selection of \textit{Kepler} stars drawn from \citet{AmyKepler} with periods within 10\% of the Ruprecht 147 sequence model. 
This figure also shows our $R_{\rm var}$ measurements for the Ruprecht 147 rotator sample, and highlights EPIC 219489683. We were surprised to find that it showed a large photometric amplitude compared to similar stars in the \textit{Kepler} field, defying expectations from \citet{Timo2019}.
Examining the magnetic activity of this star, and other Ruprecht~147 stars expected to fall in and near the gap, might shed light on the mechanism responsible for carving out this feature in the color--period distribution.

\subsection{Estimating \prot\ uncertainties} \label{a:eprot}

\citet{Lamm2004}, \citet{Agueros2018}, and others estimated \prot\ uncertainties using the width of a Gaussian fit to the primary Lomb--Scargle periodogram peak. For example, there are five rotators with similar color in the NGC 752 sample \citep{Agueros2018}, and the mean reported error is 2.4 days.
Separately, the \prot\ dispersion as a function of color combines measurement uncertainties with the degree of convergence of the slow sequence (how intrinsically tight the slow sequence is) and the range of differential rotation in these stars.
The standard deviation for this NGC 752 sample is 1.4~days, which
suggests that the periodogram width is an overestimate of the true uncertainty for that sample. 

For both the \textit{K2} and PTF rotators, the width of the periodogram peak indicates a typical uncertainty of $\lesssim$10\% for our targets. 
Focusing on the slowly rotating sequence cooler than $\teff < 5500$~K, \prot = 22.6$\pm$1.7~days (8\%).
Finally, there are three stars for which we have measured \prot\ with both \textit{K2} and PTF---the differences in the periods measured from each survey are 0.1, 0.4, and 1.6 days, or $<$7\%. For this work, given that we observe only two to four full rotation cycles in the light curves, we adopt a 10\% uncertainty for all rotators, which amounts to $\approx$15\% in gyrochronal age per star.

\subsection{Notes on additional stars} \label{a:othernotes}

\textit{EPIC 219590752}---The \textit{Gaia} DR2 RV error is 4.8~\kms, which according to our criteria would classify it as a candidate short-period binary. However, HARPS RVs collected over 400 days have a dispersion of 18~\mps, indicating that if this is a spectroscopic binary, it must have a long orbital period. We reclassify it from a short- to long-period SB1. 

\textit{EPIC 219582840---}This star has an RV variance-to-error ratio $e/i = 5.4$, 
which classifies it as a candidate short-period binary according to our $e/i > 4$ threshold.
However, two MIKE RVs taken 18 days apart differ by only 1.5~\kms, which is consistent within the measurement uncertainties. 
Considered together with Hectochelle RVs taken two years earlier, the difference of 7.5~\kms\ confirms it is an SB1. We reclassify it as a long-period SB1. We acknowledge that our classifications could be inaccurate until orbital solutions are available for these stars.

 
\section{Fundamental Properties of NGC 6819} \label{a:6819}

NGC~6819 is a rich, well-studied cluster. 
\citet{CG2018} identified 
1915 members with \textit{Gaia} DR2 data, 
and \citet{Milliman2014} reported 566 RV-single members and 93 spectroscopic binary members.

Metallicity measurements in the recent literature  
range between [Fe/H] = $-0.06$ and $+0.10$~dex 
and have an average value of $+0.03$~dex
\citep{Bragaglia2001, Friel2002, Marshall2005, Twarog6819, LeeBrown2015, Milliman2015, Ness2016, Slumstrup2017, Cummings2018, AnthonyTwarog2018, Deliyannis2019}. 
We also cross-matched the 
\citet{CG2018} cluster membership catalogs 
for NGC~6819 and M67 
with the LAMOST DR4 catalog \citep{lamost2015} and 
calculated metallicities for each cluster (median and standard deviation): 
for NGC~6819 we found [Fe/H] = $+0.05 \pm 0.08$ for 54 stars, and
for M67 we found [Fe/H] = $+0.05 \pm 0.04$ for 10 stars,
indicating they they share a similar metallicity. 
M67 is commonly found to have a metallicity and abundance pattern 
similar to the Sun \citep{M67SolarTwin, Liu2016M67}.
Assuming a solar metallicity would make the 
metallicity of NGC 6819 0.1~dex lower than our value for Ruprecht~147.
The SDSS Data Release 16 includes a value added catalog 
from the Open Cluster Chemical Analysis and Mapping survey \citep[OCCAM;][]{Donor2018, Donor2020},\footnote{\url{https://www.sdss.org/dr16/data_access/value-added-catalogs/?vac_id=open-cluster-chemical-abundances-and-mapping-catalog}}
which includes stellar properties and chemical abundances for open cluster members 
based on APOGEE spectra and \textit{Gaia} DR2 astrometric membership
\citep{APOGEE16}: 
for NGC~6819, [Fe/H] = $+0.05 \pm 0.04$~dex 
and for Ruprecht~147, [Fe/H] = $+0.12 \pm 0.02$~dex (33 stars).
The metallicities appear to be similar enough ($\lesssim$0.1~dex) 
that any minor differences should have negligible impact on this study.

The average interstellar reddening of NGC 6819 has been constrained to 
$E(B-V) =$~0.14-0.17 \citep{Bragaglia2001, Twarog6819, Cummings2018, Deliyannis2019}, 
and it varies significantly across the cluster 
\citep[$\Delta E(B-V) \approx$~0.05-0.06;][]{Platais6819,Twarog6819}.
The distance modulus is $(m - M)_0 =$~11.90-11.94 \citep{Yang2013, Twarog6819, Cummings2018}.
The age is 2.3-2.6~Gyr \citep{Yang2013, Twarog6819, Cummings2018, Brewer6819, Soydugan2020}, 
which is approximately coeval with the 2.7~Gyr Ruprecht~147.

\subsection{The relative reddening of NGC~6819 compared to Ruprecht~147 using the red clump}
Before joining the rotation samples for Ruprecht~147 and NGC~6819, we must 
determine if the mean interstellar reddening values are determined consistently. To do this, we make use of each cluster's red clump population.
Red clump giants can serve as distance indicators because of 
their low intrinsic luminosity dispersion compared to other giants. This is especially true when infrared or near-infrared magnitudes are used because these are relatively insensitive to interstellar dust and stellar metallicity \citep{Paczynski1998, Churchwell2009, Hawkins2017}. 

The red clumps in NGC~6819 and Ruprecht~147 should have essentially identical magnitudes in all bands due to 
their similar metallicity and age, 
except for the fact that NGC~6819 is much more distant and suffers greater 
interstellar reddening and extinction than the stars of Ruprecht~147.
We can use this fact to calculate the relative distance modulus and extinction for NGC~6819 by comparing its red clump to that of Ruprecht~147.

\citet{Stello2011} identified red clump members of NGC~6819 via asteroseismic analysis of \textit{Kepler} light curves
(in their table, class = clump, asteroseismic member = yes). 
We trimmed that sample by selecting those consistent with single-star membership according to the \citet{Hole2009} 
RV catalog (bin = SM), 
and rejected those identified as known or possible blends, 
leaving us with ten stars.
We list their \textit{Gaia} DR2 and KIC IDs in Table~\ref{t:clump}, 
along with photometric magnitudes from \textit{Gaia} DR2 
\citep[$G, G_{\rm BP}, G_{\rm RP}$;][]{DR2phot2},
the Two Micron All Sky Survey \citep[2MASS; $J, H, K_S$;][]{2MASS}, and 
the \textit{Wide-field Infrared Survey Explorer} \citep[\textit{WISE}; $W1, W2, W3$;][]{WISE}.
Table~\ref{t:clump} also identifies 
five RV-single members of Ruprecht~147 that appear to be red clump stars, 
based on their \textit{K2} asteroseismic power spectra (Lund et al., in prep).

\begin{deluxetable}{ccDDDDDDDDD}
\tablecaption{Data for red clump members of NGC~6819 and Ruprecht~147 \label{t:clump}}
\tablewidth{0pt}
\tablehead{
\colhead{\textit{Gaia} DR2 Source ID} & \colhead{KIC / EPIC} & 
\multicolumn2c{$G$} & \multicolumn2c{$G_{BP}$} & \multicolumn2c{$G_{RP}$} & 
\multicolumn2c{$J$} & \multicolumn2c{$H$} & \multicolumn2c{$K_S$} & 
\multicolumn2c{$W1$} & \multicolumn2c{$W2$} & \multicolumn2c{$W3$}
}
\decimals
\startdata
\sidehead{\textit{NGC 6819:}}
2076300101294355968 & 5024327 & 12.761 & 13.390 & 12.020 & 10.992 & 10.450 & 10.323 & 10.154 & 10.265 & 10.294 \\
2076392528993233024 & 5024967 & 12.803 & 13.444 & 12.053 & 11.005 & 10.471 & 10.323 & 10.067 & 10.496 & 10.324 \\
2076488323940775680 & 5111949 & 12.767 & 13.395 & 12.024 & 11.009 & 10.464 & 10.346 & 10.233 & 10.341 & 10.392 \\
2076488049062931328 & 5112288 & 12.788 & 13.423 & 12.038 & 11.026 & 10.460 & 10.337 & 10.229 & 10.324 & 10.491 \\
2076487838597288320 & 5112373 & 12.766 & 13.402 & 12.016 & 10.990 & 10.393 & 10.312 & 10.195 & 10.275 & 10.396 \\
2076487872957025408 & 5112387 & 12.786 & 13.424 & 12.036 & 11.005 & 10.448 & 10.341 & 10.237 & 10.323 & 10.387 \\
2076581919866484224 & 5112401 & 12.572 & 13.208 & 11.824 & 10.804 & 10.230 & 10.081 & 10.001 & 10.116 & 10.200 \\
2076393937742523904 & 5112730 & 12.729 & 13.365 & 11.979 & 10.939 & 10.400 & 10.250 & 10.164 & 10.254 & 10.220 \\
2076393662864626176 & 5112950 & 12.715 & 13.367 & 11.954 & 10.922 & 10.355 & 10.207 & 10.083 & 10.187 & 10.040 \\
2076582710140560000 & 5200152 & 12.678 & 13.289 & 11.952 & 10.974 & 10.425 & 10.306 & 10.193 & 10.317 & 10.528 \\
\sidehead{\textit{Ruprecht 147:}}
4084757199808535296 & 219239754 &  8.037 &  8.627 &  7.346 &  6.406 &  5.883 &  5.735 &  5.743 &  5.691 &  5.711 \\ 
4183930438518525184 & 219624547 & 8.247 & 8.878 & 7.528 & 6.537 & 6.022 & 5.834 & 5.783 & 5.772 & 5.813 \\ 
4087762027643173248 & 219310397 & 8.028 & 8.620 & 7.337 & 6.425 & 5.926 & 5.718 & 5.626 & 5.664 & 5.699 \\ 
4184125807991900928 & 219614490 & 8.124 & 8.735 & 7.416 & 6.448 & 5.939 & 5.787 & 5.666 & 5.685 & 5.733 \\ 
4184137077986034048 & 219704882 & 8.025 & 8.660 & 7.311 & 6.401 & 5.801 & 5.633 & 5.615 & 5.528 & 5.576 \\ 
\enddata
\tablecomments{Photometry from \textit{Gaia} DR2, 2MASS, and \textit{WISE}.}
\end{deluxetable}

We computed median apparent magnitudes in each band for each cluster, 
then calculated the distance and extinction values needed to balance their absolute magnitudes, 
finding $\Delta (m - M)_0 = 4.57$ and $\Delta A_V = 0.14$.
Applying our adopted values for Ruprecht~147 ($(m - M)_0 = 7.4, A_V = 0.30$), 
we find for NGC~6819 $(m - M)_0 = 11.97, A_V = 0.44$.
These values are consistent with the literature values summarized previously, and we adopt our value for this study.

\begin{figure*}
\begin{center}
\includegraphics[trim=1.1cm 0.4cm 0.0cm 0.5cm, clip=True,  width=3.5in]{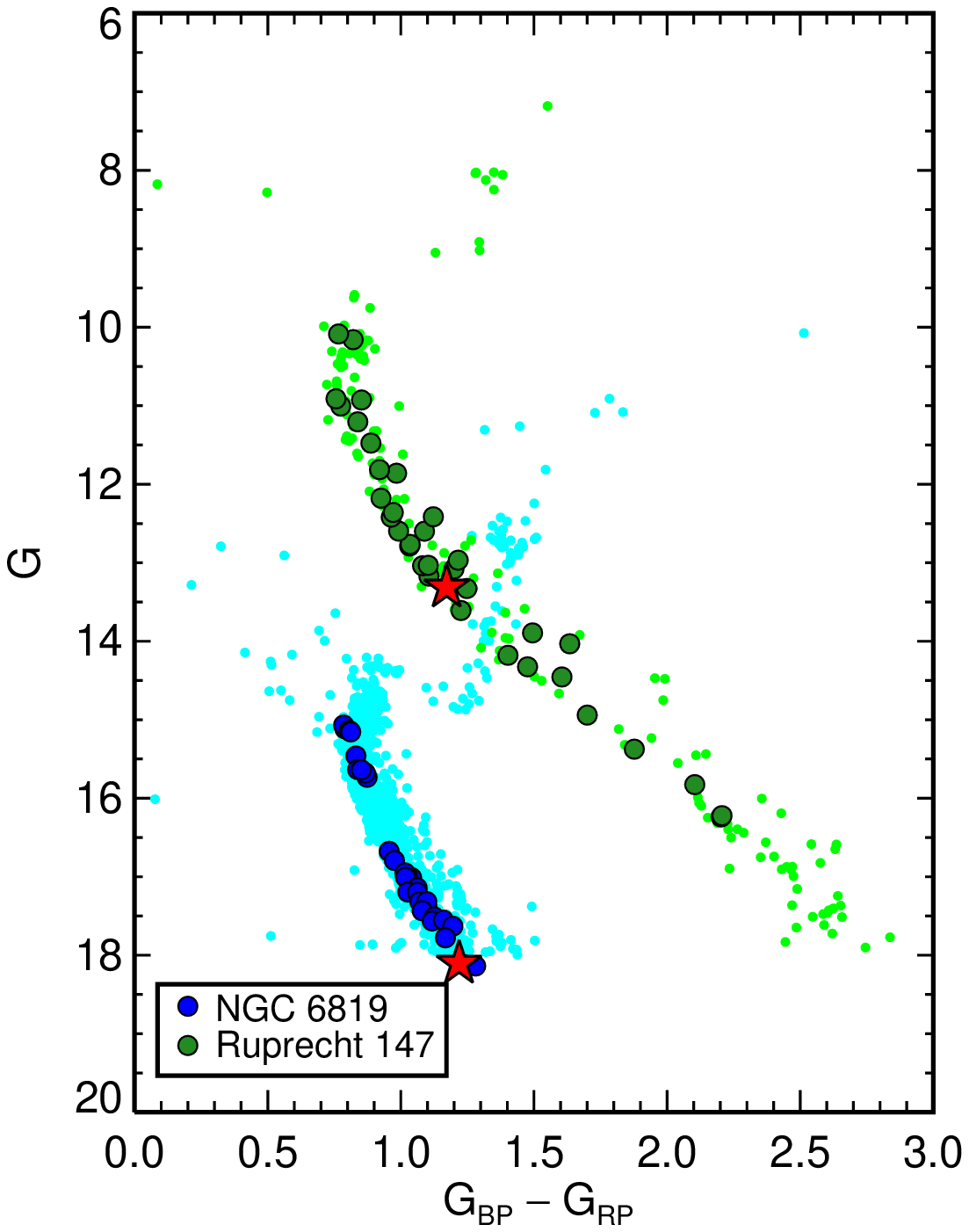}
\includegraphics[trim=0.9cm 0.4cm 0.0cm 0.5cm, clip=True,  width=3.4in]{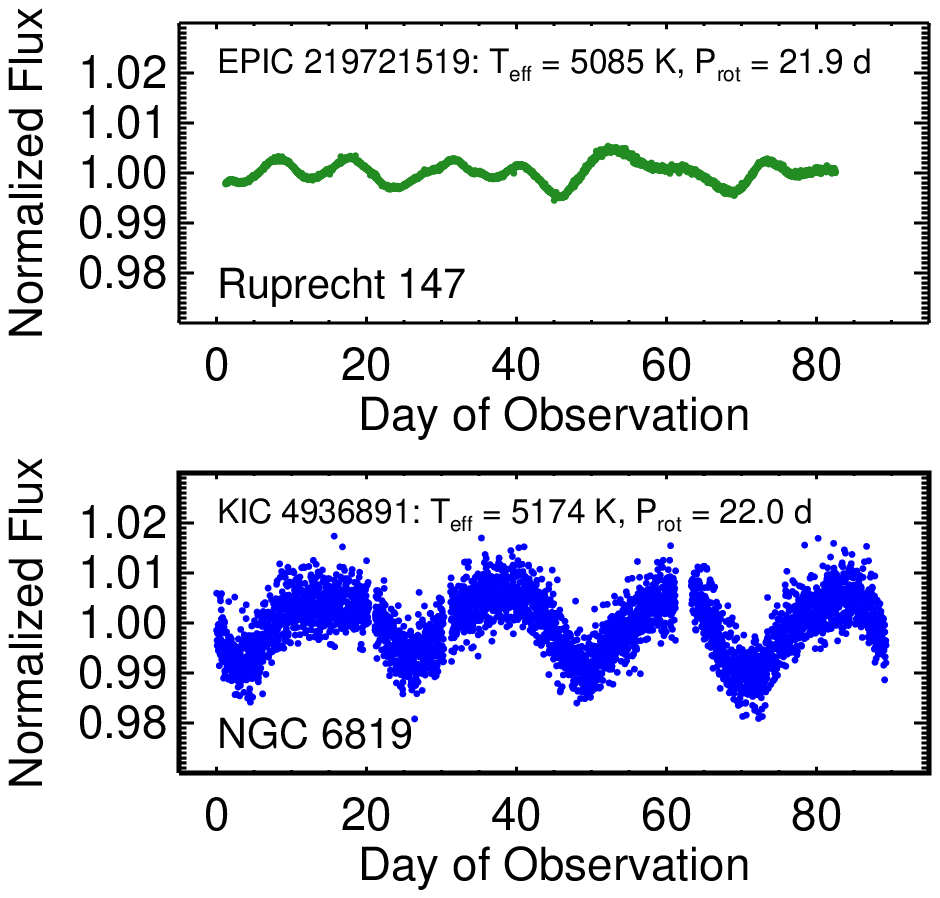}
\caption{\textit{Left---}\textit{Gaia} DR2 CMDs for NGC~6819 (cyan points) and Ruprecht~147 (bright green points), and their rotators (blue and dark green points, respectively).   Two stars with similar effective temperatures and rotation periods are marked in each cluster with red stars. \textit{Right---}The \textit{K2} light curve for EPIC~219721519, in Ruprecht~147 (\textit{top}; $\teff = 5085$~K, $\prot = 21.9$~d), and the \textit{Kepler} light curve for  KIC~4936891, in NGC~6819 (\textit{bottom}; $\teff = 5174$~K, $\prot = 22.0$~d), 
the red star symbols marked in the CMD. While the \prot\ are evident in each panel, the noise is significantly increased in the NGC~6819 light curves because stars in that cluster are much fainter and more extinguished ($\Delta (m - M)_V = 4.71$; i.e., $\approx$77$\times$ fainter). Conversely, systematics are typically much stronger in the Ruprecht~147 light curves due to the increased pointing instability. 
\label{f:compare}}
\end{center}\end{figure*}

Figure~\ref{f:compare} plots the apparent CMDs for NGC 6819 and Ruprecht 147, and highlights the stars with measured rotation periods. 
This figure panel also includes  representative light curves from each cluster, which illustrates the increased photometric noise for the much fainter members of NGC 6819 relative to our Ruprecht 147 targets.

\section{The \textit{Kepler} \prot\ dataset} \label{a:kepler}

\citet{AmyKepler} published \prot\ for 34,030 stars 
in the \textit{Kepler} field using the ACF technique.
M.~Bedell cross-matched the \textit{Kepler} Input Catalog (KIC) 
with \textit{Gaia} DR2.\footnote{\url{https://gaia-kepler.fun}}
We downloaded the \textit{Kepler}--\textit{Gaia} cross-match table produced using a 
1$''$ search radius, 
and then matched it with the \citet{AmyKepler} catalog 
according to the KIC IDs.
Next, we trimmed the sample in the following ways, 
which resulted in 16,259 nearby rotators 
cleaned of some categories of binaries and evolved stars:
\begin{itemize}
    \item Photometric binaries: We derived an empirical main sequence by fitting a cubic basis spline to the Hyades CMD, 
    and for each \textit{Kepler} rotator, 
    we calculated the difference between its absolute magnitude (using the parallax for the distance correction) and the expected $M_G$ based on its \gbr\ color and the Hyades CMD model. Stars with photometric excesses $\Delta G > 0.5$ were trimmed.
    \item Distant stars: Stars beyond 1~kpc were trimmed.
    \item Astrometric binaries: Stars with excess astrometric noise indicative of binarity following the criteria we applied for the Ruprecht~147 sample were trimmed 
    (i.e., $\epsilon_i > 0$ and $D > 6$).
    \item Spectroscopic binaries: Stars with \textit{Gaia} DR2 $\sigma_{\rm RV} > 3$~\kms\ were trimmed.
\end{itemize}

Interstellar reddening can smear out intrinsically sharp structure 
in the color--period diagram, 
which is why we have restricted the sample to $<1000$~pc. 
To assess the impact of reddening on this nearby subset, we used 
the catalog of spectroscopic parameters for the 
California--Kepler Survey stars 
observed with HIRES \citep[CKS;][]{CKS:Petigura2017}
and analyzed with SME by \citet{BrewerCKS}.
These stars tend to be much more distant than those in the CPS 
sample \citep{Brewer2016} 
that we used to construct our color--temperature relation
(664 pc vs 45 pc based on the median parallax of each sample).
However, since the spectra were collected with the same instrument and setup, 
and analyzed with the same procedure, we expect that our color--temperature 
relation can be used to accurately predict the intrinsic color for the more distant CKS stars, allowing us to estimate the amount of interstellar reddening toward each star.

\begin{figure}\begin{center}
\includegraphics[trim=0.0cm 0.0cm 0.0cm 0.0cm, clip=True,  width=3.4in]{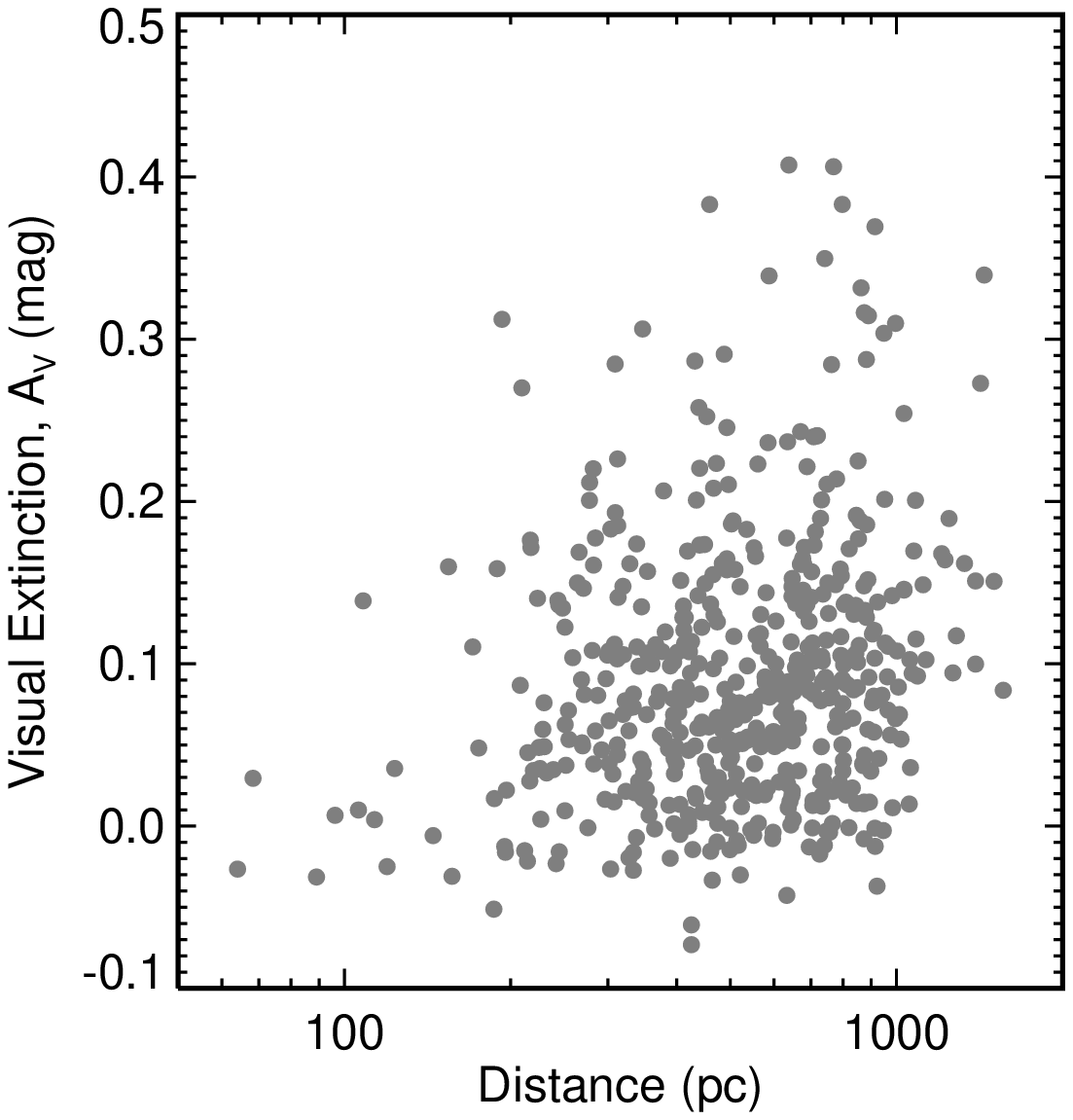}
\includegraphics[trim=0.0cm 0.0cm 0.0cm 0.0cm, clip=True,  width=3.4in]{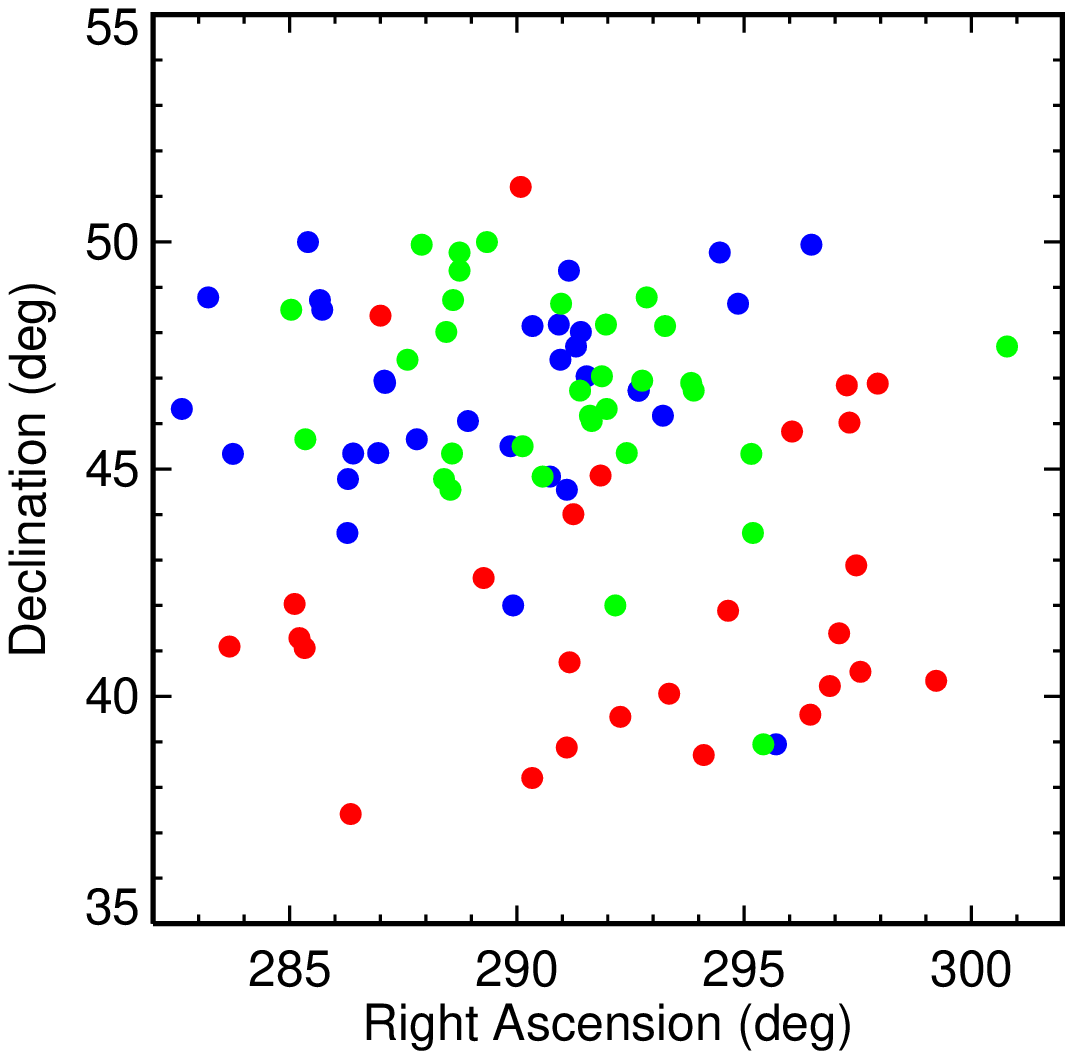}
  \caption{Interstellar reddening/extinction in the \textit{Kepler} field. 
  (\textit{left}) 
  Using spectroscopic \teff\ for CKS stars observed with HIRES \citep{BrewerCKS}, 
  we calculate the amount of reddening with our empirical color--temperature relation, then plot the results as a function of distance (using the inverted parallax). Within 150~pc, the reddening is negligible, and it 
  increases with distance. 
  (\textit{right}) The equatorial coordinates for 
  the subset of this sample at $d = 800 \pm 100$~pc are plotted
  and color-coded according to the amount of reddening 
  (blue: $A_V < 0.05$; green: $0.05 < A_V < 0.15$; red: $A_V > 0.15$).
  The reddening varies coherently across the \textit{Kepler} field, 
  indicating the presence of foreground clouds.
  The scatter in $A_V$ shown in the left panel appears to 
  be primarily caused by differential reddening, not 
  measurement uncertainties.
  The median trend is approximately $E(B-V) = 0.04$~mag\;kpc$^{-1}$, or $A_V = 0.124$ mag\;kpc$^{-1}$.
   \label{f:kepred}}
\end{center}\end{figure}

We find negligible reddening for stars within 150~pc, 
shown in Figure~\ref{f:kepred}.
At 500~pc, the median value is $E(B - V) = 0.017$ (10\ith\ and 90\ith\ percentiles are 0.00 and 0.06), corresponding to $A_V = 0.054$ (0.00, 0.19).
By 800~pc, the median reddening (extinction) has increased to 
0.031 (0.096).  
At this distance, the reddening coherently varies across the \textit{Kepler} field (right panel of the same figure), 
indicating that discrete cloud structures in the foreground are responsible 
for the larger scatter in the distance vs $A_V$ diagram. 
This median reddening is approximately $E(B-V) \approx 0.04 $~mag\;kpc$^{-1}$. The large variation across the field with this relatively small total value supports our decision to restrict the \textit{Kepler} sample to $<$1~kpc; i.e., the differential reddening pattern is real, but the impact is not large so intrinsically sharp features in the color--period distribution will not be smeared out much.

\section{A color- and time-dependent braking index}\label{a:brake}

\citet{Barnes2003} proposed that stars spin down such that 
their \prot\ could be described by a simple function that had 
separable dependencies on mass (using photometric color as a proxy) and time: 
$\prot = f(M_\star, t) = g(M_\star) \times h(t)$. 
Following \citet{skumanich1972}, $h(t) = t^n$; i.e., stars spin down 
with a common braking index that is constant in time.
Although the cluster \prot\ data demonstrate that stars do not spin down continuously, it is not yet clear if stars share a common post-stalling braking index, or if it is constant in time (i.e., Skumanich-like). 

In Section~\ref{s:resume}, we applied such a Skumanich-like braking index with the value $n=0.62$ to 
calculate the ``resume time,'' $t_R$, when stars resume spinning down after exiting the stalling phase. 
This particular value of the braking index was tuned by \citet{Douglas2019} using solar-color Praesepe stars and the Sun.
In this appendix, we will repeat this exercise using various pairings of benchmark populations, 
including Praesepe, NGC~6811, NGC~752, the joint NGC~6819 + Ruprecht~147 sample, and four old K dwarfs.
We will use polynomial fits to the \prot\ sequences for these various clusters, provided in Table~\ref{t:models}.

\subsection{NGC~752 K dwarfs compared to Ruprecht~147} \label{a:752}

For NGC~752, \citet{Agueros2018} presented rotation periods for five mid-K dwarfs 
with \textit{Gaia} DR2 astrometry consistent with membership.\footnote{Listing Gaia DR2 IDs with \prot\ in parenthesis, 
these stars are 
342523783591381504 (13.0 days),
342907478789972864 (14.0 days),
342854392994216064 (14.0 days),
342869957955643264 (16.6 days), and
342889611726120832 (13.9 days).}
They have $\teff = 4450 \pm 60$~K and $M_\star \approx$0.72-0.75~\msun, 
with $\prot = 14.0 \pm 1.4$~days. For Ruprecht~147, there are two rotators with these approximate colors, 
assuming $A_V = 0.1$ for NGC~752: 
EPIC~219651610 (featured in Figures~\ref{f:k2} and \ref{f:ptf}) and EPIC~219722781. These two stars have an average $\prot = 21.0$~days.\footnote{We can also fit a linear relation for the rotators with $1 < \gbr_0 < 1.8$ (15 stars) to estimate the 2.7~Gyr \prot\ at the median color for the NGC~752 mid-K dwarfs, 
and we find $\prot = 20.9$~days, in agreement with the two stars most similar in color to the NGC~752 stars.}

The average braking index for these stars between 1.4 and 2.7~Gyr is therefore 
$n = 0.62$$\pm$0.1. This is consistent with the value found with Praesepe and the Sun, 
although the precision is limited by uncertainties in the $\prot$ for these K~dwarfs and the uncertainties in the ages of each cluster.
$\prot$ for additional members of NGC~752, 
and/or for an even older cluster, would be helpful for further testing this hypothesis and improving the precision of this calculation.

\subsection{NGC~6819 + Ruprecht~147 relative to NGC~6811}

Comparing the eight solar analog rotators in NGC~6819 ($\teff = $ 5600-5800~K) 
to the NGC 6811 sequence (1 Gyr), we find $n = 0.59 \pm 0.02$, which is 
close to the Praesepe--Sun value. However, 
extending this analysis to the full NGC~6819 and Ruprecht~147 sample, 
we find that stars with $4600 < \teff < 5600$~K (27 stars) have $n = 0.73 \pm 0.04$, which suggests that the post-stalling braking index does have some dependence on mass, or that the post-stalling braking index varies over this time interval. 

Regarding the nine stars cooler than 4600~K, the braking index decreases away 
from the $n = 0.73$ plateau toward cooler temperatures, reaching $n \approx 0.2$ at 3700~K. 
Presumably, the time-averaged braking index between 1 and 2.7 Gyr for these cool dwarfs is still affected by stalling. 
As \citet{Curtis2019} described, such a low braking index cannot be maintained 
throughout the age of the Universe; 
otherwise, these late-K and early-M dwarfs would not be able to spin all the way down to the 40-50 day periods observed in the field. 

\subsection{Old nearby K dwarfs relative to Ruprecht~147}

Comparing the old and nearby K dwarfs  
($\alpha$~Cen\;B, and 61~Cyg\;A and B) 
to the Ruprecht~147 slow sequence, 
we calculate $n =$ 0.58, 0.63, and 0.80, respectively
(sorted from warmest to coolest).
The warmest two are consistent with the values derived 
with the following pairs: 
Praesepe and the Sun, NGC~752 and Ruprecht~147, and 
the solar analogs of NGC~6819 and NGC~6811. 
Therefore, the gyrochronology ages inferred by projecting 
the Ruprecht~147 sequence forward in time with a $n = 0.62$ braking law
are consistent with the 6~Gyr ages we adopted 
at $-$5.4\% and 1.5\%. 

However, the value we find for 61~Cyg\;B is 30\% larger than the mean for the other three K dwarfs, and its gyrochronology age is 26.3\% larger than the \citet{Kervella2008} result.
At its temperature (4044~K), the Ruprecht~147 sequence appears to dip down 
as it crosses the \textit{Kepler} intermediate period gap.

Establishing additional cool and old benchmarks for stellar rotation will be critical for deciphering the rotational evolution of such stars.
For example, is this discrepancy due to inaccurate \prot\ or age parameters 
for 61~Cyg\;B? A mean value of $n = 0.62$ for the pair can be reached by increasing the age 61~Cyg by 0.8 Gyr (within the 1 Gyr uncertainty). 
Or is the braking history more complex than we have assumed?

\subsection{The slow period edge in the \textit{Kepler} distribution relative to Ruprecht~147}

The \prot\ distribution for \textit{Kepler} field stars has a well-defined 
upper envelope. 
\citet{vanSaders2018} divided this feature into warm and cool portions, 
separated at $\teff \approx $5000~K.
The warm side is steeply sloped, and is probably shaped by 
a detection threshold and/or a final shutdown of magnetic braking 
\citep{vanSaders2016}. 

Stars cooler than 5000~K are not old enough to be affected by these 
proposals, so the upper edge might represent the oldest stars in the 
\textit{Kepler} sample.
If we assume those stars are 10 Gyr old (or 13.7 Gyr old), 
then it would take a time-averaged braking index of $n = 0.55$ (or 0.45) to project the Ruprecht~147 sequence forward in time to match this edge,
similar to the classic Skumanich law.

\subsection{Summary of calculations for the post-stalling braking index}

All values calculated from these various samples range  between $n \approx$ 0.5--0.8, summarized below:

\textit{Solar analogs---}All samples indicate $n \approx 0.6$, 
including Praesepe and the Sun, NGC 6811 and the Sun, NGC 6811 and NGC 6819.

\textit{Early-K dwarfs---}The NGC~6819 + R147 joint sample compared to NGC~6811 yielded $n = 0.73$; 
however, we found $n \approx 0.6$ from the comparison between  the old K dwarfs in the field with the R147 sequence. This suggests that stars of a given mass might not spin down with a single-valued post-stalling braking index 
that is constant in time.

\textit{Mid-K dwarfs---}The R147 stars compared to the NGC 6811 sequence, the NGC 752 stars compared to the R147 sequence, and 61~Cyg\;A compared to the R147 sequence all returned $n \approx 0.6$. 

\textit{Late-K to early-M dwarfs---}The comparison between R147 stars and the NGC 6811 sequence is likely biased by stalling ($n \approx$ 0.3-0.4). 
Comparing 61~Cyg\;B with the R147 sequence returned one of the highest values of $n \approx 0.8$. 
Finally, projecting the R147 sequence forward in time with $n = 0.5$ reaches the upper edge of the Kepler field in $\approx$11~Gyr.


\bibliographystyle{aasjournal}





\end{document}